%% file: main.tex
\newif\ifllncs
 \llncsfalse

\ifllncs
\documentclass[envcountsame]{llncs}

\else
\documentclass[letterpaper,11pt]{article}
\usepackage{amsthm}
\usepackage{fullpage}
\usepackage{authblk}
\fi

\usepackage{amsmath,amsfonts,amssymb}
\usepackage{breakcites}
\usepackage[pdfstartview=FitH,colorlinks,linkcolor=blue,citecolor=blue]{hyperref} 
\usepackage[margin=1cm]{caption}
\DeclareMathAlphabet{\mathbbold}{U}{bbold}{m}{n}

\usepackage{graphicx} 

\usepackage{mdframed}

\title{Unclonable Cryptography in Linear Quantum Memory\thanks{This work is subsumed by \cite{huang2026uncloneable}.}}

\ifllncs
\author{Omri Shmueli \and Mark Zhandry} 
\date{} 
\institute{NTT Research} 

\else
\author[1]{Omri Shmueli}
\author[2]{Mark Zhandry}
\affil[1]{NTT Research\footnote{This work was done in part while the author was a research fellow at the Simons Institute for the Theory of Computing at UC Berkeley.}}
\affil[2]{Stanford University\footnote{Work performed while also at NTT Research.}}
\date{}
\fi

\input{macros}

\begin{document}
\maketitle

\begin{abstract}
Quantum cryptography is a rapidly-developing area which leverages quantum information to accomplish classically-impossible tasks. In many of these protocols, quantum states are used as long-term cryptographic keys. Typically, this is to ensure the keys cannot be copied by an adversary, owing to the quantum no-cloning theorem. Unfortunately, due to quantum state's tendency to decohere, persistent quantum memory will likely be one of the most challenging resources for quantum computers. As such, it will be important to minimize persistent memory in quantum protocols.

In this work, we consider the case of one-shot signatures (OSS), and more general quantum signing tokens. These are important unclonable primitives, where quantum signing keys allow for signing a single message but not two. Naturally, these quantum signing keys would require storage in long-term quantum memory. Very recently, the first OSS was constructed in a classical oracle model and also in the standard model, but we observe that the quantum memory required for these protocols is quite large. In this work, we significantly decrease the quantum secret key size, in some cases achieving asymptotically optimal size. To do so, we develop novel techniques for proving the security of cryptosystems using coset states, which are one of the main tools used in unclonable cryptography.
\noindent
\end{abstract}

\ifllncs
\pagestyle{plain}
\else

\thispagestyle{empty}
\newpage
\tableofcontents
\newpage
\thispagestyle{empty}

\fi

\input{introduction}

\ifllncs
\bibliographystyle{alpha}
\bibliography{bib,abbrev0,crypto}
\appendix
\fi

\ifllncs
\else

\input{crypto_tools}

\input{oracle_construction}

\input{standard_construction}

\input{standard_model_2_to_1_reduction}

\fi

\ifllncs\else
\bibliographystyle{alpha}
\bibliography{bib,abbrev0,crypto}
\fi

\end{document}

%% file: macros.tex
\ifllncs
\else
\newtheorem{theorem}{Theorem}
\newtheorem{lemma}[theorem]{Lemma}
\newtheorem{remark}[theorem]{Remark}
\newtheorem{corollary}[theorem]{Corollary}
\newtheorem{definition}[theorem]{Definition}

\newtheorem{claim}[theorem]{Claim}
\fi
\newtheorem{construction}[theorem]{Construction}
\newmdtheoremenv{boxclaim}[theorem]{Claim}

\newcommand{\Z}{\mathbb{Z}}

\newcommand{\Id}{\mathbf{I}}

\newcommand{\As}{\mathcal{A}}

\newcommand{\Ds}{\mathcal{D}}
\newcommand{\Cs}{\mathcal{C}}

\newcommand{\Os}{\mathcal{O}}
\newcommand{\Ps}{\mathcal{P}}

\newcommand{\Ss}{\mathcal{S}}

\newcommand{\Us}{\mathcal{U}}
\newcommand{\Xs}{\mathcal{X}}
\newcommand{\Ys}{\mathcal{Y}}

\newcommand{\negl}{\mathsf{negl}}
\newcommand{\poly}{\mathsf{poly}}
\newcommand{\polylog}{\mathsf{polylog}}

\newcommand{\Span}{\mathsf{Span}}
\newcommand{\colspan}{\mathsf{ColSpan}}
\newcommand{\rowspan}{\mathsf{RowSpan}}

\newcommand{\prp}{{\Pi}}

\newcommand{\punc}{{\sf Punc}}
\newcommand{\eval}{{\sf Eval}}
\newcommand{\prf}{{\sf F}}

\newcommand{\permute}{{\sf Permute}}
\newcommand{\iO}{{\sf iO}}
\newcommand{\gen}{{\sf Gen}}
\newcommand{\OSSgen}{{\sf TokenGen}}
\newcommand{\hash}{{\sf Hash}}
\newcommand{\sign}{{\sf Sign}}
\newcommand{\ver}{{\sf Ver}}
\newcommand{\setup}{{\sf Setup}}

\newcommand{\crs}{{\sf CRS}}
\newcommand{\pk}{{\sf pk}}
\newcommand{\sk}{{\sf sk}}

\newif\ifnotes
\notesfalse

\ifnotes

\newcommand{\mz}[1]{$\ll$\textsf{\color{red} Mark: { #1}}$\gg$}

\else
\newcommand{\mz}[1]{}
\fi

\newcommand{\Hyb}{\mathsf{Hyb}}
\newcommand{\Oracle}{\mathcal{O}}
\newcommand{\Nat}{\mathbb{N}}
\newcommand{\bbZ}{\mathbb{Z}}
\newcommand{\Img}{\text{Img}}
\newcommand{\zo}{\{ 0, 1 \}}
\newcommand{\outSize}{\mathbf{out}_{L}}
\newcommand{\inSize}{\mathbf{in}_{L}}
\newcommand{\eSize}{\mathbf{E}_{L}}

\newcommand{\PRP}{\mathsf{PRP}}
\newcommand{\PRF}{\mathsf{PRF}}
\newcommand{\Adv}{\mathcal{A}}
\newcommand{\AdvB}{\mathcal{B}}
\newcommand{\secp}{\lambda}
\newcommand{\Obf}{\mathsf{O}}
\newcommand{\td}{\mathsf{td}}
\newcommand{\OWF}{{\sf OWF}}

\newcommand{\Enc}{\mathsf{Encode}}

\newcommand{\LF}{\mathsf{LF}}
\newcommand{\LFGen}{\LF\mathsf{.KeyGen}}
\newcommand{\LFF}{\LF\mathsf{.F}}

\newcommand{\hashL}{L} 
\newcommand{\hashQ}{\mathcal{Q}} 

\newcommand{\linspan}{\mathsf{Span}}

\newcommand{\matA}{\mathbf{A}}
\newcommand{\matB}{\mathbf{B}}
\newcommand{\matC}{\mathbf{C}}

\newcommand{\matI}{\mathbf{I}}
\newcommand{\matM}{\mathbf{M}}

\newcommand{\matS}{\mathbf{S}}
\newcommand{\matT}{\mathbf{T}}

\newcommand{\vecA}{\mathbf{a}}
\newcommand{\vecB}{\mathbf{b}}
\newcommand{\vecC}{\mathbf{c}}
\newcommand{\vecD}{\mathbf{d}}
\newcommand{\vecE}{\mathbf{e}}
\newcommand{\vecF}{\mathbf{f}}

\newcommand{\vecR}{\mathbf{r}}
\newcommand{\vecS}{\mathbf{s}}
\newcommand{\vecT}{\mathbf{t}}
\newcommand{\vecU}{\mathbf{u}}
\newcommand{\vecV}{\mathbf{v}}
\newcommand{\vecW}{\mathbf{w}}
\newcommand{\vecX}{\mathbf{x}}
\newcommand{\vecZ}{\mathbf{z}}

\newcommand{\ket}[1]{|{#1}\rangle}

%% file: introduction.tex
\section{Introduction}\label{sec:intro}

Quantum computers will completely upend cryptography. In the short term, quantum algorithms will render insecure all of the public key cryptosystems deployed today. In the longer term, however, quantum computers will enable never-before-possible cryptographic applications. The canonical such application is \emph{unclonable} quantum software~\cite{CCC:Aaronson09,C:ALLZZ21,TCC:LLQZ22}, which seeks to make programs uncopyable through no-cloning theorem, a uniquely quantum phenomenon which dictates that it is impossible to copy quantum states. There are numerous works considering special cases or variants of unclonable software in cryptography, where secret keys for signatures, pseudorandom functions, decryption, etc, are made unclonable (e.g.~\cite{BS23,STOC:AGKZ20,EPRINT:GeoZha20,C:Shmueli22,C:ShmZha25,C:CLLZ21,TCC:CakGoy24}). A strengthening of unclonable software is quantum one-time programs~\cite{C:BroGutSte13,STOC:GLRRV25}, which are not only unclonable, but self-destruct after evaluation. Unclonable software also arises implicitly in the study of obfuscating quantum programs~\cite{STOC:BKNY23,STOC:BarBraVai24, huang2025obfuscation, bartusek2025classical}.

\paragraph{Long-term quantum memory.} In all of the applications listed above, a key requirement is quantum memory to store the software / secret key. Importantly, this quantum memory needs to be persistent and long term, since in typical applications one aquires the software/key long before it is used. Long-term quantum memory, however, seems very difficult to achieve, as quantum states have a tendency to quickly decohere. Contrast this with short-term, or ``online'' quantum memory used during quantum computations, which just needs to last as long as the computation. We expect long-term quantum memory to eventually be possible, but we also expect it will be more expensive and more difficult to maintain than either classical or short-term quantum memory.

Therefore, an important metric for protocols leveraging quantum information, and in particular, unclonable cryptosystems, is how much long-term quantum memory is needed.

\paragraph{The quadratic memory barrier.} Unfortunately, many works, and in particular all the applications cited above, suffer from very large programs/keys. Namely, the quantum part of the program/key in these works is always at least quadratic in the security parameter $\lambda$. This is a result of these works all operating in a bit-by-bit fashion over the program's inputs. Each bit of input is assigned an unclonable quantum state that enables evaluating the program. Here, we assume that the program inputs are at least $\lambda$ bits. Moreover, in order to maintain unclonability, each of the unclonable quantum states is at least $\lambda$ bits. The resulting combined quantum states are therefore $\Omega(\lambda^2)$ bits, and other issues could make them even longer\footnote{We note that the above actually suggests quantum states of size $\Omega(n\lambda)$ where $n$ is the input length to the program. However, it is often possible to shrink the quantum state to $\Omega(\lambda^2)$ by only assigning quantum states to a $\lambda$-bit hash of the input.}.

We therefore propose breaking the quadratic long-term quantum memory barrier as an important and interesting goal.

\paragraph{The case of signature tokens and one-shot signatures.} In this work, we focus on the class of one-time programs and specifically on the case of quantum signature tokens, first defined by Ben-David and Sattath~\cite{BS23}. A quantum signature token is a secret signing key that allows for signing any message, but after signing the message, the token self-destructs. Formally, the quantum signing key $\ket{\sk}$ is sampled together with a classical public verification key $\pk$, and it is computationally impossible to find two classical signatures that are both verified with respect to the same classical key $\pk$. Such signature tokens are clearly impossible classically, but for signing tokens consisting of quantum states, the act of signing constitutes an irreversible measurement on the token, which may make it incapable of signing future messages. Here, the quantum signing key is generated by a trusted third party.

Amos, Georgiou, Kiayias, and Zhandry~\cite{STOC:AGKZ20} propose an even stronger notion called one-shot signatures (OSS). Here, the original quantum secret key may be generated by the adversary themselves, and even then, it cannot sign more than a single message. One-shot signatures in particular have numerous applications, such as blockchain-free smart contracts~\cite{STOC:AGKZ20, Sattath22}, overcoming lower-bounds in consensus protocols~\cite{Drake23}, and providing a solution to the blockchain scalability problem \cite{coladangelo2020quantum}.

In~\cite{BS23}, signature tokens are constructed in a classical oracle model. In~\cite{STOC:AGKZ20}, OSS is constructed in an oracle model, but without proof. Very recently, Shmueli and Zhandry~\cite{C:ShmZha25} construct OSS that is provably secure in a classical oracle model, or secure in the standard model under (somewhat) standard cryptographic assumptions.

Quantum signing keys inherently require $\Omega(\lambda)$ qubits, since an adversary can always brute-force search for the signing key. In all of these constructions, however, the base construction signs only a single bit. We can then upgrade this scheme to sign general messages, by signing each bit separately using a separate key. The result, is quantum signing keys of size $\Omega(\lambda^2)$. In the case of the OSS construction in~\cite{C:ShmZha25}, the signatures are even larger due to additional considerations in their construction/security proof: a token for signing a single bit takes $\Omega(\lambda^2)$ in a classical oracle model, and even larger in the standard model. This boils down to signing keys of overall size $\Omega(\lambda^3)$ for OSS to sign $\lambda$-bit messages in a classical oracle model.

\subsection{Results}
In this work, we overcome the quadratic barrier for long-term quantum memory in both signature tokens and one-shot signatures. First, in a classical oracle model we give constructions with asymptotically-optimal quantum storage $O(\lambda)$ and unconditional security:
\begin{theorem}
Relative to a classical oracle there exists secure OSS with $O(\lambda)$-sized quantum secret keys, where each quantum secret key can sign on messages of size $\lambda$ and for every $T$ and a quantum unbounded algorithm making $T$ queries to the oracles, the probability to find two different signatures corresponding to the same public verification key $\pk$ is $\frac{ \poly(T) }{ 2^{\lambda} }$, for some polynomial $\poly(\cdot)$.
\end{theorem}

In the standard model, depending on the computational assumptions used, we also obtain either $O\left( \lambda \right)$ or $O\left( \lambda^2 \right)$-sized quantum secret keys (Theorems \ref{theorem:standard_model_1}, \ref{theorem:standard_model_2} and \ref{theorem:standard_model_3}).

Our construction is an optimization of the OSS construction of \cite{C:ShmZha25}, which in turn has a specific structure that allows the sampling of signature tokens in the form of \emph{coset states}. Coset states, in turn, play a major role in quantum cryptography. For example, in all of the above-listed applications of unclonable programs, coset states are the quantum states which sit in long-term quantum memory (and take $\Omega\left( \lambda \right)$ qubits each). As a byproduct, we believe these techniques may be useful for quantum memory optimization in other applications built on similar structures, such as quantum copy protection and quantum one-time programs, and leave this as an interesting direction for future work.

\paragraph{Main Technical Contribution.}
Our results are achieved by two main technical steps, which can be viewed as two separate new techniques. First, we develop a novel quantum algorithm for signing coset states, which signs on an entire message $m \in \zo^{\lambda}$ using a single (small) quantum coset state. Our signing algorithm also signs the bits of $m$ in parallel.

The second contribution is a new hardness reduction of the OSS construction of \cite{C:ShmZha25}, which originally require their generated quantum signature tokens to have size $\Omega\left( \lambda^2 \right)$. A central object defined in the proof of \cite{C:ShmZha25} are coset partition functions (CPFs), which are the source of the size-$\Omega\left( \lambda^2 \right)$ states. Our new reduction introduces a subset of CPFs with additional algebraic structure. Our structured CPFs generally allow for a smaller quantum state ($\approx \lambda$ instead of $\approx\lambda^2$) in reductions, while maintaining the same level of provable security, thus saving a factor of $\lambda$ in quantum memory and showing that the generated coset states can be of size $O\left( \lambda \right)$.


Combining the above two steps, we are ultimately left with a OSS scheme that generates quantum signing keys of size $O\left( \lambda \right)$ for which it is $2^{\Omega(\lambda)}$-hard to forge a signature. These keys enable signing $\lambda$-bit messages, and by a parallel signing algorithm.

\subsection{Overview of Techniques} \label{section:overview}

\paragraph{What is $\lambda$, anyway?} We first make precise what it means to have $O(\lambda)$-bit quantum long-term storage. After all, we could take any scheme with $\lambda^c$ quantum storage, and simply define $\lambda'=\lambda^c$ as the new security parameter. Since the usual polynomial/negligible way of defining cryptographic notions is robust to polynomial changes in the security parameter, $\lambda'$ is a valid security parameter, and now the quantum storage is $\lambda'$, linear in the security parameter. But of course, we didn't actually change the scheme!

Instead, in order to have a robust measure of secret key size in terms of the security parameter, we follow the ``bit security'' view. Here, $\lambda$ is set to be such that there are no $2^\lambda$-time attacks. Slightly more formally, $\lambda$-bit security means that there is no attack running in time $T$ that succeeds with probability $p$ such that $T/p < 2^\lambda$. Our goal is to achieve $O(\lambda)$-qubit quantum secret keys under this notion of $\lambda$.

Put another way, the best attack time (or really, ratio of attack time to success probability) should be exponential in the quantum secret key length.

\paragraph{The OSS of~\cite{C:ShmZha25}.} We briefly recall the version of the OSS scheme of~\cite{C:ShmZha25} defined relative to an oracle. The oracle comes in three parts:
\begin{itemize}
    \item An oracle $\Ps(x)$, which outputs a string $y$ and a vector $\vecU$. There are two guarantees. First, the map $x \mapsto y$, which we will denote by $H$, is many-to-1. Second, when restricted to the set of $x$'s such that $H(x)=y$, the map $x \mapsto \vecU_{x}$ perfectly embeds the preimage set into an affine subspace $S_y$. $\Ps$ is in particular injective.
    
    \item An oracle $\Ps^{-1}\left( y,\vecU \right)$, which inverts $\Ps$. $\Ps^{-1}$ checks that $\vecU \in S_{y}$ and inverts only in that case, and outputs $\bot$ otherwise.
    
    \item An oracle $\Ds\left( y, \vecV \right)$, which checks if $\vecV \in S_y^\perp$, the subspace dual to $S_y$.
\end{itemize}
In \cite{C:ShmZha25} it is shown that given the above oracles, the function $H$ is both collision-resistant and non-collapsing. The notion of non-collapsing \cite{unruh2016computationally} here, roughly means that given the oracles $\left( \Ps, \Ps^{-1}, \Ds \right)$, uniform superpositions over preimage sets of $H$ can be distinguished from classical states. A non-collapsing CRH generically implies OSS (while the converse is not known), and generally, to obtain OSS from a non-collapsing CRH, one can consider the superposition of preimages as the quantum signing key. However, the specific construction above directly gives an OSS as well, as it samples coset states, which land themselves naturally to known techniques for quantum signing.

To set up a quantum secret key/classical public key pair, one creates the uniform superposition $\sum_{x \in \zo^n} \ket{x}$ over all possible inputs $x$, evaluates $H(x)$, and measures to get $y$. The state then collapses to the uniform superposition over pre-images of $y$. Then, one can use $\Ps$ once again to compute a superposition of vectors in the affine space $\sum_{x : H(x)=y} \ket{x, \vecU_{x}}$ (or do this already in the first call to $\Ps$), followed by an application of the inverse $\Ps^{-1}$, to un-compute the information in the left register, containing the $x$'s. We are left with the state in the form $\sum_{x : H(x)=y} \ket{\vecU_{x}} := \sum_{\vecU \in S_{y}} \ket{\vecU}$ which we will call $\ket{\psi_y}$. The public key is $\pk := y$, and the secret key is $\ket{\sk} := \ket{\psi_{y}}$.

A signature on a bit $b$ is a string $\vecU_x$ starting with $b$ such that $\vecU_{x} \in S_{y}$ for $\pk=y$. Observe that by measuring $|\sk\rangle$, the signer can obtain a signature on a random bit $b$. Here, we describe a useful view of how to sign an arbitrary bit; it is not exactly how the prior works sign\footnote{The only work using this signing technique is \cite{C:Shmueli22}.}, but will help illustrate some of our ideas. In order to sign an arbitrary bit, measure the first bit of $\vecU$. If it matches $b$, the signer measures the rest of $\vecU$ to get a signature on $b$. If it doesn't match $b$, the signer can actually correct the state back to $|\sk\rangle=|\psi_y\rangle$, and try again. This step requires the ability to project onto $|\psi_y\rangle$, which is accomplished with the oracle $\Ds$, following now-standard techniques~\cite{STOC:AarChr12}. Finally, the inability to sign more than one message follows from the collision resistance of $H$, which~\cite{C:ShmZha25} prove through a careful reduction to standard quantum collision-resistance results. In fact, the collision-resistance of $H$ proves that the OSS has strong unforgeability, where it is impossible to even find two different signatures for the same message.


\paragraph{Why the large signing keys.} We now explore why~\cite{C:ShmZha25} has large secret keys. Clearly, $\Omega(\lambda)$-qubit keys are necessary. There are two additional sources for the large secret key, resulting in $\Omega(\lambda^3)$-qubit keys:
\begin{itemize}
    \item
    To sign large messages,~\cite{C:ShmZha25} sign bit-by-bit, with each bit having its own key. For messages of length $\ell$, this requires secret keys containing at least $\Omega\left( \lambda \cdot \ell \right)$ qubits. Note that we can always hash large messages down to $\lambda$ bits before signing. Thus, we can take $\ell=\lambda$.
    
    \item
    Another source is more subtle. In the security proof, ~\cite{C:ShmZha25} reduce security to the collision-resistance of something they call a coset partition function (CPF). A CPF is a function where all the pre-image sets are affine subspaces. They in particular need a CPF where the preimage sets have size $2^\lambda$. To get such a pre-image set, they start with random 2-to-1 functions, which are automatically CPFs with pre-image sets of size 2, since any two points form a line in $\Z_2^n$. Standard query-complexity lower-bounds show that such 2-to-1 functions are collision resistant, provided the input length is $\lambda$ bits. In order to get a CPF with $2^\lambda$-sized preimage sets,~\cite{C:ShmZha25} simply apply $\lambda$ separate 2-to-1 functions in parallel. This results in a CPF which is collision-resistant but has input of size $\lambda^2$. In the security proof, the CPF is embedded inside the oracles $\left( \Ps, \Ps^{-1}, \Ds \right)$, in a way such that its input size lower-bounds the size of the signing keys.
    Then, combined with needing $\ell=\lambda$ separate secret keys, the overall key size is $\Omega\left( \lambda^3 \right)$.
\end{itemize}

We now briefly describe how we overcome both of the issues above, to obtain secret keys of size $O(\lambda)$ in the oracle model.

\paragraph{Signing multiple bits using a single quantum state.} First, we show how to modify the construction to sign $\lambda$ bits with one secret key, as opposed to needing $\lambda$ separate keys. The starting observation is that there is an easy way to extend verification to handle many bit messages: a signature on $m\in\{0,1\}^\lambda$ is just a string $\vecU$ whose first $\lambda$ bits are $m$ and such that $\vecU \in S_{y}$ for $\pk=y$. The inability to sign multiple messages still follows from the collision resistance of $H$.

The problem is that the measure-then-correct signing algorithm does not work. The natural generalization of the algorithm is to measure the first $\lambda$ bits of $\vecU$, and hope that they match the message; if not, then correct the state back to $|\sk\rangle=|\psi_y\rangle$, and try again. The problem is that there is no granularity here: either the first $\lambda$ bits of $\vecU$ match $m$, or if \emph{any} bit of $\vecU$ fails to match $m$, we have to start all over again. The probability that the first $\lambda$ bits of the measured $\vecU$ match $m$ is $2^{-\lambda}$, meaning in expectation $2^\lambda$ trials are necessary to sign a message\footnote{An additional issue is that the time to correct also grows exponentially with $\lambda$.}.
However, we show that by adding extra information to our dual oracle, we are able to sign in polynomial time.

To illustrate, for $i \in [\lambda]$ and an $i$-bit string $x$, let $|\psi_{y,x}\rangle$ be the uniform superposition over $S_y$ whose initial $i$ bits are exactly $x$. Then $|\psi_{y,m}\rangle$ is the state we want to ultimately obtain, as a measurement on it will give us a signature for $m$. The state that we start with, which equals to our quantum signing key, is $|\psi_{y,\emptyset}\rangle$. Then by the same measure-then-correct approach, we can obtain the state $|\psi_{y,m_1}\rangle$, where $m_1$ is the first bit of $m$. Now, we want to perform a fresh round of measure-then-correct, but where we start from $|\psi_{y,m_1}\rangle$ and move to $|\psi_{y,m_1 m_2}\rangle$. The probability of success in the measuring step is $1/2$, so the expected number of trials is $2$, which is efficient. We can continue in this way, gradually moving to $|\psi_{y,m_1m_2m_3}\rangle,\cdots$ all the way to $|\psi_{y,m}\rangle$, and each step is efficient since in each step, we are only measuring one bit, and so in expectation only need two trials per step.

The issue is that, in order to correct to the states $|\psi_{y,m_1}\rangle,|\psi_{y,m_1m_2}\rangle,\cdots$, we need a mechanism to project onto those states efficiently. Unfortunately, the existing oracles do not provide this. However, we can modify the oracles to accomplish this. Namely, to project onto $|\psi_y\rangle$, we needed two oracles, one for $S_y$ and one for $S_y^\perp$. In order to project onto $|\psi_{y,m_1}\rangle$, we also need two oracles. One is for $S_{y,m_1}$, the affine subspace of $S_y$ whose first bit is $m_1$. The other is $S_{y,m_1}^\perp$, the dual to $S_{y,m_1}$.

Fortunately, we actually already have $S_{y,m_1}$ for free: membership in $S_{y,m_1}$ is checked by checking membership in $S_y$ and then checking that the first bit equals $m_1$. However, $S_{y,m_1}^\perp$ is a super-space of $S_y^\perp$, and we do not (yet) have access to this oracle. Our solution is to provide it. Now, as we go to more and more bits of $m$, there will be more and more subspaces $S_{y,m}^\perp$ we would have to provide oracles for. Eventually this will be exponentially many subspaces. This may actually be fine to provide access to all these subspaces, but we do not know how to prove it. Instead, we observe that the different choices of subspace $S_{y,m}^\perp$ as $m$ varies actually have predictable relationships to each other. It actually suffices to give out membership checks for $S_{y,0}^\perp,S_{y,00}^\perp,\cdots, S_{y,0^\lambda}^\perp$, and this will allow for checking membership in all the $S_{y,m}^\perp$ subspaces. Note that adding these oracles allows for efficient signing, but we can no longer use the security proof from~\cite{C:ShmZha25} as a black box due to the extra oracles. In fact, our actual signing algorithm, which we describe next, even needs more information, so we will later touch upon how we modify this.



\paragraph{Parallel Signing.}
One drawback of the above approach is that it makes the signing process sequential, in that the signing of bit $i$ of $m$ has to be processed before the next bit $i + 1$ can be processed. This is because, for the projection on the $(i + 1)$-th state $|\psi_{y, m_1 \cdots  m_{i + 1} }\rangle$ to have constant success probability, we need to already be in the state $|\psi_{y, m_1 \cdots  m_{i} }\rangle$. Contrast with the original signing procedure which just signs each bit in parallel (but using many quantum states). 

We overcome this by actually modifying the signing process above to make it highly parallelizable. We measure \emph{all} the first $\lambda$ bits of $\vecU$. With overwhelming probability, roughly half of them will match the corresponding bit of $m$. So we keep those bits, but correct all the bits that did not match, and do this correction in parallel. This requires significant care, and there are two main problems to overcome. First, the iterative signing process outlined above, needed to give out membership checks for the subspaces $S_{y,0}^\perp, S_{y,00}^\perp, S_{y,000}^\perp, \cdots$. But these subspaces, by depending on ever-larger prefixes, \emph{only} allow for signing the bits in sequential order. When we measure the first $\lambda$ bits of $\vecU$, the positions that do not match $m$ will be in random positions, and the existing structure does not allow for directly correcting those bits.
Instead, we give out a small collection of large subspaces, whose \emph{intersections} allow us to simulate the exponentially-many subspaces, which are needed in order to fix any possible bit positions. Specifically, we give access to $\lambda$ distinct subspaces of $S_{y,0^{\lambda}}^\perp$, all of which have dimension one smaller than $S_{y,0^{\lambda}}^\perp$. The differing intersections of these, turn out to be sufficient to correct any error.

So, we have the information we need for correcting, and now turn to computational complexity, which is where the second difficulty. Naively correcting the $\approx \lambda/2$ bits that didn't match the message-to-sign, and transforming them back into the uniform superposition seems to require exponential time, since the time to correct grows exponentially with the number of bits that need correcting. However, we show that we can efficiently correct them to uniform superposition with a random choice of phase; the phase turns out not to matter, since the next step is to just measure the bits anyway. What is important is that measuring after correcting re-randomizes these bits. Therefore we measure the bits again, and at this point, roughly $3\lambda/4$ of the bits will match $m$. Another round gives $7\lambda/8$, and so on.

We can continue in this way, until we get all the bits of $m$ to match. This would require roughly $\log\lambda$ sequential rounds of queries. But we can do better: instead of signing $m$ directly, we instead map $m$ to a codeword $c$ of a good error correcting code, and sign $c$. We then relax the condition on $\vecU$ being a valid signature to only require the first $\lambda$ bits of $\vecU$ to have a Hamming distance at most, say, $\lambda/6 > \lambda/8$ away from the codeword $c$. With overwhelming probability, our signing process above, when run for sequential 3 steps, will produce such a valid signature. 

As for security, assuming the code has minimum distance strictly greater than $\lambda/3$, for any $\vecU$, there will be at most one codeword $c$ within the required distance of $\vecU$. Thus, valid signatures on two  distinct messages will lead to two distinct $\vecU$ that collide, contradicting the collision resistance of $H$. Note that a random linear code will obtain such a minimum distance with high probability, as long as we set the rate to be an appropriately small constant.

Finally, as mentioned, the information our oracles provide is significantly more than just giving out membership access to $S_{y}$, $S_{y}^{\bot}$, and the previous security analysis of \cite{C:ShmZha25} cannot be used as is. By a new analysis we show that even with this extra information, it is impossible to find a collision in $H$. In the process, we also define and prove the security of a generalization of subspace-hiding obfuscation, which we refer to as subspace-hiding functions in this work (see Section \ref{section:subspace_hiding_function}).

\paragraph{Overcoming the CPF input-size Problem.}
We go back to thinking about the size of the quantum signing key for signing $\lambda$-bit messages. Until now, we explained how to go from keys of size $\Omega\left( \lambda^3 \right)$ qubits, due to the original construction of \cite{C:ShmZha25}, to $O\left( \lambda^2 \right)$ in our optimized construction. Next, we turn to overcoming the issue with the coset partition function (CPF) input size. Looking at the final step of the security reduction of \cite{C:ShmZha25}, what is shown is that given only the oracles $\Ps$, $\Ps^{-1}$ (that is, without the dual oracle $\Ds$), the function $H$ is collision resistant. The central tool of the reduction is a CPF $\hashQ$ from $n$ to $r$ bits, where the size of each preimage coset of $\hashQ$ is $2^\lambda$ and $\hashQ$ is collision-resistant. Very roughly, the reduction embeds $\hashQ$ inside both oracles $\Ps, \Ps^{-1}$, in an indistinguishable way, to later claim that a collision in $H$ constitutes a collision in $\hashQ$. In order to claim indistinguishability between the original oracles $\Ps, \Ps^{-1}$ and the oracles where $\hashQ$ is embedded in, there are two primary arguments where the structure of $\hashQ$ comes up.

First, when computing $y$ i.e., the first part of the output of $\Ps$, the original $\Ps$ executes a random permutation $\Pi$ on the input $x \in \zo^n$ and $H(x) := y \in \zo^r$ is given by the first $r$ output bits of the permutation's output. In order to embed $\hashQ$ indistinguishably, \cite{C:ShmZha25} observe that a CPF from $n$ to $r$ bits can always be extended to a permutation in the following way: the first $r$ bits are simply the output of $\hashQ$ which tells us what coset the input element belonged to, and the remaining $n - r$ bits can serve as the coordinates vector $\vecZ \in \bbZ_{2}^{n - r}$ of that element with respect to the coset of the output $y$ -- recall that because $\hashQ$ is a CPF, the preimage set of each $y$ is an affine subspace, and thus every element has a coordinates vector. Finally, composing a random permutation with any fixed permutation is statistically equivalent to just a random permutation, so, the permutation that we extended $\hashQ$ to, can be composed with the random permutation $\Pi$ inside the original $\Ps$. This constitutes that for the first part of the output of $\Ps$ (and in fact, also for the inverse $\Ps^{-1}$), $\hashQ$ can be put inside the oracles in an indistinguishable manner.

The second place where the structure of $\hashQ$ is used is to simulate the vectorial part of the output of $\Ps$, that is, the vectors $\vecU_{x} \in S_{y}$. Here, the original $\Ps$ samples a random affine subspace $S_{y}$ given each output $y$. Now, recall that due to the inverse oracle $\Ps^{-1}$ there should be an \emph{invertible} mapping between the inputs $x \in \zo^n$ and the vectors $\vecU_{x} \in S_{y}$, so, the affine subspaces in the security reduction to the collision-resistance of $\hashQ$ cannot just be sampled independently of $\hashQ$. There needs to be an efficient invertible mapping between the simulated affine subspaces and the actual preimage sets of $\hashQ$. Fortunately, the structure of the CPF can be used again, but in a different manner: in order to simulate a random affine subspace, it turns out it is enough to take the preimage set elements of $\hashQ$ for some output $y$, and multiply them by the same random full-rank matrix (that is, a random i.i.d. matrix for every output $y$ of $\hashQ$). Intuitively, the multiplied element can now be thought of as an element in a random affine subspace, and the reason is that the original preimage set of $\hashQ$ was an affine subspace already, and multiplying any affine subspace by a random matrix gives a \emph{random} affine subspace. Finally, since the reduction is the one sampling the random matrix, the mapping between affine subspace elements from the simulated $S_{y}$, and elements from the preimage set containing the $x$'s, is efficiently computable.

Let us denote by $n$ the input size to $\Ps$ and by $k$ the size of the vectors in the affine subspaces $S_{y}$. Note that in the above reduction, the input size to $\hashQ$, which we denote by $n'$, lower bounds both $n$ and $k$. This means that, if we had a way to construct a CPF with input size $n' \approx \lambda$ such that $\hashQ$ is $2^\lambda$-collision-resistant\footnote{That is, where the probability for any quantum algorithm making a polynomial number of queries to find a collision is $2^{-\Omega(\lambda)}$}, we would be done. Unfortunately, we only know how to build a CPF with input size $\lambda^2$ that is $2^\lambda$-collision-resistant. As mentioned earlier in the introduction, this is done by taking the parallel repetition of a random 2-to-1 function. It seems plausible that such CPFs may exist, and in fact~\cite{STOC:AGKZ20} provide a CPF which could reasonably be conjectured to have the desired properties. However, we do not know how to prove such a result.

\paragraph{Folding Coset Partition Functions.}
We show a new technique to construct a CPF such that even if its input is large, for every input there is a \emph{folded version} of that input. Furthermore there is an invertible mapping between folded and original inputs, which does not break the collision-resistance of $\hashQ$. Concretely, our folding CPF is given by two algorithms 
$$
\hashQ : \bbZ_{2}^{\lambda^2} \rightarrow \left( \bbZ_{2}^{\lambda^2 - \lambda} \times \bbZ_{2}^{2\cdot \lambda - 1} \right), \enspace \enspace 
\hashQ^{-1} : \left( \bbZ_{2}^{\lambda^2 - \lambda} \times \bbZ_{2}^{2\cdot \lambda - 1} \right) \rightarrow \bbZ_{2}^{\lambda^2} 
\enspace .
$$

In the previous work, the CPF $\hashQ : \bbZ_{2}^{\lambda^2} \rightarrow \bbZ_{2}^{\lambda^2 - \lambda}$ was given by taking a parallel repetition of $\lambda$ functions, each is a random 2-to-1 function from $\lambda$ to $\lambda - 1$ bits. Here, the first step to our solution will be not to take arbitrary 2-to-1 functions but \emph{claw-free permutations}. A claw-free permutation $\hashL : \zo^\lambda \rightarrow \zo^{\lambda - 1}$ is given by a pair of random permutations $\Pi_{0}, \Pi_{1}$ such that $\hashL\left( b \in \zo, x \in \zo^{\lambda - 1} \right) := \Pi_{b}(x)$. Claw-free permutations can easily be proved as collision-resistant in a classical oracle model (and we give a proof in the body of our paper). A nice property of these functions is that collisions always differ in the first bit. In particular, taking the view of claw-free permutations as 1-dimensional CPFs, one can think about the first bit as containing the information of the coordinates vector of that coset element in the preimage set. This coordinate information is computable publicly for a claw-free permutation, and this will be useful for our new reduction.

Our function $\hashQ : \bbZ_{2}^{\lambda^2} \rightarrow \left( \bbZ_{2}^{\lambda^2 - \lambda} \times \bbZ_{2}^{2\cdot \lambda - 1} \right)$ will be computed as follows: The first part of the output will be a parallel repetition of a claw-free permutation, and accordingly takes $\lambda \cdot (\lambda - 1) = \lambda^2 - \lambda$ bits. The second part of the output will itself contain two parts: the first part is $\lambda$ bits, and contains the first input bit of each of the $\lambda$ inputs to the claw-free permutations. The second part (of the second part) is the \emph{sum} of the rest of the inputs. That is, if $\vecW = \left( b_{1}, \vecW_{1}, \cdots, b_{\lambda}, \vecW_{\lambda} \right)$ is the input for $\hashQ$ broken down to the $\lambda$ inputs to the claw-free permutations, then we are thinking of $\overline{\vecW} := \sum_{j \in [\lambda]} \vecW_{j}$.  The overall output of $\hashQ$ is
$$
\left( y_{1}, \cdots, y_{\lambda} \right) \in \zo^{\lambda^2 - \lambda}, \left( b_1, \cdots, b_\lambda, \overline{\vecW} \right) \in \zo^{2\cdot \lambda - 1} \enspace .
$$

The inverse function $\hashQ^{-1}$ can be computed as follows. We have the permutation pairs
$$
\left( \left( \Pi_{1, 0}, \Pi_{1, 1} \right), \cdots, \left( \Pi_{\lambda, 0}, \Pi_{\lambda, 1} \right) \right)
$$
defining our function $\hashQ$, and we can think about the inverse permutations of these pairs 
$$
\left( \left( \Pi^{-1}_{1, 0}, \Pi^{-1}_{1, 1} \right), \cdots, \left( \Pi^{-1}_{\lambda, 0}, \Pi^{-1}_{\lambda, 1} \right) \right)
$$
as the "secret key". We use this secret key in order to implement $\hashQ^{-1}$ in a straightforward way: For each $j \in [\lambda]$, we have the $j$-th output $y_{j} \in \zo^{\lambda - 1}$ and the bit $b_{j}$, so to invert we compute $b_{j}, \Pi^{-1}_{j, b_{j}}(y_{j})$. For this, we do not even use the sum part $\overline{\vecW}$ of the input to $\hashQ^{-1}$.

The key insight to our reduction, is that the oracles $\hashQ$, $\hashQ^{-1}$ can both be perfectly simulated \emph{even if one of the instances of $\hashL$ does not contain the secret key}. In a nutshell, assume that you want to simulate $\hashQ$, $\hashQ^{-1}$, and for some instance $i^* \in [\lambda]$ of the claw pairs, you do have the forward computation $\Pi_{i^*, 0}, \Pi_{i^*, 1}$ but not the inverse pair $\Pi^{-1}_{i^*, 0}, \Pi^{-1}_{i^*, 1}$. We will want to find a collision in that instance. Computing $\hashQ$ in this setting is still easy: we just compute all claw pairs in the forward direction and then for the second part of the output we output the first bits of the corresponding inputs, and also sum the rest of the parts, as described above. The simulation of $\hashQ^{-1}$ shows where the sum of the inputs becomes handy: for all instances $j \in [\lambda] \setminus \{ i^* \}$ we compute the inverses as before -- thanks to the claw-free permutations only needing the output $y_{j}$ and the bit telling us which of the two inputs was it, we recover the corresponding $\{ \vecW_{j} \}_{j \in [\lambda] \setminus \{ i^* \} }$. Finally, all we need to do to get the last $\vecW_{i^*}$ is by subtracting all of the $\lambda - 1$ elements we obtained, from the sum we got as input to $\hashQ^{-1}$: 
$$
\vecW_{i^*} \gets \overline{\vecW} - \sum_{ j \in [\lambda] \setminus \{ i^* \} } \vecW_{j} \enspace .
$$

The last observation to make is that the folded part actually structures a coset as well. In the body of the paper we further explain how the folded inputs do not only constitute an invertible map (while keeping $\hashQ$ collision-free), but these sets also form affine spaces, just smaller ones. In particular, note that the second output of $\hashQ$ takes only $2\lambda - 1 \in O(\lambda)$ bits. Circling back to the reduction of the collision-resistance of $\Ps, \Ps^{-1}$, we still take the $\lambda^2$-bit outputs of $\Ps$ as the input of $\hashQ$, but now can use the folded part, taking only $O(\lambda)$ bits as vectors to simulate the affine spaces $S_{y}$. Since we have the invertible mapping $\hashQ$, $\hashQ^{-1}$ we also can invert between $\Ps$ and $\Ps^{-1}$, but we shaved a factor of $\lambda$ from the vectorial part. Since our signing keys are ultimately the vectorial part, we get signature tokens of size $O(\lambda)$.

\paragraph{Standard Model Construction.} We now turn to constructing OSS in the standard model, with small quantum secret keys. It is straightforward to adapt all of our techniques to the standard-model construction of~\cite{C:ShmZha25} to get a secure construction. However, it turns out it is not immediate that the construction has linear-sized quantum secret keys. 

To see the issue, recall that we are setting $\lambda$ to be such that all $T$ time attacks with success probability $p$ have $T/p \geq 2^\lambda$. In particular, if secret keys are length $O(\lambda)$, this means all attacks must take exponential time and/or have exponentially-small success probability, as functions of the length of the secret key.

In the oracle version above, we were able to achieve this due to exponential bounds on the success probability of bounded-query algorithms. In the standard model, achieving such a result requires in particular the hardness of each of the cryptographic building blocks against attacks running in time $2^\lambda$. 

Note that the various building blocks may be instantiated with different security parameters. For example, we may set the security parameter for indistinguishability obfuscation to $\lambda_{\iO}=\lambda^c$ for some large constant $c$\enspace\footnote{In fact, this already happens in the existing proof of~\cite{C:ShmZha25}.}. In this case, we only need that the $\iO$ is secure against attacks running in time $2^\lambda=2^{\lambda_{\iO}^{1/c}}$; that is, we only need subexponential-time security of the iO. However, setting the security parameter for components in this way will affect the sizes of certain parameters. For iO, fortunately setting the security parameter in this way only affects the size of the obfuscated program that is a part of the common reference string, and not the quantum secret key size. 

The takeaway, however, is that (1) all the primitives need to be at a bare minimum subexponentially-secure\footnote{If a primitive $P$ is only $2^{\lambda_P^{o(1)}}$-secure, then in order to achieve $2^{\lambda_P^{o(1)}}\geq 2^\lambda$, we must set $\lambda_P$ to be super-polynomial in $\lambda$. This would cause some system components to have super-polynomial run-time and size.}, and (2) we need to be extremely careful about how the various components interact with the parameter sizes.

\medskip

Recall the assumptions used to instantiate~\cite{C:ShmZha25}:
\begin{itemize}
    \item Indistinguishability obfuscation (iO). They already need sub-exponential security for their proof, and as discussed above, sub-exponential security suffices for us.
    \item A sub-exponentially-secure one-way function. The one-way function arises primarily in two places. One is for constructing an object called a permutable PRP, an analog of punctured PRFs for permutations. We note that for this application sub-exponential security still suffices for us, as it will blow up the key but not the block size of the PRP; the key blows up the size of the obfuscated program, which only affects the CRS.

    The other application is to implement what they call a sparse trigger. Here, there is a bad event (elaborated on below) which occurs on an $\alpha$-fraction of inputs. The bad inputs are hidden by mapping this to a $\log(1/\alpha)$-bit string, which in turn serves as input to a pseudorandom generator. Remember that the PRG needs to be secure against $2^\lambda$-time attacks. If the PRG is exponentially secure, we can handle $\alpha=2^{-O(\lambda)}$, but if the PRG is only sub-exponentially secure, this means $\alpha$ must be correspondingly smaller. This will come up when we discuss the next assumption.

    \item The polynomial hardness of LWE with sub-exponential noise/modulus ratio. Here, we certainly need to upgrade this to subexponential hardness\footnote{As in, we need to assume that LWE with a subexponential noise/modulus ration is subexponentially hard.}. But it actually gets worse. LWE is used in two places. First, it is used to implement a lossy function in order to ``bloat'' the subspaces $\Ds$ accepts. In this step, we incur a loss in the security reduction equal to the range of the lossy function in lossy mode. On the other hand, we are reducing to the hardness of a one-way function whose effective security parameter is limited by the dimension of the ambient space. But remember that our secret keys are now superpositions of vectors that live in this space. As such, for linear-sized secret keys, we need the lossy function to have range size $2^{O(\lambda)}$ while still having security against algorithms running in time $2^{O(\lambda)}$. Unfortunately, LWE is not secure in this regime, due to attacks running in time $2^{O(\lambda/\log\lambda)}$ where $\lambda$ represents the bit-length of the secret key\footnote{LWE attacks run in time $2^{O(n)}$ for dimension $n$. But the secret is a vector in $\Z_q^n$ which has bit-length $n\log q$. Since $q$ is typically polynomial in $\lambda$, this gives $n=\Omega(\lambda\log \lambda)$.}. Fortunately, by slightly blowing up the key size to be quasi-linear $\tilde{O}(\lambda)$, we can account for such attacks, though we need the very strong assumption of near-exponentially-secure LWE.

    The more problematic part is where LWE is used to realize a collision-resistant approximately 2-to-1 function. This construction has errors (namely, places where it is not 2-to-1) at a rate of roughly the noise-modulus ratio of LWE. Thus, in light of the sparse trigger requirements, we need the noise-modulus ratio to be on the order of $2^\lambda$. It turns out that setting the noise modulus ratio this high blows of the input/output size of the 2-to-1 function without increasing hardness. The result is that the input/output size of the 2-to-1 function is on the order at least $\Omega(\lambda^3)$, even assuming a security level for LWE that exactly matches the best-known attacks. By carefully modifying the proof we can get the ``effective'' size to be $O(\lambda^2)$, but we do not know how to shave off the last factor of $\lambda$, preventing us from obtaining linear-sized quantum secret keys.
\end{itemize}

It turns out this issue with the 2-to-1 functions is the only barrier to achieving quasi-linear-sized quantum secret keys in the standard model. If we just accept this limitation, we are able to obtain:

\begin{theorem} \label{theorem:standard_model_1}
Assume all of the following: (1) sub-exponentially- secure indistinguishability obfuscation, (2) exponentially-secure one-way functions, and (3) ``optimally-secure'' LWE. Then, there exists OSS with $O(\lambda^2)$-sized quantum secret keys, that can sign on messages of size $\lambda$ and such that for every $T$-time quantum adversary, the probability to break strong unforgeability is bounded by $\frac{T}{2^{\Omega(\lambda)}}$. 
\end{theorem}
Here, ``optimal'' LWE means (1) a noise modulus ratio of $2^{O(\sqrt{n\log q})}$ such that the constant hidden by the Big-Oh is sufficiently small to block known attacks, and (2) there are no attacks with $T/p$ smaller than $2^{O(n)}$. We will set $n=O(\lambda)$. Note that for polynomial noise, we have $q\approx 2^{O(\sqrt{n\log q})}$, and then (1) translates to $q=2^{O(n)}=2^{O(\lambda)}$, which is needed for the sparse trigger. (2) is needed for simply to obtain security against algorithms running in time $2^\lambda=2^{O(n)}$.

Alternatively, if we just assume the existence of an exponentially-secure perfect 2-to-1 trapdoor function, we avoid the 2-to-1 function issue entirely, and obtain:

\begin{theorem} \label{theorem:standard_model_2}
Assume all of the following: (1) sub-exponentially-secure indistinguishability obfuscation, (2) exponentially-secure one-way functions, (3) $2^{O(\lambda_{LWE}/\log \lambda_{LWE})}$-secure LWE with a polynomial noise-modulus ratio, and (4) perfect 2-to-1 trapdoor functions with exponentially-secure collision-resistance, which are ``permutation-indistinguishable''. Then, there exists OSS with $\tilde{O}(\lambda)$-sized quantum secret keys, that can sign on messages of size $\lambda$ and such that for every $T$-time quantum adversary, the probability to break strong unforgeability is bounded by $\frac{T}{2^{\Omega(\lambda)}}$. 
\end{theorem}

Note that (4) implies (2). Here, $2^{O(\lambda_{LWE}/\log\lambda_{LWE})}$-secure LWE is a relaxation of optimal LWE to the case of polynomial $q$, requiring that security scales exponentially in the dimension $n$, while the secrets have size $\lambda_{LWE}=n\log q=O(n\log n)$. We will set $n=O(\lambda)$, giving $\lambda_{LWE}=O(\lambda\log\lambda)$, so that we achieve security against $2^\lambda$-time attacks.

Being ``permutation-indistinguishable'' is a mild technical condition required to be compatible with the proof of security using permutable PRPs, which is formalized in Definition \ref{definition:permutation_indistinguishable_tdf}. We do not know any provably secure constructions satisfying (4), but a simple conjectured such function is the following. Take two PRPs $P_{k_0},P_{k_1}$ with random keys $k_0,k_1$, and obfuscate the function mapping $(b,x)\mapsto P_{k_b}(x)$. This is a 2-to-1 function. $k_0,k_1$ is also a trapdoor, since applying $P_{k_0}^{-1}(y)$ and $P_{k_1}^{-1}(y)$ gives the two pre-images of $y$. We could in particular set $P$ to be a permutable PRP constructed in~\cite{C:ShmZha25}, which is permutation-indistinguishable. Assuming ideal obfuscation, such an obfuscation of $P$ would in fact be exponentially collision-resistant (assuming the underlying one-way functions are sub-exponentially secure). A common heuristic is to assume that indistinguishability obfuscation (iO) actually acts as an ideal obfuscator. While this heuristic is false in general~\cite{C:BGIRSV01}, it seems reasonable to apply it to ``natural'' programs that were not explicitly designed to demonstrate an impossibility. So it isn't immediately impossible that using iO would also give this level of security\footnote{Note that we certainly need sub-exponential iO, but do not necessarily need exponential iO, since again it only affects the obfuscated program size and not the size of the quantum secret key.}. However, it seems unlikely that we could actually prove such a construction is secure just based on iO and one-way functions, since in particular this implies collision-resistance, which is likely not provable from these assumptions alone~\cite{FOCS:AshSeg15}. Note that still, it might be possible to prove the security of this exact construction as is, using more assumptions. For example, assuming collision-resistant primitives which could possibly be hidden, indistinguishably under the iO, and come up only in later hybrids (which is the same technique used to prove the collision-resistance of our one-shot signature scheme, based on LWE).

Finally, we can even get truly linear secret keys if we eliminate the use of LWE all together, and additionally assume an exponentially-secure \emph{lossy} function:

\begin{theorem} \label{theorem:standard_model_3}
Assume all of the following: (1) sub-exponentially-secure indistinguishability obfuscation, (2) exponentially-secure one-way functions, (3) exponentially-secure lossy functions, and (4) perfect 2-to-1 trapdoor functions with exponentially-secure collision-resistance, which are ``permutation-indistinguishable''. Then, there exists OSS with $O(\lambda)$-sized quantum secret keys, that can sign on messages of size $\lambda$ and such that for every $T$-time quantum adversary, the probability to break strong unforgeability is bounded by $\frac{T}{2^{\Omega(\lambda)}}$.
\end{theorem}
A lossy function~\cite{STOC:PeiWat08} is a function that comes in two modes; an injective mode and a lossy mode. In the injective mode the function is injective, while in the lossy mode the function has a small image of size at most $2^{O(\lambda)}$. It is straightforward to construct lossy trapdoor function using ideal obfuscation and sub-exponentially-secure PRFs: the injective mode is just the obfuscation of an injective PRF, and the lossy mode is the composition $F\circ G$ of two PRFs, where the image of $G$ has size $2^{O(\lambda)}$. Therefore, we can plausibly instantiate (3) by applying iO. As before, while the construction itself may have a security proof, it seems unlikely that this could be proved without using more assumptions, given that lossy functions imply collision-resistance.

\subsection{Concurrent and Independent Work}

In a concurrent and independent work,~\cite{HuaVai25} also consider the size of the quantum secret key in~\cite{C:ShmZha25}. Their work contains a few results, some of which partially overlap with ours.

\paragraph{OSS in the oracle setting.}
In the oracle setting, they are able to get the key size down to $O(\lambda^2)$. They shave one multiplicative factor $\lambda$ relative to \cite{C:ShmZha25}, by signing all the bits of the message using a single token rather than signing individual bits using many tokens. However, their result got stuck at $\lambda^2$ because they were not able to overcome the CPF size bound as we do. Their signing algorithm is sequential, unlike our parallel algorithm.

\begin{remark}
They actually claim that the oracle-model construction of~\cite{C:ShmZha25} requires secret keys of size $\Omega(\lambda^4)$, while it requires $\Theta\left( \lambda^3 \right)$. The reason for the $\lambda^4$ is when one uses the generic transformation from non-collapsing collision-resistant hash functions to OSS~\cite{TQC:DalSpo23}. However, while~\cite{C:ShmZha25} cite~\cite{TQC:DalSpo23} as a motivation in the overview, the actual construction signs messages directly as opposed to going through~\cite{TQC:DalSpo23}. As such, this factor of $\lambda$ is not actually relevant.
\end{remark}

\paragraph{OSS in the plain model.}
Then, moving from the oracle model to the plain model, \cite{HuaVai25} claim that they achieve the same $O\left( \lambda^2 \right)$-qubit secret keys in the plain model, by adapting the security proof of~\cite{C:ShmZha25} to their modifications. The result is a claim of OSS with $O\left( \lambda^2 \right)$-qubit secret keys under the same assumptions as~\cite{C:ShmZha25}:

\newpage
\begin{boxclaim}[Claimed in~\cite{HuaVai25}]\label{claim:hv}Assuming (1) subexponentially-secure indistinguishability obfuscation, (2) subexponentially- secure one-way functions, and (3) (polynomially-secure) LWE with a subexponential noise-modulus ratio, there exists a secure OSS scheme with perfect correctness and $O(\lambda^2)$-bit signatures and $O(\lambda^2)$-qubit signing keys for $\poly(\lambda)$-bit messages.\end{boxclaim}

Unfortunately, Claim~\ref{claim:hv} is false: while the security proof is valid for justifying the security of the OSS scheme against polynomial-time attacks, it does not actually justify $O(\lambda^2)$-qubit secret keys. The reason is that they only prove that the construction is polynomially-secure (as was done in~\cite{C:ShmZha25}), without paying attention to the fact that claiming $\lambda$-bit security -- as required to meaningfully discuss key sizes as a function of $\lambda$ -- requires security against $2^\lambda$-time attackers. In particular, Claim~\ref{claim:hv} cannot possibly be true, for \emph{any} fixed polynomial-length secret key, for the simple reason that any such claim inherently requires all building blocks to be sub-exponentially secure,\footnote{Since, even if you set the security parameter of the underlying component to be some polynomial $\kappa$ in $\lambda$, this still means the attack runs in time $2^{\kappa^c}$. Polynomial security may only guarantee a security against a running time of $2^{\log^2\kappa}$.} but Claim~\ref{claim:hv} only assumes the polynomial hardness of LWE. Digging deeper, they also miss the fact that using sub-exponential hardness for the sparse trigger blows up the secret key size. Moreover, even upgrading to ``optimal'' LWE still further blows up the secret key size, as discussed above.

A straightforward adaptation of their proof, using the assumptions as we do -- namely, exponentially-secure one-way functions and same ``optimal'' LWE assumption -- results in their quantum secret keys having size $\Omega(\lambda^4)$, as opposed to our $O(\lambda^2)$. Again, one factor of $\lambda$ improvement in our work comes from the same factor of $\lambda$ improvement from the oracle setting, and the other factor of $\lambda$ comes from our trick of only using ``effective'' input size in the LWE-based 2-to-1 functions.

\paragraph{Perfect correctness.} ~\cite{HuaVai25} show how to make the OSS construction perfectly correct, which we do not consider. Our parallel signing algorithm has an exponentially-small correctness error of $2^{-\Omega\left( \lambda \right)}$. An interesting open question is whether our parallel signing algorithm can be adapted to be perfectly correct.

\paragraph{Quantum fire in the oracle model.} Lastly,~\cite{HuaVai25} also discuss applications to quantum fire -- quantum states that can be cloned but not sent over classical communication -- which were recently explored in~\cite{ITCS:NehZha24,STOC:BosNehZha25,CGS25}. We do not consider this setting in the current work.

%% file: crypto_tools.tex
\section{Cryptographic Tools} \label{section:crypto_tools}
We use the following known primitives and notions. Both are implicitly classical primitives with security holding against quantum algorithms.

\begin{definition} [Puncturable PRFs] \label{def:pprf}
A puncturable pseudorandom function (P-PRF) is a pair of efficient algorithms $\left( \prf, \punc, \eval \right)$ with associated output-length function $m\left( \lambda \right)$ such that:
\begin{itemize}

  \item $\prf : \{0,1\}^\lambda \times \{0,1\}^* \rightarrow \{0,1\}^{m\left( \lambda \right)}$ is a deterministic polynomial-time algorithm.
  
  \item $\punc\left( k, S \right)$ is a probabilistic polynomial-time algorithm which takes as input a key $k \in \{0,1\}^\lambda$ and a set of points $S \subseteq \{0,1\}^*$. It outputs a punctured key $k^S$.
  
  \item $\eval\left( k^S, x \right)$ is a deterministic polynomial-time algorithm.
  
  \item {\bf Correctness:}
  For any $\lambda \in \Nat$, $S \subseteq \{ 0, 1 \}^*$, $k \in \{0,1\}^\lambda$, $x \notin S$, and $k^S$ in the support of $\punc\left( k, S \right)$, we have that $\eval\left( k^S, x \right) = \prf\left( k, x \right)$.
  
  \item {\bf Security:}
  For any quantum polynomial-time algorithm $\As$, there exists a negligible function $\epsilon$ such that the following experiment with $\As$ outputs 1 with probability at most $\frac{1}{2} + \epsilon\left( \lambda \right)$:
  \begin{itemize}
    \item $\As(1^\lambda)$ produces a set $S \subseteq \{0,1\}^*$.
    
    \item The experiment chooses a random $k\gets\{0,1\}^\lambda$ and computes $k^S \gets \punc\left( k, S \right)$. For each $x\in S$, it also sets $y_x^0 := \prf\left( k, x \right)$ and samples $y_x^1 \gets \{0,1\}^{m(\lambda)}$ uniformly at random. Then it chooses a random bit $b$. It finally gives $k^S, \{\left( x, y_x^b \right) \}_{ x \in S }$ to $\As$.
    
    \item $\As$ outputs a guess $b'$ for $b$. The experiment outputs 1 if $b'=b$.
  \end{itemize}
  
\end{itemize}
\end{definition}

\paragraph{Different security levels.}
For arbitrary functions $f_{0}, f_{1} : \Nat \rightarrow \Nat$, we say that the P-PRF is $\left( f_{0}, \frac{1}{f_{1}} \right)$-secure, if in the above security part of the definition, we ask that the indistinguishability holds for every adversary of size $\leq f_{0}(\secp)$ and we swap $\epsilon(\secp)$ with $\frac{1}{ f_{1}(\secp) }$. Concretely, a sub-exponentially secure P-PRF scheme would be one such that there exists a positive real constant $c > 0$ such that the scheme is $\left( 2^{\lambda^c}, \frac{1}{2^{\lambda^c}} \right)$-secure.

\begin{definition} [Indistinguishability Obfuscation (iO)] \label{def:iO}
An indistinguishability obfuscator (iO) for Boolean circuits is a probabilistic polynomial-time algorithm $\iO\left( \cdot, \cdot, \cdot \right)$ with the following properties:
\begin{itemize}

  \item {\bf Correctness:}
  For all $\lambda, s \in \Nat$, Boolean circuits $C$ of size at most $s$, and all inputs $x$ to $C$,
  \[
  \Pr\left[
  \Obf_{C}(x) = C(x) \: : \: \Obf_{C} \leftarrow \iO\left( 1^\lambda, 1^s, C \right)
  \right] = 1
  \enspace .
  \]

  \item {\bf Security:}
  For every polynomial $\poly(\cdot)$ there exists a negligible function $\epsilon$ such that the following holds.
  Let $\secp, s \in \Nat$, and let $C_{0}$, $C_{1}$ two classical circuits of (1) the same functionality (i.e., for every possible input they have the same output) and (2) both have size $\leq s$.
  $$
  \biggl\{
  \Obf_{C_{0}} \: : \: \Obf_{ C_{0} } \leftarrow \iO\left( 1^\lambda, 1^s, C_{0} \right)
  \biggr\}
  $$
  $$
  \approx_{ \left( \poly(\secp), \epsilon(\secp) \right) }
  \biggl\{
  \Obf_{C_{1}} \: : \: \Obf_{ C_{1} } \leftarrow \iO\left( 1^\lambda, 1^s, C_{1} \right)
  \biggr\}
  \enspace .
  $$ 
\end{itemize}
\end{definition}

\paragraph{Different security levels.}
For arbitrary functions $f_{0}, f_{1} : \Nat \rightarrow \Nat$, we say that an iO scheme is $\left( f_{0}, \frac{1}{f_{1}} \right)$-secure, if in the above security part of the definition, we swap $\poly$ with a specific function $f_{0}$ and the negligible function with the function $\frac{1}{f_{1}}$. Concretely, a sub-exponentially secure iO scheme would be one such that there exists a positive real constant $c > 0$ such that the scheme is $\left( 2^{\lambda^c}, \frac{1}{2^{\lambda^c}} \right)$-secure.

\begin{definition} [Lossy Functions] \label{definition:lf}
A lossy function (LF) scheme consists of classical algorithms $(\LFGen$, $\LFF)$ with the following syntax.

\begin{itemize}
    \item
    $\pk \gets \LFGen\left( 1^\secp, b, 1^{\ell} \right)$: a probabilistic polynomial-time algorithm that gets as input the security parameter $\secp \in \Nat$, a bit $b$ and a lossyness parameter $\ell \in \Nat$, $\secp \geq \ell$. The algorithm outputs a public key.

    \item
    $y \gets \LFF\left( \pk, x \right)$: a deterministic polynomial-time algorithm that gets as input the security parameter $\secp \in \Nat$, the public key $\pk$ and an input $x \in \{ 0, 1 \}^\secp$ and outputs a string $y \in \{ 0,1 \}^{m}$ for some $m \geq \secp$.
\end{itemize}

The scheme satisfies the following guarantees.

\begin{itemize}
    \item {\bf Statistical Correctness for Injective Mode:}
    There exists a negligible function $\negl(\cdot)$ such that for every $\secp, \ell \in \Nat$, 
    $$
    \Pr_{
    \pk \gets \LFGen\left( 1^\secp, 0, 1^{\ell} \right)
    }
    \left[
    \Big| \Img\left( \LFF\left( \pk, \cdot \right) \right) \Big| = 2^\secp
    \right]
    \geq
    1 - \negl(\secp) \enspace .
    $$

    \item {\bf Statistical Correctness for Lossy Mode:}
    There exists a negligible function $\negl(\cdot)$ such that for every $\secp, \ell \in \Nat$, 
    $$
    \Pr_{
    \pk \gets \LFGen\left( 1^\secp, 1, 1^{\ell} \right)
    }
    \left[
    \Big| \Img\left( \LFF\left( \pk, \cdot \right) \right) \Big| \leq 2^\ell
    \right]
    \geq
    1 - \negl(\secp) \enspace .
    $$
    
    \item {\bf Security:}
    For every polynomial $\poly(\cdot)$ there exists a negligible function $\epsilon$ such that the following holds.
    Let $\secp, \ell \in \Nat$, then (note that in the following computational indistinguishability, the security parameter is $\ell$ and not $\secp$),
    $$
    \biggl\{
    \pk_{0} \: : \: \pk_{0} \gets \LFGen\left( 1^\lambda, 0, \ell \right)
    \biggr\}
    $$
    $$
    \approx_{ \left( \poly(\ell), \epsilon(\ell) \right) }
    \biggl\{
    \pk_{1} \: : \: \pk_{1} \gets \LFGen\left( 1^\lambda, 1, \ell \right)
    \biggr\}
    \enspace .
    $$ 
\end{itemize}

\end{definition}

\paragraph{Different security levels.}
For arbitrary functions $f_{0}, f_{1} : \Nat \rightarrow \Nat$, we say that an LF scheme is $\left( f_{0}, \frac{1}{f_{1}} \right)$-secure, if in the above security part of the definition, we swap $\poly$ with a specific function $f_{0}$ and the negligible function with the function $\frac{1}{f_{1}}$. Concretely, a sub-exponentially secure LF scheme would be one such that there exists a positive real constant $c > 0$ such that the scheme is $\left( 2^{\ell^c}, \frac{1}{2^{\ell^c}} \right)$-secure.


\subsection{Learning With Errors}

Let $\chi_{\sigma}$ be the distribution over $\Z$ where $\Pr[x\gets\chi_{\sigma}]\propto e^{2\pi x^2/\sigma^2}$. 

\begin{definition}Let $T,p,m,q,\sigma $ be functions in $n$ where $m$, $\log(q)$, and $\log(\sigma)$ are bounded by polynomials in $n$. The $(T,p,m,q,\sigma)$-LWE assumption holds if, for any $T$-time adversary $\As$, 
$$
\left|
\Pr\left[
\As(A,v)=1:\substack{A\gets\Z_q^{m\times n}\\v\gets\Z_q^m}
\right]
-
\Pr\left[
\As(A,v)=1:\substack{A\gets\Z_q^{m\times n}\\s\gets\Z_q^n, e\gets\chi_\sigma^m\\v\gets A\cdot s+e\bmod q}
\right]
\right|
\leq p
$$
\end{definition}

We say that LWE is $2^{O(\lambda/\log\lambda)}$-secure with a polynomial noise-modulus ratio if the $(T,p,m,q,\sigma)$-LWE assumption holds for $\sigma=n^{O(1)}$, $q=\sigma\times n^{O(1)}$, $m=\Omega(n\log q)$, and for any $T/p\leq 2^{O(n)}$. Here, we are implicitly setting $\lambda = O(n\log q)=O(n\log n)$, which corresponds to the bit-length of the LWE secret vector $s$.

We say that LWE is optimally-secure if the $(T,p,m,q,\sigma)$-LWE holds for some $\sigma=n^{O(1)}$, $q=2^{O(n)}$, and $m=\Omega(n\log q)$, and for any $T/p\leq 2^{O(n)}$.

Both of these assumptions appear consistent with known lattice attacks, with optimal LWE being the stronger of the two. For optimal LWE in particular, attacks running in time sub-exponential in $n$ seen to require $q/\sigma \geq 2^{\Omega(\sqrt{n\log q})}$, which in the regime of polynomial $\sigma$ translates to $q\geq 2^{\Omega(n)}$. By setting $q=2^{O(n)}$ where the constant in the $O(n)$ is sufficiently small, we hope to avoid such attacks.

\paragraph{Lossy functions from LWE.} In this paper, we will need a lossy function that is $(f_0,1/f_1)$-secure for $f_0=f_1=2^{\ell/\polylog\ell}$. We briefly discuss how to achieve this from our strong variants of LWE.

In~\cite{C:WatZha24}, building on prior works~\cite{STOC:PeiWat08,C:AKPW13}, lossy functions are considered from LWE. We will not give the whole construction here, but just remark that in their construction, they are able to obtain $\ell$ set to be proportional to the bit-length of the LWE secret, namely $\ell=n\log q$. They only need polynomial noise and modulus. As such, by the $2^{O(\lambda/\log \lambda)}$-secure LWE assumption, we can set $n=O(\lambda)$ and hence $\ell=O(\lambda\log\lambda)$, giving the desired lossiness:

\begin{theorem}[Implicit in~\cite{C:WatZha24}] Assuming LWE is $2^{O(\lambda/\log\lambda)}$-secure, there exists lossy functions that are $(2^{O(\lambda/\log\lambda)},2^{-O(\lambda/\log\lambda)})$-secure.
\end{theorem}

\subsection{Two iO Techniques}
Here, we recall two standard iO techniques that we will abstract as useful lemmas.

\paragraph{Sparse Random Triggers.}
Let $P$ be some program and $P'$ an arbitrary different program. Let $R$ be a function with range $[N]$ for some $N$ that is exponential in the security parameter. Let $J_y$ (for `join') be the following program:
$$
J_y(x) =
\begin{cases}
P(x) \text{ if } R(x) \neq y \\
P'(x) \text{ if } R(x) = y
\end{cases}
\enspace .
$$

\begin{lemma} \label{lem:iopuncture}
Suppose one-way functions exist. For sufficiently large polynomial $s$ and for $y$ chosen uniformly in $\{0,1\}^\lambda$, $\iO(1^\lambda,1^s,P)$ and $\iO(1^\lambda,1^s,J_y)$ are computationally indisitnguishable even given the description of $P$. Moreover, $y$ is computationally unpredictable given $P,\iO(1^\lambda,1^s,J_y)$
\end{lemma}

\paragraph{Swapping distributions.}
We now move to the next standard technique. Let $\{ D^x_0 \}_x, \{ D^x_1 \}_x$ be two families of distributions over the same domain $\Ys$, which can also be thought of as deterministic functions $D_0(x;r)$, $D_1(x; r)$ that take as input an index $x$ and some random coins $r$. Let $P$ be a program that makes queries to an oracle $O: \Xs \rightarrow \Ys$ for some set $\Xs$. Then we have the following:

\begin{lemma} \label{lem:distswap}
Let $\left( \prf, \punc \right)$ be a $\left( f_{\prf}, \delta_\prf \right)$-secure puncturable PRF and $\iO$ be a $\left( f_{\iO}, \delta_\iO \right)$-secure iO. Let $\Xs$ a finite set and let $D_0 := \{ D_{0, x} \}_{x \in \Xs}$,$D_1 := \{ D_{1, x} \}_{x \in \Xs}$ two ensembles of distributions, such that for every $x \in \Xs$, $D_{0, x}$, $D_{1, x}$ are $\left( f_{D}, \delta_{D} \right)$-indistinguishable. Let $E_0(x) = D_0\left( x ; \prf(k,x) \right)$ and $E_1(x) = D_1\left( x ; \prf\left( \overline{k}, x \right) \right)$. Then for a sufficiently large polynomial $s$, $\iO\left( 1^\lambda, 1^s, P^{E_0} \right)$ and $\iO\left( 1^\lambda, 1^s, P^{E_1} \right)$ are $\left( \min\left( f_{\prf}, f_{\iO}, f_{D} \right), O\left( |\Xs|\cdot \left( \delta_\prf + \delta_\iO + \delta_{D} \right) \right) \right)$-computationally indistinguishable, where $k,\overline{k}\gets\{0,1\}^\lambda$ are uniformly random keys. 
\end{lemma}

\subsection{Subspace Hiding Functions} \label{section:subspace_hiding_function}
We prove a generalization of subspace-hiding obfuscation, which we call subspace-hiding function. Basically, the idea is this: We have a small subspace $S_{0}$, and a (possibly significantly) bigger subspace $S$, which contains $S_{0}$. Then, we have a partition $P$ of the bigger space $S$ into cosets, all of them have the same size and all are parallel cosets to $S_{0}$ (that is, each of the cosets in the partition $P$ is in the form of $x + S_{0}$ for some $x \in S$). Then, there is access to the function $F_{S}$ which given $\vecV \in \bbZ_{2}^{k}$, checks whether it is in $S$ and if it does then it outputs the index of the coset it belongs to inside the partition $P$.

The statement intuitively shows that $F$ hides $S_{0}$ as long as the dimension of $S$ (and thus in particular the dimension of $S_0$) isn't too big. The proof extends the proof of subspace hiding from \cite{EC:Zhandry19b} and generalizes the statement. We also allow to get $\ell$-duplicated the access to $F$ (for a general $\ell$), which is sometimes helpful for using the statement as a black box. The formal statement follow. 

\begin{lemma} [Subspace Hiding Function] \label{lemma:subspace_hiding_function}
Let $k, r, s, \lambda \in \Nat$ such that $r + \lambda + s \leq k$, let $S \subseteq \bbZ_{2}^{k}$ a subspace of dimension $r + \lambda$ and let $S_{0} \subset S$ a subspace of dimension $r$. Let $P$ any partition of $S$, the bigger space, into $2^\lambda$ cosets that are all parallel to $S_{0}$, the smaller space. Let $F_{S} : \bbZ_{2}^{k} \rightarrow \left( \{ \bot, \top \} \times \bbZ_{2}^{\lambda} \right)$ be the function that given input $\vecV \in \bbZ_{2}^{k}$, first checks whether $\vecV \in S$ and outputs $\bot$ if not, and otherwise outputs the index $\vecC_{\vecV} \in \bbZ_{2}^{\lambda}$ of the coset of $S_{0}$ (inside the big space $S$), such that $\vecV$ belongs to it.

\vspace{1mm}
Let $\Ss_{s}$ the uniform distribution over subspaces $T_{0}$ of dimension $r + s$ such that $S_0 \subset T_{0} \subset \bbZ_{2}^{k}$. We define $T$ as the linear span of $T_{0}$ and $S$ and define $F_{T}$ as the function which computes the index of parallel cosets of $T_{0}$ inside $T$, with respect to the same partition $P$. More formally, let $S_{0, \vecC}$ the $\vecC$-th parallel coset of $S_{0}$ inside the space $S$, for $\vecC \in \bbZ_{2}^{\lambda}$. Let $T_{0, \vecC}$ the linear sum of $T_{0}$ and $S_{0, \vecC}$. Then $T_{0, \vecC}$ is the $\vecC$-th parallel coset of $T_{0}$ inside the space $T$. Then, we have the following guarantees.

\vspace{1mm}
\textbf{Information theoretical indistinguishability in the oracle model:}
For every $\ell \in \Nat$, for every oracle-aided quantum algorithm $\Adv$ making at most $q$ quantum queries, we have the following indistinguishability between oracle distributions:
$$
\{
\Oracle_{F_{S}}^{ \otimes \ell }
\}
\approx_{ \frac{ \lambda \cdot \ell \cdot q \cdot s }{ \sqrt{ 2^{k - r - \lambda - s} } } }
\{
\Oracle_{F_{T}}^{ \otimes \ell }
\;
:
\;
T_{0} \gets \Ss_{s} 
\} \enspace .
$$

\textbf{Computational indistinguishability in the standard model:}
Denote $\lambda' := \lambda_{\OWF} := k - r - \lambda - s$, $\lambda_{\iO} := \poly(\secp')$ security parameters for OWF and iO, and let $\iO$ an indistinguishability obfuscation scheme that is $\left( f_{\iO}(\cdot), \epsilon_{\iO}(\cdot) \right)$-secure, and assume that $\left( f_{\OWF}(\cdot), \epsilon_{\OWF}(\cdot) \right)$-secure injective one-way functions exist. Then, for a sufficiently large $p := p(\secp')$ polynomial in the security parameter, we have the following indistinguishability for $f(\secp') := \min\left( f_{\OWF}(\secp'), f_{\iO}\left( \lambda_{\iO} \right) \right)$, $\epsilon(\secp') := \max\left( \epsilon_{\OWF}(\secp'), \epsilon_{\iO}\left( \lambda_{\iO} \right) \right)$ and for every $\ell \in \Nat$
$$
\{
\Obf_{F_{S}}
\;
:
\;
\Obf_{F_{S}} \gets \iO\left( 1^{\secp_{\iO}}, 1^{p}, C_{F_{S}} \right)
\}^{ \otimes \ell }
\approx_{\left( \frac{f(\secp') - \poly(\secp')}{ \lambda \cdot \ell }, \; s \cdot \epsilon(\secp') \right)}
$$
$$
\{
\{
\Obf_{F_{T}}
\;
:
\; 
\Obf_{F_{T}} \gets \iO\left( 1^{\secp_{\iO}}, 1^{p}, C_{F_{T}} \right)
\}^{ \otimes \ell }
\;
:
\;
T_{0} \gets \Ss_{s}
\}
\enspace .
$$
\end{lemma}

\begin{proof}
    We prove the claim by a hybrid argument with $s$ steps, increasing the dimension of the random superspace $T_{0}$ by $1$ at each step.

    \vspace{2mm}
    \noindent
    We start with algebraically characterizing the subspaces and cosets involved in the proof. Specifically, the goal of the first step of the proof will be to reduce the task of indistinguishability between the functions to indistinguishability in the form of subspace hiding. Observe that for the subspaces $S_0, S$ and the partition $P$ of parallel cosets to $S_{0}$, there exists a matrix $\matS \in \bbZ_{2}^{k \times (k - r)}$ such that
    $$
    S_{0} = \colspan\left( \matS \right)^{\bot} \enspace , \enspace
    S = \colspan\left( \matS^{\left( k - r - \lambda \right)} \right)^{\bot} \enspace ,
    $$
    where $\matS^{\left( k - r - \lambda \right)} \in \bbZ_{2}^{k \times (k - r - \lambda)}$ are the $k - r - \lambda$ leftmost columns of $\matS$. Also, $\matS$ satisfies that for the partition $P$ and every input $\vecV \in \bbZ_{2}^{k}$ to $F$, the output can be computed using $\matS$ as follows: check that $\vecV^{T} \cdot \matS^{\left( k - r - \lambda \right)} = 0^{k - r - \lambda}$ and output $\bot$ if not. Otherwise, the output is the inner product between $\vecV$ and the remaining $\lambda$ rightmost columns of $\matS$.

    We next define $\lambda$ subspaces as follows: For $j \in [\lambda]$ we have $S_{j} = \colspan\left( \matS^{\left( k - r - \lambda \right)}, \matS_{j} \right)^{\bot}$, where $\matS_{j}$ is $j$-th rightmost column of $\matS$. So, for every $j \in [\lambda]$, the subspace $S_{j}$ is the set of vectors $\vecV \in \bbZ_{2}^{k}$ such that $\vecV^{T} \cdot \left( \matS^{\left( k - r - \lambda \right)}, \matS_{j} \right) = 0^{k - r - \lambda + 1}$.

    We next describe a further equivalent way for computing $F$ using only black-box access to the membership checks for $S$ and the $\lambda$ additional subspaces: First check that the input vector $\vecV$ is in $S$ and output $\bot$ if not. Then, the $j$-th output bit of $F$ is exactly the negation of the membership of $\vecV$ inside $S_{j}$. So, at this point we described the computation of $F$ using black-box access to a small number of subspaces on the one hand, but also there's a matrix dominating this computation, inside the black boxes of the membership checks.

    \paragraph{Increasing the dimension of all subspaces all at once.}
    Consider the sequence of subspaces $\left( S^{0}_0, \cdots, S^{0}_{\lambda}, S^{0} \right)$ defined as $\left( S_0, \cdots, S_{\lambda}, S \right)$, and we will increase the dimension of all $\lambda + 2$ subspaces simultaneously by $1$, indistinguishably and randomly. We do this as follows: we sample a uniformly random $\vecA \in \bbZ_{2}^{k - r}$ and consider the modified subspaces $S^{1}_0, \cdots, S^{1}_{\lambda}, S^{1}$ such that for $j \in [\lambda]$, the subspace $S^{1}_{j}$ is the set of vectors $\vecV \in \bbZ_{2}^{k}$ such that $\vecV^{T} \cdot \left( \matS^{\left( k - r - \lambda \right)}, \matS_{j} \right)$ either equals $0^{k - r - \lambda + 1}$ or to the corresponding elements of $\vecA$: Formally, the leftmost $k - r - \lambda$ elements of $\vecA$ should equal $\vecV^{T} \cdot \matS^{\left( k - r - \lambda \right)}$ and also we should have the equality $\vecA_{j} = \vecV^{T} \cdot \matS_{j}$, where $\vecA_{j}$ is the $j$-th rightmost element of $\vecA$. Similarly, the subspace $S_{0}^{1}$ is the set of vectors $\vecV \in \bbZ_{2}^{k}$ such that $\vecV^{T} \cdot \matS$ either equals $0^{k - r}$ or to the entirety of $\vecA$. The subspace $S^{1}$ is the set of vectors $\vecV \in \bbZ_{2}^{k}$ such that $\vecV^{T} \cdot \matS^{\left( k - r - \lambda \right)}$ either equals $0^{k - r - \lambda}$ or to the corresponding leftmost elements of $\vecA$.
    
    We have the following guarantees:
    \begin{itemize}
        \item 
        For all $j \in [\lambda]$, conditioned on the leftmost $k - r - \lambda$ elements of $\vecA$ do not equal $0^{k - r - \lambda}$ (which happens with overwhelming probability), the subspace $S^{1}_{j}$ is a random superspace of $S^{0}_{j}$, with one more dimension.

        \item 
        \textbf{Simulation of all other membership checks $S_{0}^{b}, \cdots, S_{\lambda}^{b}$ using oracle access to the biggest subspace $S^{b}$:}
        We consider oracle access to $\Oracle$, which is a membership check for $S^{b}$, for either $b=0$ or $b=1$. We use it to implement all other $\lambda + 1$ membership checks $S_{0}^{b}, \cdots, S_{\lambda}^{b}$. We are given oracle access to membership check in $S^{b}$ and recall that it is given by the inner product with the matrix $\matS^{(k - r - \lambda)}$ and the last $k - r - \lambda$ elements of the uniformly random $\vecA$. We first sample the remaining $\lambda$ elements of $\vecA$ ourselves. In order to implement oracle access to $S_{0}^{b}$ for input $\vecV \in \bbZ_{2}^{k}$, we check membership in $S^{b}$, if the answer was negative we reject. If the answer was positive, we check whether $\vecV^{T} \cdot \matS^{(k - r - \lambda)} = 0^{k - r - \lambda}$: if so, we accept iff $\vecV^{T} \cdot \matS_{j} = 0$, and if $\vecV^{T} \cdot \matS^{(k - r - \lambda)} \neq 0^{k - r - \lambda}$, we accept iff $\vecV^{T} \cdot \matS_{j}$ equals the element we sampled (ourselves) for the $j$-th element of $\vecA$. The simulation of the smallest membership check $S_{0}^{b}$ is done analogously (by checking all $\lambda$ columns instead of a single one). One can verify that for $b \in \{ 0, 1 \}$, if the oracle access was for $S^{b}$ then we simulated access for $S_{0}^{b}, \cdots, S_{\lambda}^{b}$. this means that overall a parallel query to all $\lambda + 1$ oracles can be simulated by $\lambda + 2$ oracle queries to the biggest membership check. We obtain the following indistinguishability statements given this simulation.

        \item 
        \textbf{Information theoretical indistinguishability:}
        Due to the randomness of $\vecA$ (and in particular its leftmost $k - r - \lambda$ elements), by standard quantum lower bounds for unstructured search, for every $\ell \in \Nat$ we have that for any $q$-query algorithm,
        $$
        \{ \left( \Oracle_{S^{0}_{j}} \right)_{ j \in \{ 0, 1, \cdots, \lambda \} }, \Oracle_{ S^{0} } \}^{ \otimes \ell }
        \approx_{\frac{ \lambda \cdot \ell \cdot q }{ \sqrt{ 2^{k - r - \lambda} } }}
        \{ \left( \Oracle_{S^{1}_{j}} \right)_{ j \in \{ 0, 1, \cdots, \lambda \} }, \Oracle_{ S^{1} } \}^{ \otimes \ell }
        \enspace .
        $$

        \item 
        \textbf{Computational indistinguishability:}
        We can think of the leftmost $k - r - \lambda$ bits of the uniformly random $\vecA \in \bbZ_{2}^{k - r}$ as a secret image of an injective one-way function. By known techniques in indistinguishability obfuscation and one-way functions (subspace-hiding obfuscation \cite{zhandry2021quantum} and extractability obfuscation \cite{boyle2014extractability}) we have,
        $$
        \{
        \Obf_{S^{0}_{0}}, \cdots, \Obf_{S^{0}_{\lambda}}, \Obf_{S^{0}}
        \;
        :
        \;
        \Obf_{S^{0}_{i}} \gets \iO\left( 1^{\secp_{\iO}}, 1^{p}, C_{S^{0}_{i}} \right)
        \}^{ \otimes \ell }
        \approx_{\left( \frac{f(\secp') - \poly(\secp')}{ \lambda \cdot \ell }, \; \epsilon(\secp') \right)}
        $$
        $$
        \{
        \{
        \Obf_{S^{1}_{0}}, \cdots, \Obf_{S^{1}_{\lambda}}, \Obf_{S^{1}}
        \;
        :
        \;
        \Obf_{S^{1}_{i}} \gets \iO\left( 1^{\secp_{\iO}}, 1^{p}, C_{S^{1}_{i}} \right)
        \}^{ \otimes \ell }
        \;
        :
        \;
        \vecA \gets \bbZ_{2}^{k - r}
        \}
        $$
    \end{itemize}
    Finally, note that if we compute $F$ using the (indistinguishable) membership checks for $S^{1}_0, \cdots, S^{1}_{\lambda}, S^{1}$ instead of the original $S^{0}_0, \cdots, S^{0}_{\lambda}, S^{0}$, we get a corresponding partition and specifically, we get exactly a function $F_{T}$ such that the random superspace $T_{0}$ of $S_{0}$ is one dimension bigger.

    \paragraph{Generalizing to an arbitrary $s$.}
    The above proves our claim for $s = 1$. For a general $s$ we continue making dimension additions in the exact same manner. At each of the steps $i \in [s]$ we have the subspaces $\{ S^{i - 1}_{j} \}_{j \in \{ 0, 1, \cdots, \lambda \} }, S^{i - 1}$ and we move to subspaces $\{ S^{i}_{j} \}_{j \in \{ 0, 1, \cdots, \lambda \} }, S^{i}$ that are one dimension bigger. At the beginning of every step $i$ we also re-initialize the matrix for a different size $\matS_{i} \in \bbZ_{2}^{k \times (k - r - (i - 1))}$. Note that at step $i \in [s]$ the oracle indistinguishability is (for any $q$-query algorithm),
    $$
        \{
        \{ \Oracle_{S^{i - 1}_{j}} \}_{j \in \{ 0, 1, \cdots, \lambda \} }, \Oracle_{ S^{i - 1} }
        \}^{ \otimes \ell }
        \approx_{\frac{ \lambda \cdot \ell \cdot q }{ \sqrt{ 2^{k - r - \lambda - (i - 1)} } }}
        \{
        \{ \Oracle_{S^{i}_{j}} \}_{j \in \{ 0, 1, \cdots, \lambda \} }, \Oracle_{ S^{i} }
        \}^{ \otimes \ell }
        \enspace ,
    $$
    and the computational indistinguishability is 
    $$
        \{
        \Obf_{S^{i - 1}_{0}}, \cdots, \Obf_{S^{i - 1}_{\lambda}}, \Obf_{S^{i - 1}}
        \;
        :
        \;
        \Obf_{S^{i - 1}_{j}} \gets \iO\left( 1^{\secp_{\iO}}, 1^{p}, C_{S^{i - 1}_{j}} \right)
        \}^{ \otimes \ell }
        \approx_{\left( \frac{f(\secp') - \poly(\secp')}{\lambda \cdot \ell}, \; \epsilon(\secp') \right)}
    $$
    $$
        \{
        \{
        \Obf_{S^{i}_{0}}, \cdots, \Obf_{S^{i}_{\lambda}}, \Obf_{S^{i}}
        \;
        :
        \;
        \Obf_{S^{i}_{j}} \gets \iO\left( 1^{\secp_{\iO}}, 1^{p}, C_{S^{i}_{j}} \right)
        \}^{ \otimes \ell }
        \;
        :
        \;
        \vecA \gets \bbZ_{2}^{k - r}
        \}
        \enspace .
    $$

    Overall we get that for a $q$-query algorithm $\Adv$, the distinguishing advantage between
    $$
    \{ \Oracle_{S_0}, \cdots, \Oracle_{S_{\lambda}}, \Oracle_{ S } \}^{ \otimes \ell }
    \enspace , \enspace
    \{ \Oracle_{T_0}, \cdots, \Oracle_{T_{\lambda}}, \Oracle_{ T } \; : \; T_0 \gets \Ss_{s} \}^{ \otimes \ell }
    $$
    is 
    $$
    \sum_{ i \in \{ 0, 1, \cdots, s - 1 \} } \frac{ \lambda \cdot \ell \cdot q }{ \sqrt{ 2^{ k - r - \lambda - i } } }
    \leq
    \frac{ \lambda \cdot \ell \cdot q \cdot s }{ \sqrt{ 2^{ k - r - \lambda - s } } }
    \enspace ,
    $$
    as needed for the oracle case.

    As for the computational case, we have a similar indistinguishability
    $$
    \{
    \Obf_{S_{0}}, \cdots, \Obf_{S_{\lambda}}, \Obf_S
    \;
    :
    \;
    \Obf_{S_{j}} \gets \iO\left( 1^{\secp_{\iO}}, 1^{p}, C_{S_{j}} \right)
    \}^{ \otimes \ell }
    \approx_{\left( \frac{f(\secp') - \poly(\secp')}{ \lambda \cdot \ell }, \; s \cdot \epsilon(\secp') \right)}
    $$
    $$
    \{
    \{
    \Obf_{T_{0}}, \cdots, \Obf_{T_{\lambda}}, \Obf_T
    \;
    :
    \;
    T \gets \Ss_{s}, 
    \Obf_{T_{j}} \gets \iO\left( 1^{\secp_{\iO}}, 1^{p}, C_{T_{j}} \right)
    \}^{ \otimes \ell }
    \;
    :
    \;
    T_{0} \gets \Ss_{s}
    \}
    \enspace .
    $$

    To end the proof, it remains to observe that in the same manner that we implemented $F_{S}$ using membership checks for the subspaces $S_{0}, S_{1}, \cdots, S_{\lambda}, S$, we can use the membership checks for $T_{0}, T_{1}, \cdots, T_{\lambda}, T$ to implement the function $F_{T}$. This finishes our proof.
\end{proof}

\vspace{1mm}
\paragraph{Indistinguishability between static and independent repetitions of $F_{T}$.}
We next state the following Corollary below, which will be useful for followup supporting lemmas. The Corollary intuitively says that the following two distributions are indistinguishable. The first is where we sample $T_{0} \gets \Ss_{s}$ once and then give $\ell$-duplicate access to $F_{T}$ (in the case of standard-model obfuscated circuits, this means we sample $\ell$ i.i.d. obfuscations of $F_{T}$). The second distribution is where we generate $\ell$ i.i.d. samples, that is, for every $i \in [\ell]$ we sample a fresh $T_{0}^{i} \gets \Ss_{s}$, and give access to the $\ell$ functions $F_{T^{1}}, \cdots, F_{T^{\ell}}$. The formal statement and proof follow.

\begin{corollary} \label{corollary:subspace_hiding_function_main_corollary} [Indistinguishability between static and independent repetitions of $F_{T}$]
Let $k, r, s, \lambda \in \Nat$ such that $r + \lambda + s \leq k$, let $S \subseteq \bbZ_{2}^{k}$ a subspace of dimension $r + \lambda$ and let $S_{0} \subset S$ a subspace of dimension $r$. Let $P$ any partition of $S$, the bigger space, into $2^\lambda$ cosets that are all parallel to $S_{0}$, the smaller space. Let $F_{S} : \bbZ_{2}^{k} \rightarrow \left( \{ \bot, \top \} \times \bbZ_{2}^{\lambda} \right)$ be the function that given input $\vecV \in \bbZ_{2}^{k}$, first checks whether $\vecV \in S$ and outputs $\bot$ if not, and otherwise outputs the index $\vecC_{\vecV} \in \bbZ_{2}^{\lambda}$ of the coset of $S_{0}$ (inside the big space $S$), such that $\vecV$ belongs to it.

\vspace{1mm}
Let $\Ss_{s}$ the uniform distribution over subspaces $T_{0}$ of dimension $r + s$ such that $S_0 \subset T_{0} \subset \bbZ_{2}^{k}$. We define $T$ as the linear span of $T_{0}$ and $S$ and define $F_{T}$ as the function which computes the index of parallel cosets of $T_{0}$ inside $T$, with respect to the same partition $P$. More formally, let $S_{0, \vecC}$ the $\vecC$-th parallel coset of $S_{0}$ inside the space $S$, for $\vecC \in \bbZ_{2}^{\lambda}$. Let $T_{0, \vecC}$ the linear sum of $T_{0}$ and $S_{0, \vecC}$. Then $T_{0, \vecC}$ is the $\vecC$-th parallel coset of $T_{0}$ inside the space $T$. Then, we have the following guarantees.

\vspace{1mm}
\textbf{Information theoretical indistinguishability in the oracle model:}
For every $\ell \in \Nat$, for every oracle-aided quantum algorithm $\Adv$ making at most $q$ quantum queries, we have the following indistinguishability between oracle distributions:
$$
\{
\Oracle_{F_{T}}^{ \otimes \ell }
\;
:
\;
T_{0} \gets \Ss_{s} 
\}
\approx_{ \frac{ 2 \cdot \lambda \cdot \ell \cdot q \cdot s }{ \sqrt{ 2^{k - r - \lambda - s} } } }
\{
\Oracle_{ F_{T^{1}} }, \cdots, \Oracle_{ F_{T^{\ell}} }
\;
:
\;
\forall i \in [\ell], T_{0}^{i} \gets \Ss_{s} 
\}
\enspace .
$$

\textbf{Computational indistinguishability in the standard model:}
Denote $\lambda' := \lambda_{\OWF} := k - r - \lambda - s$, $\lambda_{\iO} := \poly(\secp')$ security parameters for OWF and iO, and let $\iO$ an indistinguishability obfuscation scheme that is $\left( f_{\iO}(\cdot), \epsilon_{\iO}(\cdot) \right)$-secure, and assume that $\left( f_{\OWF}(\cdot), \epsilon_{\OWF}(\cdot) \right)$-secure injective one-way functions exist. Then, for a sufficiently large $p := p(\secp')$ polynomial in the security parameter, we have the following indistinguishability for $f(\secp') := \min\left( f_{\OWF}(\secp'), f_{\iO}\left( \lambda_{\iO} \right) \right)$, $\epsilon(\secp') := \max\left( \epsilon_{\OWF}(\secp'), \epsilon_{\iO}\left( \lambda_{\iO} \right) \right)$ and for every $\ell \in \Nat$
$$
\{
\{
\Obf_{F_{T}}
\;
:
\; 
\Obf_{F_{T}} \gets \iO\left( 1^{\secp_{\iO}}, 1^{p}, C_{F_{T}} \right)
\}^{ \otimes \ell }
\;
:
\;
T_{0} \gets \Ss_{s}
\}
\approx_{\left( \frac{f(\secp') - \poly(\secp')}{2\cdot \lambda \cdot \ell}, \; s \cdot \epsilon(\secp') \right)}
$$
$$
\{
\Obf_{ F_{T^{i}} }, \cdots , \Obf_{ F_{T^{\ell}} }
\;
:
\;
\forall i \in [\ell], T_{0}^{i} \gets \Ss_{s}
,
\Obf_{ F_{T^{i}} } \gets \iO\left( 1^{\secp_{\iO}}, 1^{p}, C_{F_{T^{i}}} \right)
\}
\enspace .
$$
\end{corollary}

\begin{proof}
    The proof is simple and is given by two steps. We start with the distribution on the left, whether it is the oracle model case or standard-model obfuscations. It is indistinguishable from $\ell$-access to just $F_{S}$, with parameters given by Lemma \ref{lemma:subspace_hiding_function}. Next, in the second step of the proof we do an $\ell$-step hybrid argument, invoking the same Lemma \ref{lemma:subspace_hiding_function}, each time using the "$\ell$" in the Lemma \ref{lemma:subspace_hiding_function} equal $1$, but use it $\ell$ times here. Overall, we get the indistinguishability parameters (in the oracle model or standard model) as in this Corollary \ref{corollary:subspace_hiding_function_main_corollary}.
\end{proof}

\subsection{Hardness of Concentration in Hidden Random Subspace}
We re-prove and generalize results from \cite{C:ShmZha25, C:Shmueli22, shmueli2022public}, regarding hardness of finding vectors that are always inside some hidden subspace $T$. Specifically, we generalize them to the setting including subspace-hiding functions (which generalize subspace-hiding of a single subspace membership check). The formal statement and proof follow.

\begin{lemma} [IO Dual Subspace Anti-Concentration] \label{lemma:dual_subspace_concentration}
Let $k, r, s, \lambda \in \Nat$ such that $r + \lambda + s \leq k$, let $S \subseteq \bbZ_{2}^{k}$ a subspace of dimension $r + \lambda$ and let $S_{0} \subset S$ a subspace of dimension $r$. Let $P$ any partition of $S$, the bigger space, into $2^\lambda$ cosets that are all parallel to $S_{0}$, the smaller space. Let $F_{S} : \bbZ_{2}^{k} \rightarrow \left( \{ \bot, \top \} \times \bbZ_{2}^{\lambda} \right)$ be the function that given input $\vecV \in \bbZ_{2}^{k}$, first checks whether $\vecV \in S$ and outputs $\bot$ if not, and otherwise outputs the index $\vecC_{\vecV} \in \bbZ_{2}^{\lambda}$ of the coset of $S_{0}$ (inside the big space $S$), such that $\vecV$ belongs to it.

\vspace{1mm}
Let $\Ss_{s}$ the uniform distribution over subspaces $T_{0}$ of dimension $r + s$ such that $S_0 \subset T_{0} \subset \bbZ_{2}^{k}$. We define $T$ as the linear span of $T_{0}$ and $S$ and define $F_{T}$ as the function which computes the index of parallel cosets of $T_{0}$ inside $T$, with respect to the same partition $P$. More formally, let $S_{0, \vecC}$ the $\vecC$-th parallel coset of $S_{0}$ inside the space $S$, for $\vecC \in \bbZ_{2}^{\lambda}$. Let $T_{0, \vecC}$ the linear sum of $T_{0}$ and $S_{0, \vecC}$. Then $T_{0, \vecC}$ is the $\vecC$-th parallel coset of $T_{0}$ inside the space $T$. Then, we have the following guarantees.

\vspace{1mm}
\textbf{Information theoretical hardness in the oracle model:}
Assume there is an oracle-aided quantum algorithm $\Adv$ making at most $q$ quantum queries and outputting a vector $\vecU \in \bbZ_{2}^{k}$ at the end of its execution, such that 
$$
\Pr
\left[
\Adv^{\Oracle_{ F_{T} }} \in \left( T_{0}^{\bot} \setminus \{ 0 \} \right)
\;
:
\;
T_{0} \gets \Ss_{s}
\right]
\geq
\epsilon
\enspace .
$$
Also, denote $t := k - r - s$, $\ell := \frac{ k\left( t + 1 \right) }{ \epsilon }$ and assume the following conditions:
\begin{enumerate}
    \item \label{subspace_concentration_condition_1_unconditional}
    $\frac{t}{2^{ s - t }} \leq \frac{\epsilon}{2}$, and

    \item \label{subspace_concentration_condition_2_unconditional}
    $\frac{ 3 \cdot \lambda \cdot \ell^2 \cdot q \cdot s }{ \sqrt{ 2^{k - r - \lambda - s} } } \leq \frac{1}{2}$.
\end{enumerate}
Then, it is necessarily the case that 
$$
\Pr
\left[
\Adv^{\Oracle_{ F_{T} }} \in \left( S_{0}^{\bot} \setminus T_{0}^{\bot} \right)
\;
:
\;
T_{0} \gets \Ss_{s}
\right]
\geq
\frac{\epsilon}{16 \cdot k \cdot (t + 1)}
\enspace .
$$

\textbf{Computational hardness in the standard model:}
Denote $\lambda' := \lambda_{\OWF} := k - r - \lambda - s$, $\lambda_{\iO} := \poly(\secp')$ security parameters for OWF and iO, and let $\iO$ an indistinguishability obfuscation scheme that is $\left( f_{\iO}(\cdot), \epsilon_{\iO}(\cdot) \right)$-secure, and assume that $\left( f_{\OWF}(\cdot), \epsilon_{\OWF}(\cdot) \right)$-secure injective one-way functions exist. Then, for a sufficiently large $p := p(\secp')$ polynomial in the security parameter, we have the following guarantees for $f(\secp') := \min\left( f_{\OWF}(\secp'), f_{\iO}\left( \lambda_{\iO} \right) \right)$, $\epsilon(\secp') := \max\left( \epsilon_{\OWF}(\secp'), \epsilon_{\iO}\left( \lambda_{\iO} \right) \right)$. Denote by $\Oracle_{ \secp, p, s }$ the distribution over obfuscated circuits that samples $T_{0} \gets \Ss_{s}$ and then $\Obf_{ F_{T} } \gets \iO\left( 1^{\secp_{\iO}}, 1^{p}, C_{ F_{T} } \right)$.

Assume there is a quantum algorithm $\Adv$ of complexity $T_{\Adv}$ such that,
$$
\Pr
\left[
\Adv\left( \Obf_{ F_{T} } \right) \in \left( T_{0}^{\bot} \setminus \{ 0 \} \right)
\;
:
\;
\Obf_{ F_{T} } \gets \Oracle_{ \secp, p, s }
\right]
\geq \epsilon
\enspace .
$$
Also, denote $t := k - r - s$, $\ell := \frac{ k\left( t + 1 \right) }{ \epsilon }$ and assume the following conditions:
\begin{enumerate}
    \item \label{subspace_concentration_condition_1_computational}
    $\frac{t}{2^{ s - t }} \leq \frac{\epsilon}{2}$, and

    \item \label{subspace_concentration_condition_2_computational}
    $\ell \cdot T_{\Adv} + \ell \cdot k^{3} \leq \frac{f(\secp') - \poly(\secp')}{2\cdot \lambda \cdot \ell}$, and

    \item \label{subspace_concentration_condition_3_computational}
    $(s + 1) \cdot \epsilon(\lambda') \leq 1/2$.
\end{enumerate}

Then, it is necessarily the case that 
$$
\Pr
\left[
\Adv\left( \Obf_{ F_{T} } \right) \in \left( S_{0}^{\bot} \setminus T_{0}^{\bot} \right)
\;
:
\;
\Obf_{ F_{T} } \gets \Oracle_{ \secp, p, s }
\right]
\geq
\frac{\epsilon}{16 \cdot k \cdot (t + 1)}
\enspace .
$$
\end{lemma}

\begin{proof}
We start with defining the following reduction $\AdvB$, that will use the circuit $\Adv$ as part of its machinery.

\paragraph{The reduction $\AdvB$.}
The input to $\AdvB$ contains $\ell := \frac{ k \cdot \left( t + 1 \right) }{ \epsilon }$ samples of an access to some $F$, $\left( \Obf^{(1)}, \cdots, \Obf^{(\ell)} \right)$, for $t := k - r - s$. These samples can be oracle access (in the classical oracle model) or obfuscated circuits (in the standard model). Given the $\ell$ samples, execute $\Adv\left( \Obf^{(i)} \right)$ for every $i \in [\ell]$ and obtain $\ell$ vectors $\{ u_{1}, \cdots, u_{\ell} \}$. Then, take only the vectors $\{ v_{1}, \cdots, v_{m} \}$ that are inside $S_{0}^{\bot}$, and then compute the dimension of their span, $D := \dim\left( \linspan\left( v_{1}, \cdots, v_{m} \right) \right)$. Note the following properties of the reduction:
\begin{itemize}
    \item 
    In the oracle model, the number of queries that $\AdvB$ makes is $q \cdot \ell$.

    \item
    In the standard model, the running time of $\AdvB$ is $\ell \cdot T_{\Adv} + \ell \cdot k^{3}$, where $\ell \cdot T_{\Adv}$ is for producing the $\ell$ outputs of $\Adv$ and $\ell \cdot k^{3}$ is for (naively) executing Gaussian elimination $\ell$ times, to repeatedly check whether the new vector $v_{i}$ adds a dimension i.e., whether it is outside of the span $\linspan\left( v_{1}, \cdots, v_{i - 1} \right)$ of the previous vectors.
\end{itemize}

From hereon we will use the notations for the standard model, and these implicitly mean we are giving black-box oracle access in the classical oracle model. We will be careful to make the distinction, when it is needed, between what happens in the oracle model and what happens in the standard model, when the adversary gets the obfuscated circuits themselves.

\paragraph{Executing $\AdvB$ on the distribution $\Ds_{1}$.}
Consider the following distribution $\Ds_{1}$: Sample $\ell$ i.i.d superspaces $T_{0}^{1}, \cdots, T_{0}^{\ell}$, and for each of them, send a sample of the function $F_{T^{i}}$: $\Obf_{ F_{T^{1}} }, \cdots, \Obf_{ F_{T^{\ell}} }$. Observe that this is one of the two distributions proven to be indistinguishable in \ref{corollary:subspace_hiding_function_main_corollary}. Let us see what happens when we execute $\AdvB$ on a sample from the distribution $\mathcal{D}_{1}$.

Consider the $\ell$ vectors $\{ u_{1}, \cdots, u_{\ell} \}$ obtained by executing $\Adv$ on each of the input samples. Recall that $\ell := \frac{1}{\epsilon} \cdot k \cdot \left( t + 1 \right)$ and consider a partition of the vectors into $t + 1$ consecutive sequences (or buckets), accordingly, each of length $\frac{1}{\epsilon} \cdot k$. In order to show that the probability for the reduction $\AdvB$ to have $D \geq t + 1$ is high, we show that with high probability, in each bucket $j \in [t + 1]$ there is a vector $u_{i}$ that's inside the corresponding dual $\left( T_{0}^{i} \right)^{\bot}$, but such that also the intersection between $\left( T_{0}^{i} \right)^{\bot}$ and the linear span of the previous $j - 1$ vectors that hit their duals by $\Adv$, is only the zero vector $0^{k}$. Note that the last condition indeed implies $D \geq t + 1$. 

For every $i \in [\ell]$ we define the probability $p_{i}$. We start with defining it for the indices $i$ from the first bucket, and then proceed to define it recursively for the rest of the buckets. For indices $i \in [\frac{1}{\epsilon} \cdot k]$ in the first bucket, $p_{i}$ is the probability, over sampling $T_{0}^{i}$ and over the execution of $\Adv$, for the event that given $\Obf_{ F_{T^{i}} }$, the output of $\Adv$ is $u_{i} \in \left( \left( T_{0}^{i} \right)^{\bot} \setminus \{ 0 \} \right)$, and in such case we define the $i$-th execution as successful.
We denote by $u_{(1)}$ the first vector in the first bucket where a successful execution happens (and define $u_{(1)} := \bot$ if no success happened over the entire bucket).
For any $i$ inside any bucket $j \in \left( [t + 1] \setminus \{ 1 \} \right)$ other than the first bucket, we define $p_{i}$ as the probability that (1) as before, $u_{i} \in \left( \left( T_{0}^{i} \right)^{\bot} \setminus \{ 0 \} \right)$, and also (2) the intersection between $\left( T_{0}^{i} \right)^{\bot}$ and the linear span of the first vectors (from each respective bucket so far) is only $\{ 0^k \}$, that is,
$$
\left( T_{0}^{i} \right)^{\bot} \cap \linspan\left( u_{(1)}, \cdots, u_{(j - 1)} \right) = \{ 0^k \} \enspace .
$$
In words, $p_{i}$ is the probability that the output of the adversary hits the dual subspace, and also the (randomly-sampled) dual does not have a non-trivial intersection with the space of vectors that we aggregated so far as the first to be successful. We prove that with high probability, all $t + 1$ buckets have at least one successful execution. 

To see this, we define the following probability $p'$ which we show lower bounds $p_{i}$, and is defined as follows. First, let $\overline{T}$ any subspace of dimension bounded by $t$. $p'_{ \overline{T} }$ is the probability that (1) when sampling $T_{0}^{\bot}$, the intersection of $T_{0}^{\bot}$ with $\overline{T}$ is only the zero vector, and also (2) the output of the adversary $\Adv$ was inside $T_{0}^{\bot}$. $p'$ is defined as the minimal probability taken over all possible choices of $\overline{T}$. After one verifies that indeed for every $i$ we have $p' \leq p_{i}$, it is sufficient to lower bound $p'$.

\paragraph{Lower bound for the probability $p'$.}
The probability $p'$ is for an event that's defined as the logical AND of two events, and as usual, equals the product between the probability $p'_{0}$ of the first event (the trivial intersection between the subspaces), times the conditional probability $p'_{1}$ of the second event (that $\Adv$ hits a non-zero vector in the dual $T_{0}^{\bot}$), conditioned on the first event.

First we lower bound the probability $p'_{0}$ by upper bounding the complement probability, that is, we show that the probability for a non-trivial intersection is small. Consider the random process of choosing a basis the dual $T_{0}^{\bot}$. The process of choosing a basis for the dual has $t$ steps, and in each step we choose a random vector in $S_{0}^{\bot}$, that's outside the span we aggregated so far (the first vector can be any non-zero vector in $S_{0}^{\bot}$, the second needs to be outside of the span of the vector that was picked in the first step, etc). Given a subspace $\overline{T}$ of dimension $t$, what is the probability for the two subspaces to have only a trivial intersection? It is exactly the sum over $z \in [t]$ (which we think of as the steps for sampling $T_{0}^{\bot}$) of the following event: In the $t$-step process of choosing a basis for $T_{0}^{\bot}$, index $z$ was the first to cause the subspaces to have a non-zero intersection. Recall that for each $z \in [t]$, the probability that $z$ was such first index to cause an intersection, equals the probability that the $z$-th sampled basis vector for $T_{0}^{\bot}$ is a vector that's inside the unified span of $\overline{T}$ and the aggregated span of $T_{0}^{\bot}$ so far, after $z - 1$ samples. This amounts to the probability
$$
\sum_{z \in [t]} \frac{ |\overline{T}| \cdot 2^{z - 1} }{ |S_{0}^{\bot}| }
=
\sum_{z \in [t]} \frac{ 2^{t} \cdot 2^{z - 1} }{ 2^{k - r} }
=
2^{ -s } \cdot \sum_{z \in \{ 0, 1, \cdots, t - 1 \}} 2^{z}
$$
$$
=
2^{ -s } \cdot \left( 2^{t} - 1 \right)
< 
2^{t - s}
\enspace .
$$
This means that $p'_{0} \geq 1 - 2^{t - s}$.

The lower bound for the conditional probability $p'_{1}$ is now quite easy: Note that since $\Pr\left[ A | B \right] \geq \Pr\left[ A \right] - \Pr\left[ \lnot B \right]$, letting the event $A$ be the event that the adversary $\Adv$ outputs a vector in the dual $T_{0}^{\bot}$ and $B$ the event that $T_{0}^{\bot}$ has only a trivial intersection with $\overline{T}$, we get $p'_{1} \geq \epsilon - 2^{t - s}$. By Condition \ref{subspace_concentration_condition_1_computational} in our statement, it follows that $\frac{t}{2^{ s - t }} \leq \frac{\epsilon}{2}$ and thus $p'_{1} \geq \frac{\epsilon}{2}$. Overall we got $p' := p'_{0} \cdot p'_{1} \geq \left( 1 - 2^{t - s} \right) \cdot \frac{\epsilon}{2} > \frac{\epsilon}{4}$.

Finally, to see why we get an overall high probability for $D \geq t + 1$ on a sample from $\Ds_{1}$, observe the following. In each bucket there are $\frac{k}{\epsilon}$ attempts and each succeeds with probability at least $\frac{\epsilon}{4}$ (as we just seen) and thus the overall success probability in a bucket is $\geq 1 - e^{-\Omega(k)}$. Accordingly, the probability to succeed at least once in each of the $t + 1$ buckets (and thus to satisfy $D \geq t + 1$) is $\geq 1 - (t + 1)\cdot e^{-\Omega(k)}$, by considering the complement probability and applying union bound. Overall, when executing $\AdvB$ on a sample from $\Ds_{1}$, the probability for $D \geq t + 1$ is $\geq 1 - e^{ -\Omega(k) }$, in both the oracle model and the standard model using obfuscations.

\paragraph{Executing $\AdvB$ on the distribution $\Ds_{2}$.}
Consider a different distribution $\Ds_{2}$: Sample $T_{0}$ once, then sample $\ell$ i.i.d. access to the same function $C_{ F_{T} }$, that is (in the standard model) $\Obf^{(1)}_{ F_{T} }, \cdots, \Obf^{(\ell)}_{ F_{T} }$. By Corollary \ref{corollary:subspace_hiding_function_main_corollary} we have the following indistinguishabilities between $\Ds_{1}$ and $\Ds_{2}$, in the oracle and standard model.
\begin{itemize}
    \item 
    In the oracle model the reduction $\AdvB$ executes $q \cdot \ell$ queries, so we have the indistinguishability,
    $$
    \{
    \Oracle_{ F_{T^{1}} }, \cdots, \Oracle_{ F_{T^{\ell}} }
    \;
    :
    \;
    \forall i \in [\ell], T_{0}^{i} \gets \Ss_{s} 
    \}
    \approx_{ \frac{ 2 \cdot \lambda \cdot \ell^2 \cdot q \cdot s }{ \sqrt{ 2^{k - r - \lambda - s} } } }
    \{
    \Oracle_{F_{T}}^{ \otimes \ell }
    \;
    :
    \;
    T_{0} \gets \Ss_{s} 
    \}
    \enspace .
    $$
    Since given a sample from $\Ds_{1}$, the algorithm $\AdvB$ outputs $D \geq t + 1$ with probability $\geq 1 - e^{ -\Omega(k) }$, by the above indistinguishability, whenever we execute $\AdvB$ on a sample from $\Ds_{2}$, then with probability at least
    $$
    \geq
    1 - e^{ -\Omega(k) } - \frac{ 2 \cdot \lambda \cdot \ell^2 \cdot q \cdot s }{ \sqrt{ 2^{k - r - \lambda - s} } }
    \geq
    1 - \frac{ 3 \cdot \lambda \cdot \ell^2 \cdot q \cdot s }{ \sqrt{ 2^{k - r - \lambda - s} } }
    $$
    we have $D \geq t + 1$. By our Condition \ref{subspace_concentration_condition_2_unconditional} in the Lemma (for the oracle-model case) that $\frac{ 3 \cdot \lambda \cdot \ell^2 \cdot q \cdot s }{ \sqrt{ 2^{k - r - \lambda - s} } } \leq \frac{1}{2}$, with probability at least $\frac{1}{2}$ we have $D \geq t + 1$ given a sample from $\Ds_{2}$, in the oracle model.

    \item
    In the standard model, we have the indistinguishability
    $$
    \{
    \Obf_{ F_{T^{i}} }, \cdots , \Obf_{ F_{T^{\ell}} }
    \;
    :
    \;
    \forall i \in [\ell], T_{0}^{i} \gets \Ss_{s}
    ,
    \Obf_{ F_{T^{i}} } \gets \iO\left( 1^{\secp_{\iO}}, 1^{p}, C_{F_{T^{i}}} \right)
    \}
    \approx_{
    \left( \frac{f(\secp') - \poly(\secp')}{2\cdot \lambda \cdot \ell}, \; s \cdot \epsilon(\secp') \right)
    }
    $$
    $$
    \{
    \{
    \Obf_{F_{T}}
    \;
    :
    \; 
    \Obf_{F_{T}} \gets \iO\left( 1^{\secp_{\iO}}, 1^{p}, C_{F_{T}} \right)
    \}^{ \otimes \ell }
    \;
    :
    \;
    T_{0} \gets \Ss_{s}
    \}
    \enspace .
    $$
    Recall that the running time of $\AdvB$ is $\ell \cdot T_{\Adv} + \ell \cdot k^{3}$ and by our Lemma's Condition \ref{subspace_concentration_condition_2_computational}, this (abovementioned) complexity of $\AdvB$ is $\leq \frac{f(\secp') - \poly(\secp')}{2\cdot \lambda \cdot \ell}$. Since given a sample oracle from $\Ds_{1}$, the algorithm $\AdvB$ outputs $D \geq t + 1$ with probability $\geq 1 - e^{ -\Omega(k) }$, by the above indistinguishability, whenever we execute $\AdvB$ on a sample from $\Ds_{2}$, then with probability at least
    $$
    \geq
    1 - e^{ -\Omega(k) } - s \cdot \epsilon(\secp')
    \geq
    1 - (s + 1) \cdot \epsilon(\secp')
    $$
    we have $D \geq t + 1$. By our Condition \ref{subspace_concentration_condition_3_computational} in the Lemma, that $(s + 1) \cdot \epsilon(\secp') \leq \frac{1}{2}$, with probability at least $\frac{1}{2}$ we have $D \geq t + 1$ given a sample from $\Ds_{2}$, in the standard model.
\end{itemize}

By an averaging argument, it follows that with probability at least $\frac{1}{2} \cdot \frac{1}{2} = \frac{1}{4}$ over sampling $T_{0}$, the probability $p_{T}$ for the event where $D \geq t + 1$, is at least $\frac{1}{2} \cdot \frac{1}{2} = \frac{1}{4}$. Let us call this set of superspaces $T_{0}$, "the good set" of samples, which by definition has fraction at least $\frac{1}{4}$. Recall two facts: (1) the dimension of $\left( T_{0} \right)^{\bot}$ is $t$, (2) the dimension $D$ aggregates vectors inside $S_{0}^{\bot}$. The two facts together imply that in the event $D \geq t + 1$, it is necessarily the case that there exists an execution index $i \in [\ell]$ in the reduction $\AdvB$ where $\Adv$ outputs a vector in $\left( S_{0}^{\bot} \setminus T_{0}^{\bot} \right)$, when given a sample from $\Ds_{2}$.

For every $T_{0}$ inside the good set we thus know that with probability $\frac{1}{4}$, one of the output vectors of $\Adv$ will be in $\left( S_{0}^{\bot} \setminus T_{0}^{\bot} \right)$. Since these are $\ell$ i.i.d. executions of $\Adv$, by union bound, for every $T_{0}$ inside the good set, when we prepare access of $F_T$ and execute $\Adv$, we will get a vector in $\left( S_{0}^{\bot} \setminus T_{0}^{\bot} \right)$ with probability $\geq \frac{1}{4 \cdot \ell}$. We deduce that for a uniformly random $T_{0}$, the probability for the output of $\Adv$ to be in $\left( S_{0}^{\bot} \setminus T_{0}^{\bot} \right)$, is at least the probability for this event (i.e., the output vector is in $\left( S_{0}^{\bot} \setminus T_{0}^{\bot} \right)$) intersecting with the event that $T_{0}$ is inside the good set, which in turn is at least
$$
\frac{1}{4} \cdot \frac{1}{4 \cdot \ell}
=
\frac{1}{16 \cdot \ell}
:=
\frac{\epsilon}{16 \cdot k \cdot (t + 1)}
\enspace ,
$$
which finishes our proof.
\end{proof}

%% file: oracle_construction.tex
\section{Short One-Shot Signatures Relative to a Classical Oracle} \label{sec:oss_oracle}
In this section we present our construction and security proof with respect to a classical oracle. We construct one-shot signatures that can sign on $\lambda$-bit messages and has quantum signing keys of size $\Theta\left( \lambda \right)$ qubits, and where the probability to forge is bounded by $2^{-\Omega\left( \lambda \right)}$, for every quantum algorithm making a polynomial number of queries $q$. We first describe our core construction in \ref{constr:main}, and then our OSS scheme itself in \ref{construction:OSS_oracle_model}, by relying on the core construction.

\begin{construction} \label{constr:main}
Let $\secp \in \Nat$ the statistical security parameter. Define $s := 18 \cdot \secp$ and let $n, r, k \in \Nat$ such that $r := s \cdot \secp$, $n := r + \frac{3}{2} \cdot s$, $k := \frac{3}{2}\cdot s + \lambda$.

Let $\Pi: \{ 0, 1 \}^n \rightarrow \{ 0, 1 \}^n$ be a random permutation and let $F : \{ 0, 1 \}^{r} \rightarrow \{ 0, 1 \}^{\poly(r)}$ a random function. Let $H(x)$ denote the first $r$ output bits of $\Pi(x)$, and $J(x)$ denote the remaining $n - r$ bits, which are interpreted as a vector in $\Z_{2}^{n - r}$. For each $y \in \{ 0, 1 \}^r$, using the output randomness of $F(y)$, we sample $\matA(y) \in \Z_2^{k \times (n-r)}$ and $\vecB(y) \in \Z_2^k$. The matrix $\matA(y) \in \Z_2^{k \times (n-r)}$ is random with full column rank $n - r$, and also such that its bottom $\lambda$ rows have full row rank $\lambda$. $\vecB(y) \in \Z_2^k$ is a uniformly random vector. Then, let
$$
\Ps : \{ 0, 1 \}^n \rightarrow \left( \{ 0, 1 \}^r \times \Z_2^k \right), \enspace 
\Ps^{-1} : \left( \{ 0, 1 \}^r \times \Z_2^k \right) \rightarrow  \{ 0, 1 \}^n, \enspace
\Ds_{1}: \left( \{0,1\}^r \times \Z_2^k \right) \rightarrow \bbZ_{2}^{\lambda} \enspace,
$$
be the following oracles:
\begin{align*}
    \Ps(x) &= \left( \; y \; , \; \matA(y) \cdot J(x) + \vecB(y) \; \right) \text{ where } y = H(x)
    \\
    \\
    \Ps^{-1}\left( y, \vecU \right) & =
    \begin{cases}
    \Pi^{-1}\left( y, \vecZ \right) \;\;
    & \exists \: \vecZ \in \bbZ_{2}^{n - r} : \matA(y) \cdot \vecZ + \vecB(y) = \vecU
    \\
    \bot
    &\text{ else }
    \end{cases}
    \\
    \\
    \Ds\left( y, \vecV \right) &=
    \begin{cases}
    \vecC_{y, \vecV} \;\;
    &\text{ if } \vecV^{T} \cdot \matA(y) \in \rowspan\left( \matA(y)_{[(k - \lambda) + 1]}, \cdots,  \matA(y)_{[(k - \lambda) + \lambda]} \right)
    \\
    \bot
    &\text{ otherwise }
    \end{cases}
\end{align*}
where for $\matA \in \bbZ_{2}^{k \times (n - r)}$ and $j \in [k]$, $\matA_{[j]} \in \bbZ_{2}^{n - r}$ is the $j$-th row of $\matA$ (e.g., $k$-th row is bottom), and $\vecC_{y, \vecV} \in \bbZ_{2}^{\lambda}$ is the coordinates vector, of the vector $\vecV^{T} \cdot \matA(y) \in \bbZ_{2}^{n - r}$ with respect to the basis $\matA(y)_{[(k - \lambda) + 1]}, \cdots,  \matA(y)_{[(k - \lambda) + \lambda]}$ (i.e., element $j \in [\lambda]$ of $\vecC_{y, \vecV}$ is the coefficient of $\matA(y)_{[(k - \lambda) + j]}$).
\end{construction}

We next describe our one-shot signature scheme explicitly. As part of our scheme we will use the following property of error-correcting codes:
\begin{theorem}[\cite{Pinsker1965}] \label{theorem:ecc}
There exists a universal constant $C>0$ such that, for any constant $\epsilon>0$, a random linear binary code with rate $C\cdot \epsilon^2$ has, except with probability negligible in the length of the code, a minimum distance at least $1/2-\epsilon$. For a constant $\epsilon>0$ we denote by $\Enc_{\epsilon}$ the ECC's encoding algorithm, which is efficient and just samples a random linear code with rate $C\cdot \epsilon^2$ and encodes any message $m \in \{ 0, 1 \}^{\lambda}$ to $\Enc_{\epsilon}\left( m \right) \in \zo^{ \frac{\lambda}{ C \cdot \epsilon^2 } }$.
\end{theorem}

\begin{construction} [One-Shot Signature Construction Relative to a Classical Oracle] \label{construction:OSS_oracle_model}
Our one-shot signature (OSS) scheme $\left( \setup, \OSSgen, \sign, \ver \right)$, is defined for security parameter $\lambda \in \Nat$ and can sign on messages in $\{ 0, 1 \}^{\lambda'}$, for $\lambda' := C \cdot \epsilon^2 \cdot \lambda$ such that $C > 0$ is the constant from Theorem \ref{theorem:ecc} and for $\epsilon := \frac{1}{3}$.
Our scheme is defined as follows.

\begin{itemize}
    \item
    $\left( \Ps, \Ps^{-1}, \Ds \right) \gets \setup\left( 1^\lambda \right)$: The oracle that the setup algorithm samples is exactly the oracles sampled by Construction \ref{constr:main}.
    
    \item $\left( \pk, \ket{\sk} \right) \gets \OSSgen^{\Ps, \Ps^{-1}}$: To sample a signature token, start with a uniform superposition $\ket{+}^{\otimes n}$, apply $\Ps$ to obtain
    $$
    \frac{1}{2^{ \frac{n}{2}} }
    \sum_{ x \in \{ 0, 1 \}^{n} }
    \ket{x, H(x), \vecU_{x}} \enspace ,
    $$
    where $H(\cdot)$ is the hash function from Construction \ref{constr:main} and for $x \in \{ 0, 1 \}^{n}$, the vector $\vecU_{x} \in \bbZ_{2}^{k}$ is the output vector of the oracle $\Ps$.
    
    Next, use $\Ps^{-1}$ to un-compute the register holding $x$ and then measure the register holding $H(x) \in \{ 0, 1 \}^{r}$, to obtain the state
    $$
    \ket{y} \otimes 
    \frac{1}{2^{ \frac{n - r}{2}} }
    \sum_{ \vecU \in \left( \colspan\left( \matA(y)  \right) + \vecB(y) \right) }
    \ket{ \vecU } 
    \enspace .
    $$
    Finally, set $\pk := y$ and $\ket{\sk} := \frac{1}{2^{ \frac{n - r}{2}} }
    \sum_{ \vecU \in \left( \colspan\left( \matA(y)  \right) + \vecB(y) \right) }
    \ket{ \vecU }$.
    
    \item
    $\sigma \gets \sign^{\Ps^{-1}, \Ds}\left( \pk, \ket{\sk}, m \in \{ 0, 1 \}^{\secp'} \right)$: We describe our signing algorithm in \ref{construction:quantum_signing_algorithm}.
    
    \item
    $\ver^{\Ps^{-1}}\left( \pk, m \in \zo^{\lambda'}, \sigma \right) \in \{ 0, 1 \}$: 
    Parse $\pk$ as $y \in \{ 0, 1 \}^{r}$ and parse $\sigma$ as $\vecU_{\sigma} \in \bbZ_{2}^{k}$. Encode $m \in \zo^{\secp'}$ into $\Enc_{ \epsilon }(m) := m^* \in \zo^{ \lambda }$ using the ECC from Theorem \ref{theorem:ecc}. Output $1$ iff both, $\Ps^{-1}\left( y, \vecU_{\sigma} \right) \neq \bot$ and $m^*$ has Hamming distance bounded by $\lambda\frac{ \left( \frac{1}{2} - \epsilon \right) }{2} := \frac{\lambda}{12}$ to the last $\lambda$ bits of $\sigma$. 
\end{itemize}
\end{construction}

\paragraph{Signing Multi-bit Messages.}
We next describe a quantum signing algorithm that can sign $\Omega(\lambda)$-bit messages, and furthermore does so in "parallel". Specifically, the algorithm works in iterations, and in each iteration, the algorithm makes two queries to $\Ds$ and elsewhere executes in constant parallel time. In particular, it is parallel in the sense that signing for bit $j + 1$ does not need to wait for the signing of bit $j$. Our algorithm takes constant iterations and produces a valid signature with probability exponentially close to $1$ (in the security parameter). 

\vspace{1mm}
The intuition behind the algorithm originates in the "measure and correct" signing algorithm of \cite{C:Shmueli22}, which in turn is inspired by the OSS signing procedure of \cite{STOC:AGKZ20}. Our algorithm first encodes $m$ to a large message $m^* \in \zo^\lambda$ using an ECC. Then, the algorithm tries to sign by collapsing all $\lambda$ qubits to the wanted value $m^*$. Each qubit falls to its right value with probability $1/2$, independently of the other qubits, so we expect roughly half the qubits to collapse to the right value when we do this. Then, the signing algorithm "corrects" by returning the corresponding qubits (i.e., the ones that need correction) to full superposition using the dual oracle. This correction will let us retry to collapse the qubits that did not collapse correctly for the first time, to their correct value with accordance to $m$. After some tries, we expect some minimal fraction of qubits to fall to their correct classical value. The formal procedure follows.

\begin{construction} [A many-bit parallel quantum signing algorithm for coset states] \label{construction:quantum_signing_algorithm}
    We describe the signing algorithm $\sign^{ \Ps^{-1}, \Ds }\left( \pk, \ket{\sk}, m \in \{ 0, 1 \}^{\secp'} \right)$ of the OSS scheme.

    \begin{enumerate}
        \item 
        Denote by $\Us$ the $k$-qubit register holding $\ket{\sk}$. Parse $\pk$ as $y \in \{ 0, 1 \}^{r}$ and continue only if $\Ps^{-1}\left( y, \Us \right) \neq \bot$.
    
        \item 
        Encode the message $m$ into the codeword $\Enc_{ \epsilon }(m) \rightarrow m^* \in \zo^{\lambda}$ using the ECC.

        \item
        For $t$ going from $1$ until $4$:
        \begin{enumerate}
            \item \label{signing_algo_measure_step}
            Make a measurement on the last $\lambda$ qubits of $\Us$, denote it by $m^{t} \in \zo^{\lambda}$.

            \item \label{signing_algo_hadamard_first}
            Execute $H^{\otimes k}$ on the register $\Us$.

            \item \label{signing_algo_measure_coset}
            Initialize a $\lambda$-qubit register $\Cs$ with zeros. Apply $\Ds\left( y, \cdot \right)$ to $\Us$, putting the output in $\Cs$.
            Measure only the qubits $\Cs_{i}$ such that $m^*_{i} \oplus m^{t}_{i} = 1$. Discard the measurement results and un-compute the information that's inside the register $\Cs$ by using the oracle $\Ds\left( y, \cdot \right)$ again.

            \item \label{signing_algo_hadamard_second}
            Execute $H^{\otimes k}$ again on the register $\Us$. Increment the loop variable $t$ by $1$.
        \end{enumerate}
        
        \item measure the rest of the register $\Us$ and let $\sigma \in \zo^{k}$ the measurement result.
    \end{enumerate}
\end{construction}

\begin{claim} [Correctness of Signing Algorithm]
    With probability $1 - 2^{ -\Omega\left( \lambda \right) }$ over sampling the linear code in the ECC scheme, we have that the probability for the event of getting a valid signature over the random experiment of (1) sampling $\left( \pk, \ket{\sk} \right) \gets \OSSgen^{\Ps, \Ps^{-1}}$ and then (2) executing the signing algorithm, is $1 - 2^{ -\Omega\left( \lambda \right) }$.
\end{claim}

\begin{proof}
    The correctness proof has two steps. The first and main part, proves that in every iteration of the algorithm, for the qubits of $\Us$ that did not collapse to the correct value of $m^*$, the algorithm will return these to uniform superposition at the end of that same iteration. In the second part, we explain how a simple concentration bound proves that we get a valid signature with high probability, by using the first part of correctness.

    \noindent
    \paragraph{Correction of faulty collapsed bits back to superposition.}
    We will show that the algorithm can make repeated tries for the target bits that were not signed correctly, by "re-entropization" of the qubits that collapsed to the wrong value, without re-entropizing (or changing the value) of the bits that collapsed to the correct value. The correctness of the algorithm is conditioned on the sampled matrix $\matA(y) \in \bbZ_{2}^{k \times (n - r)}$ having full rank (i.e., column-rank $n - r$), its bottom $\lambda$ rows having full rank as well (i.e. row-rank $\lambda$), and on the ECC having minimal distance of $\lambda \cdot \left( \frac{1}{2} - \epsilon \right) := \frac{\lambda}{6}$. The probability that any of these do not happen is $2^{-\Omega\left( \lambda \right)}$.

    Next, for $y \in \zo^{r}$,
    \begin{itemize}
        \item
        Let $S_{y, 0} := \colspan\left( \matA(y) \right)$, which has dimension $n - r$.

        \item 
        Let $S_{y}$ the subspace of $S_{y, 0}$ with vectors such that their last $\lambda$ bits are all $0$. Accordingly, $S_{y}$ has dimension $n - r - \lambda$.

        \item 
        For $j \in [\lambda]$, let $S_{y, j}$ the subspace of $S_{y, 0}$ with vectors such that their last $\lambda$ bits are all $0$, except bit $(k - \lambda) + j$, which may be arbitrary. Note that since the bottom $\lambda$ rows of $\matA(y)$ are full rank, it follows that $S_{y, j}$ has dimension $n - r - \lambda + 1$ and thus satisfies $S_{y} \subsetneq S_{y, j} \subsetneq S_{y, 0}$.
    \end{itemize}

    Formally, considering our signing algorithm from \ref{construction:quantum_signing_algorithm}, for every iteration $t$, observe that at the end of Step \ref{signing_algo_measure_step}, the state we have in register $\Us$ is of the form 
    \begin{equation} \label{equation:state_after_signing_attempt}
        \sum_{ \vecU \in S_{y} } (-1)^{\langle z^{t}, \vecU \rangle} \cdot \ket{ x^{t} + \vecB(y) + \vecU }
    \end{equation}
    for some $z^{t} \in S_{y}^{\bot}$ and $x^{t} \in S_{y, 0}$. As an example, after the first iteration, we get the above state for $z^{1} = 0^{k}$ and $x^{1} \in S^{y, 0}$ such that for the measurement result $m^{1} \in \zo^{\lambda}$ and for every $j \in [\lambda]$ we have $[x^{1} \oplus \vecB(y)]_{(k - \lambda) + j} = m^{1}_{j}$.

    We will show that at the end of the iteration $t$ (after Step \ref{signing_algo_hadamard_second}), after re-entropizing the bits $j \in [\lambda]$ such that $m^*_{j} \oplus m^{t}_{j} = 1$, the state in $\Us$ will be of the form
    \begin{equation} \label{equation:state_after_entropization}
        \sum_{ \vecU \in \Ss_{\left( m^*, \; m^{t} \right)} } (-1)^{\langle z^{t + 1}, \vecU \rangle} \cdot \ket{ x^{t} + \vecB(y) + \vecU } \enspace .
    \end{equation}
    for some $z^{t + 1} \in S_{y}^{\bot}$, and where $\Ss_{\left( m^*, \; m^{t} \right)} := \Span\left( \{ S_{y, j} \}_{j \; : \; [ m^* \oplus m^{t} ]_{j} = 1} \right)$. It can be verified by the reader that if we can correct to the above state, this lets us proceed for attempting to signing the rest of the bits of $m$ correctly, while keeping the already-correctly-collapsed bits.

    So, assume we have a state as per Equation \ref{equation:state_after_signing_attempt}, which happens at iteration $t$ after Step \ref{signing_algo_measure_step}. Next, after executing Step \ref{signing_algo_hadamard_first} our state is 
    $$
    \sum_{ \vecV \in S_{y}^{\bot} } (-1)^{\langle x^{t} + \vecB(y), \vecV \rangle} \cdot \ket{ z^{t} + \vecV } \enspace ,
    $$
    and after the next Step \ref{signing_algo_measure_coset}, the state makes a partial collapse to the form 
    $$
    \sum_{ \vecV \in \Ss_{\left( m^*, \; m^{t} \right)}^{\bot} } (-1)^{\langle x^{t} + \vecB(y), \vecV \rangle} \cdot \ket{ z^{t + 1} + \vecV } \enspace ,
    $$
    such that $z^{t + 1} \in S_{y}^{\bot}$.
    
    Specifically (and for the interested reader), if we look at the $| m^* \oplus m^{t} |$ measurement results (which we discarded) from Step \ref{signing_algo_measure_coset}, $z^{t + 1}$ is such that for the vector $\left( z^{t + 1} \right)^{T} \cdot \matA(y) \in \bbZ_{2}^{n - r}$, its coordinates vector $\vecC_{ y, z^{t + 1} } \in \bbZ_{2}^{\lambda}$ with respect to the basis $\matA(y)_{[(k - \lambda) + 1]}, \cdots,  \matA(y)_{[(k - \lambda) + \lambda]}$ has the same values as what was measured in Step \ref{signing_algo_measure_coset}. Finally, since after Step \ref{signing_algo_hadamard_second} the state in register $\Us$ is indeed of the form described by Equation \ref{equation:state_after_entropization}, this finishes the first part of our proof.

    \noindent
    \paragraph{Probability analysis of getting a valid signature.}
    The remainder of the proof simply uses Chernoff's concentration bound on the event that a bit in uniform superposition collapsed to its right value w.r.t. $m^*$. Specifically, we proves that at the beginning of every iteration $t$, for each of the last $\lambda$ qubits of $\Us$, if is either in uniform superposition (and independently of all of the other $\lambda$ last qubits of $\Us$), or it is classical and have the value of the corresponding bit of $m^*$. So, since it is in particular in uniform superposition, when measured its probability for collapsing to the correct value of $m^*$ is exactly $1/2$. Accordingly, due to the fact that these are statistically independent events (thanks to the fact that the row rank of the last $\lambda$ rows of $\matA(y)$ is full and equals to $\lambda$), with probability $1 - 2^{ -\Omega(\lambda) }$, roughly half of the qubits that were in superposition collapse to their correct value of $m^*$.

    This means that after $4$ iterations of the algorithm, with probability $1 - 2^{ -\Omega(\lambda) }$, Chernoff's bound implies that at most a $\frac{1}{16} + \frac{1}{100}$ fraction of the bits of $m^t$ (for $t = 4$) do not match the corresponding bits of $m^*$. It follows in particular that with probability $1 - 2^{ -\Omega(\lambda) }$, after $4$ iterations the resulting signature $\sigma \in \zo^{k}$ has less than $\frac{\lambda}{12}$ Hadamard distance to $m^*$, which means that the signature will be verified successfully by the signature verification algorithm.
\end{proof}

\paragraph{Security, and comparison to the security reduction from \cite{C:ShmZha25}.}
In order to prove security (that is, strong unforgeability of the OSS), it will be enough to prove the collision resistance of the function $H$ from the oracle $\Ps$. The structure of the proof and high-level strategy remains the same as in \cite{C:ShmZha25}: In the first part we show that a collision finder given $\left( \Ps, \Ps^{-1}, \Ds \right)$ can be transformed into a collision finder against the dual-free setting $\left( \Ps, \Ps^{-1} \right)$, this is proved in our Theorem \ref{thm:dualtodualfree}. The second part proves that collision finding in $\left( \Ps, \Ps^{-1} \right)$ is hard, by reducing to the collision-resistance of other functions. There are two points of meaningful difference from the previous proof -- one change to each of the above conceptual parts of the proof.

The first change to the proof (compared to \cite{C:ShmZha25}) is due to the adversary getting more information from the oracle $\Ds$: $\lambda$ bits instead of $1$. We need to prove that also here, the function $H$ (computed inside $\Ps$) is collision resistant. The new proof handles this by executing the same high-level ideas from the reduction from previous work, but slightly more carefully, and by adding a new generalized variation of subspace-hiding (given formally in Section \ref{section:subspace_hiding_function}), which we call subspace-hiding functions. After completing the first part of the proof, we got rid of the dual $\Ds$ and all that remains is to prove that the dual-free oracles $\left( \Ps, \Ps^{-1} \right)$ are collision-resistant.

The second change is that the dimension $k$, which is the dimension of the space that the random cosets $\colspan\left( \matA(y) \right) + \vecB(y)$ live in, is now asymptotically smaller (and even optimal). Formally, $k$ is now on the order of $\Theta\left( \lambda \right)$, where in previous work \cite{C:ShmZha25} it was on the order of $\Theta\left( \lambda^2 \right)$. Here, things are conceptually different: the previous technique for proving the collision resistance of $\left( \Ps, \Ps^{-1} \right)$ relies solely on a parallel repetition of random 2-to-1 functions, and it breaks down in this setting (see an elaborated intuitive explanation in Section \ref{section:overview}). Nonetheless, in Section \ref{subsection:dual_free_to_two_to_one_oracle} we show a new reduction and technique that resolves this.

\vspace{2mm}
Overall, we obtain the following main security Theorem.

\begin{theorem} [Collision Resistance of $H$] \label{theorem:oracle_main_security}
Let $\Oracle_{ n, r, k }$ the distribution over oracles defined in Construction \ref{constr:main}. Let $\Adv$ an oracle aided $q$-query (computationally unbounded) quantum algorithm. Then,
\[
\Pr
\left[
x_0 \neq x_1
\land 
H(x_0) = H(x_1) 
\;
:
\begin{array}
{rl}
\left( \Ps, \Ps^{-1}, \Ds \right) & \gets \Oracle_{n,r,k} \\
\left( x_0, x_1 \right) & \gets \Adv^{\Ps, \Ps^{-1}, \Ds}
\end{array}\right]
\leq
O\left( \frac{ \secp \cdot k^{2} \cdot q^{3} }{ 2^{ \secp } } \right)
\enspace .
\]
\end{theorem}

\begin{proof}
    Assume towards contradiction that there is an oracle aided quantum algorithm $\Adv$, making $q$ queries, that given a sample oracle $\left( \Ps, \Ps^{-1}, \Ds \right) \gets \Oracle_{ n, r, k }$ outputs a collision $(x_{0}, x_{1})$ in $H$ with probability $\epsilon$, such that $\epsilon \geq \omega\left( \frac{ \secp \cdot k^{2} \cdot q^{3} }{ 2^{ \secp } } \right)$.

    Note that by our parameter choices in Construction \ref{constr:main} and by our assumption towards contradiction $\epsilon \geq \omega\left( \frac{ \secp \cdot k^{2} \cdot q^{3} }{ 2^{ \secp } } \right)$, one can verify through calculation that for $s' := s - (n - r - s)$ we have (1) $\frac{ \lambda \cdot k^{8} \cdot q^8 \cdot s }{ \sqrt{ 2^{n - r - \secp - s} } } \leq o\left( \epsilon^4 \right)$, and also (2) $\frac{ k^{2} \cdot q^3 \cdot (n - r - s) }{ 2^{s'} } \leq o\left( \epsilon^2 \right)$. This means that the conditions of Theorem \ref{thm:dualtodualfree} are satisfied, and it follows there is a $q$-query algorithm $\AdvB$ that gets access only to $\left( \Ps, \Ps^{-1} \right)$, sampled from $\left( \Ps, \Ps^{-1}, \Ds \right) \gets \Oracle_{ r + s, \: r, \: k - (n - r - s) }$ that finds collisions in $H$ with probability $\geq \frac{\epsilon}{2^{6} \cdot k^{2}}$.
    
    Now, consider the statement of Theorem \ref{thm:dualfreetocol}, with the following interface: "$n$" in the Theorem will be $r + s$ here, "$r$" in the theorem stays the same and is $r$ here, and "$k$" in the Theorem is $k - (n - r - s)$ here. Note that our parameter choices in Construction \ref{constr:main} imply that $\frac{r + s}{(r + s) - r}$ and thus "$\frac{n}{n - r}$" from the theorem is an integer, and also $k - (n - r - s) \geq (r + s) - r + \lambda$ and thus "$k \geq n  - r + \lambda$" from the theorem. It follows by Theorem \ref{thm:dualfreetocol} that there is a $q$-query algorithm $\AdvB'$ that given oracle access to random claw-free permutation $H^* : \zo^{\lambda + 1} \rightarrow \zo^{\lambda}$ (as in Definition \ref{definition:claw_free_permutation}), finds a collision in $H^*$ with probability $\frac{\epsilon}{2^{6} \cdot k^{2} \cdot (n - r)}$.
    
    However, we know that finding collisions in $H^*$ is hard: It follows by Lemma \ref{lemma:claw_free_permutation_collision_resistant} that 
    $$
    \frac{\epsilon}{2^{6} \cdot k^{2} \cdot (n - r)}
    \leq
    O\left( \frac{ q^{3} }{ 2^{ \frac{ r }{ n - r } } } \right)
    \enspace ,
    $$
    which in turn implies
    $$
    \epsilon
    \leq
    O\left( \frac{ \lambda \cdot k^{2} \cdot q^{3} }{ 2^{ \secp } } \right)
    \enspace ,
    $$
    in contradiction to $\epsilon \geq \omega\left( \frac{ \secp \cdot k^{2} \cdot q^{3} }{ 2^{ \secp } } \right)$.
\end{proof}

\subsection{Bloating the Dual} \label{subsection:bloating_dual_oracle}
Let $\Oracle'_{n,r,k,s}$ denote the following distribution over $\Ps,\Ps^{-1},\Ds'$. The oracles $\Ps,\Ps^{-1}$ are defined identically to $\Oracle_{n,r,k}$, and the oracle $\Ds$ will change. Now, for $s \leq n - r - \secp$, we let $\matA(y)^{(0)} \in \bbZ_{2}^{k \times s}$ denote the first (i.e. leftmost) $s$ columns of $\matA(y) \in \bbZ_{2}^{k \times (n - r)}$ and $\matA(y)^{(1)} \in \bbZ_{2}^{k \times (n - r - s)}$ denote the remaining $n-r-s$ (rightmost) columns. In the bloated dual, we do the same things as in the original $\Ds$, but with respect to $\matA(y)^{(1)} \in \bbZ_{2}^{k \times (n - r - s)}$ rather than with respect to $\matA(y) \in \bbZ_{2}^{k \times (n - r)}$. Formally:
\[
\Ds'\left( y, \vecV \right) =
    \begin{cases}
    \vecC_{y, \vecV} \;\;
    &\text{ if } \vecV^{T} \cdot \matA(y)^{(1)} \in \rowspan\left( \matA(y)^{(1)}_{[(k - \lambda) + 1]}, \cdots,  \matA(y)^{(1)}_{[(k - \lambda) + \lambda]} \right)
    \\
    \bot
    &\text{ otherwise }
    \end{cases}
\]
where here, similarly to the construction (but not quite the same), $\vecC_{y, \vecV} \in \bbZ_{2}^{\lambda}$ is the coordinates vector, of the vector $\vecV^{T} \cdot \matA(y)^{(1)} \in \bbZ_{2}^{n - r - s}$ with respect to the basis $\matA(y)^{(1)}_{[(k - \lambda) + 1]}, \cdots,  \matA(y)^{(1)}_{[(k - \lambda) + \lambda]}$.

\begin{lemma} \label{lemma:bloating_dual_oracle}
Suppose there is an oracle aided $q$-query quantum algorithm $\Adv$ such that
\[
\Pr
\left[
\left( y_{0} = y_{1} \right)
\land
\left( x_0 \neq x_1 \right) \; :
\begin{array}{rl}
\left( \Ps, \Ps^{-1}, \Ds \right) & \gets \Oracle_{n,r,k} \\
(x_0, x_1) & \gets \Adv^{\Ps,\Ps^{-1},\Ds} \\
(y_b, \vecU_b) & \gets \Ps(x_b)
\end{array}
\right]
\geq
\epsilon \enspace .
\] 
Also, let $s \in \Nat$ such that $s \leq n - r - \secp$, define $s' := s - (n - r - s)$, and assume
\begin{enumerate}
    \item \label{bloating_dual_oracle_condition_1}
    $\frac{ \lambda \cdot k^{8} \cdot q^8 \cdot s }{ \sqrt{ 2^{n - r - s - \secp} } } \leq o\left( \epsilon^4 \right)$,

    \item \label{bloating_dual_oracle_condition_2}
    $\frac{ k^{2} \cdot q^3 \cdot (n - r - s) }{ 2^{s'} } \leq o\left( \epsilon^2 \right)$.
\end{enumerate}
Then,
\[
\Pr
\left[
\begin{array}
{rl}
& \left( y_0 = y_1 := y \right) \land \\
& \left( \vecU_0 - \vecU_1 \right) \notin \colspan \left( \matA(y)^{(1)} \right)
\end{array}
\;
:
\begin{array}
{rl}
\left( \Ps,\Ps^{-1},\Ds' \right) & \gets \Oracle'_{n,r,k,s} \\
\left( x_0, x_1 \right) & \gets \Adv^{\Ps,\Ps^{-1},\Ds'} \\
(y_b, \vecU_b) & \gets \Ps(x_b)
\end{array}
\right]
\geq
\frac{ \epsilon }{ 2^{6} \cdot k^{2} }
\enspace .
\]
\end{lemma}

\begin{proof}
Assume there is an oracle-aided $q$-query quantum algorithm $\Adv$ that given oracle access to $\left( \Ps,\Ps^{-1},\Ds \right) \gets \Os_{n,r,k}$ outputs a pair $\left( x_{0}, x_{1} \right)$ of $n$-bit strings. Denote by $\epsilon$ the probability that $x_{0}$, $x_{1}$ are both distinct and collide in $H(\cdot)$ (i.e., their $y$-values are identical).
We next define a sequence of hybrid experiments, outputs and success probabilities for them, and explain why the success probability in each consecutive pair is statistically close. 

\begin{itemize}
    \item $\Hyb_{0}$: The original execution of $\Adv$.
\end{itemize}
The process $\Hyb_{0}$ is the above execution of $\Adv$ on input oracles $\left( \Ps, \Ps^{-1}, \Ds \right)$. We define the output of the process as $(x_0, x_1)$ and the process execution is considered as successful if $x_{0}$, $x_{1}$ are both distinct and collide in $H(\cdot)$. By definition, the success probability of $\Hyb_{0}$ is $\epsilon$.

\begin{itemize}
    \item $\Hyb_{1}$: Simulating the oracles using only a bounded number of cosets $\left( \matA(y), \vecB(y) \right)$, by using small-range distribution.
\end{itemize}
Consider the function $F$ which samples for every $y \in \bbZ_{2}^{r}$ the i.i.d. coset description $\left( \matA(y), \vecB(y) \right)$. These cosets are then used in all three oracles $\Ps$, $\Ps^{-1}$ and $\Ds$. The difference between the current hybrid and the previous hybrid is that we swap $F$ with $F'$ which is sampled as follows: We set $R := \left( 300 \cdot q^{3} \right) \cdot \frac{ 2^{7} \cdot k^{2} }{ \epsilon }$ and for every $y \in \bbZ_{2}^{r}$ we sample a uniformly random $i_{y} \gets [R]$, then sample for every $i \in [R]$ a coset $\left( \matA(i) \in \bbZ_{2}^{k \times (n - r)}, \vecB(i) \in \bbZ_{2}^{k} \right)$ as usual. For $y \in \bbZ_{2}^{r}$ we define $F'(y) := \left( \matA(i_y), \vecB(i_y) \right)$ instead of the previous $F(y) := \left( \matA(y), \vecB(y) \right)$.

By the small-range distribution technique \cite{zhandry2021construct} (and concretely, Theorem A.6 from \cite{ananth2022pseudorandom}), it follows that for every quantum algorithm making at most $q$ queries and tries to distinguish between $F$ and $F'$, the distinguishing advantage is bounded by $\frac{300\cdot q^{3}}{R} < \frac{\epsilon}{8}$, which means in particular that the outputs of this hybrid and the previous one has statistical distance bounded by $\frac{\epsilon}{8}$. It follows that the success probability of the current hybrid is $:= \epsilon_{1} \geq \epsilon - \frac{\epsilon}{8} = \frac{7 \cdot \epsilon}{8}$.

\begin{itemize}
    \item $\Hyb_{2}$: Relaxing dual verification oracle to accept a larger subspace, by subspace hiding functions.
\end{itemize}
The change we make in this hybrid is that we make the membership check dual oracles more relaxed. Formally, we invoke Lemma \ref{lemma:subspace_hiding_function} with the following interface, and for every $i \in [R]$. The smallest oracle "$S_{0}$" from the Lemma \ref{lemma:subspace_hiding_function} statement is simply $S_{i, 0}^{\bot} := \colspan\left( \matA(i) \right)^{\bot}$ here and "$S$" from the lemma is $S_{i}^{\bot}$, defined as the total set of vectors that $\Ds$ does not output $\bot$ on. The partition "$P$" from the lemma statement is the partition defined by the output of the oracle $\Ds$ (observe that this is indeed a partition into cosets of $S$ that are all parallel to $S_{0}$). That is, the oracle $\Ds$ computes the function $F_{S}$ for these subspaces.

Now, for every $i \in [R]$, after sampling $\left( \matA(i), \vecB(i) \right)$, we sample superspace $T_{i, 0}^{\bot}$ and by the statement of Lemma \ref{lemma:subspace_hiding_function}, (1) $T_{i, 0}^{\bot}$ is a random superspace of $S_{i, 0}^{\bot}$ (which itself has dimension $k - (n - r)$) with $k - (n - r - s)$ dimensions, and (2) $T^{\bot}_{i}$ is the joint span of $T_{i, 0}^{\bot}$ and $S_{i}^{\bot}$. For concreteness and because it will later be useful to our proof, assume that our process of sampling $T_{i, 0}^{\bot}$ is by sampling its dual $T_{i, 0}$, which in turn is sampled by sampling a uniformly random invertible $\matM_{i} \in \bbZ_{2}^{(n - r) \times (n - r)}$, multiplying by $\matA(i)$ and then taking the rightmost $n - r - s$ columns of the generated matrix.

By the guarantees of Lemma \ref{lemma:subspace_hiding_function}, for every $i \in [R]$, changing $\Ds$ to compute $F_{T}$ is $\left( \frac{ \lambda \cdot q \cdot s }{ \sqrt{ 2^{n - r - \secp - s} } } \right)$-indistinguishable, for any $q$-query algorithm. Since we use the above indistinguishability $R$ times, we get $R \cdot \left( \frac{ \lambda \cdot q \cdot s }{ \sqrt{ 2^{n - r - \secp - s} } } \right)$-indistinguishability. It follows that the success probability of the current hybrid is $:= \epsilon_{2} \geq \epsilon_{1} - R \cdot \left( \frac{ \lambda \cdot q \cdot s }{ \sqrt{ 2^{n - r - \secp - s} } } \right) \geq \frac{7 \cdot \epsilon}{8} - O\left( \frac{ \lambda \cdot k^{2} \cdot q^4 \cdot s \cdot \frac{1}{\epsilon} }{ \sqrt{ 2^{n - r - \secp - s} } } \right)$, which in turn by Condition \ref{bloating_dual_oracle_condition_1} in our Lemma's statement, is at least $\frac{3 \cdot \epsilon}{4}$.

\begin{itemize}
    \item $\Hyb_{3}$: For every $i \in [R]$, asking for the sum of collisions to be outside of $T_{i, 0}$, by using dual-subspace anti-concentration.
\end{itemize}
In the current hybrid we change the success predicate of the experiment. Recall that as part of sampling the oracles in the previous hybrid, we sample $R$ i.i.d. cosets $\left( \matA(i), \vecB(i) \right)_{i \in [R]}$ which are used in all three oracles $\left( \Ps, \Ps^{-1}, \Ds \right)$. We then sample $R$ i.i.d. $(k - n + r + s)$-dimensional superspaces $\left( T_{i, 0}^{\bot} \right)_{i \in [R]}$ of the $R$ corresponding duals $\left( S_{i, 0}^{\bot} \right)_{i \in [R]}$, where for every $i$, $S_{i, 0} := \colspan\left( \matA(i) \right)$. The change we make to the success predicate in the current hybrid is the following: at the end of the execution we get a pair $(x_{0}, x_{1})$ from $\Adv$. We define the process as successful if $y_{0} = y_{1} := y$ and also $\left( \vecU_{0} - \vecU_{1} \right) \notin T_{i_{y}, 0}$, rather than only asking that $x_{0} \neq x_{1}$.

Note that $\Adv$ finds collisions with probability $\epsilon_{2}$ in the previous hybrid $\Hyb_{2}$ (and since this hybrid is no different, the same goes for the current hybrid), which means it finds collisions $(x_{0}, x_{1})$ such that $y_{0} = y_{1} := y$ and $\left( \vecU_{0} - \vecU_{1} \right) \in S_{i_{y}, 0}$. For every value $i \in [R]$ denote by $\epsilon_{2}^{(i)}$ the probability to find a collision in index $i$, or formally, to find $x_{0} \neq x_{1}$ such that $y_{0} = y_{1} := y$, $i_{y} = i$ and $\left( \vecU_{0} - \vecU_{1} \right) \in S_{i_{y}, 0}$. We deduce $\sum_{i \in [R]} \epsilon_{2}^{(i)} = \epsilon_{2}$. Let $L$ be a subset of indices $i \in [R]$ such that $\epsilon_{2}^{(i)} \geq \frac{\epsilon_{2}}{2 \cdot R}$ and note that $\sum_{i \in L} \epsilon_{2}^{(i)} \geq \frac{ \epsilon_{2} }{ 2 }$. Let $\epsilon_{3}$ be the success probability of the current hybrid. For every value $i \in [R]$ also denote by $\epsilon_{3}^{(i)}$ the probability to find a collision such that $y_{0} = y_{1} := y$, $\left( \vecU_{0} - \vecU_{1} \right) \notin T_{i_{y}, 0}$ and also $i_{y} = i$. We deduce $\sum_{i \in [R]} \epsilon_{3}^{(i)} = \epsilon_{3}$.

We would now like to use Lemma \ref{lemma:dual_subspace_concentration}, specifically on $S^{\bot}_{i, 0}$ as the static subspace and $T^{\bot}_{i, 0}$ as its random superspace, and use the lemma for the output of the adversary to avoid the dual $T_{i, 0}$ of the superspace $T^{\bot}_{i, 0}$. So, we make sure that we satisfy its requirements. Let any $i \in L$, we know that by definition $\epsilon_{2}^{(i)} \geq \frac{\epsilon_{2}}{2 \cdot R}$ and also recall that $\epsilon_{2} \geq \frac{3 \cdot \epsilon}{4}$, $R := \left( 300 \cdot q^{3} \right) \cdot \frac{ 2^{7} \cdot k^{2} }{ \epsilon }$ and thus
$$
\epsilon_{2}^{(i)}
\geq
\frac{\epsilon_{2}}{2 \cdot R}
\geq
\frac{3 \cdot \epsilon}{8} \cdot \frac{1}{R}
\geq 
\Omega
\left(
\frac{ \epsilon^2 }{ q^3 \cdot k^2 }
\right)
\enspace .
$$
Note the following parameter interfaces between our setting and the setting of the Lemma: "$k$", "$s$" and "$\lambda$" in the lemma are the same $k$, $s$ and $\lambda$ here and "$r$" in the lemma is $k - (n - r)$ here.
Let $s' := s - (n - r - s)$ and for any $i \in L$ let $\ell_{i} := \frac{k^2}{\epsilon_{2}^{(i)}} \leq O\left( \frac{ k^4 \cdot q^3 }{ \epsilon^2 } \right)$. Note that by our Lemma \ref{lemma:bloating_dual_oracle} statement's conditions, by Condition \ref{bloating_dual_oracle_condition_2} we have (1) $\frac{n - r - s}{2^{s'}} \leq \frac{\epsilon_{2}^{(i)}}{2}$, and by Condition \ref{bloating_dual_oracle_condition_1} we have (2) $\frac{ \lambda \cdot q \cdot \ell_{i}^{2} \cdot s }{ \sqrt{ 2^{k - (k - (n - r)) - \secp - s} } } \leq \frac{1}{2}$. Since this satisfies Lemma \ref{lemma:dual_subspace_concentration}, it follows that for every $i \in L$ we have $\epsilon_{3}^{(i)} \geq \frac{ \epsilon_{2}^{(i)} }{ 16 \cdot k^{2} }$. It follows that
$$
\epsilon_{3}
=
\sum_{i \in [R]} \epsilon_{3}^{(i)}
\geq 
\sum_{i \in L} \epsilon_{3}^{(i)}
\geq
\sum_{i \in L} \frac{ \epsilon_{2}^{(i)} }{ 16 \cdot k^{2} }
\geq
\frac{ \left( \frac{ \epsilon_{2} }{ 2 } \right) }{ 16 \cdot k^{2} }
\geq
\frac{ 3 \cdot \epsilon }{ 2^{7} \cdot k^{2} }
\enspace .
$$

\begin{itemize}
    \item $\Hyb_{4}$: For every $i \in [R]$, de-randomizing $T_{i, 0}$ and defining it as the column span of the $n - r - s$ rightmost columns of the matrix $\matA(i)$, by using the random permutation $\Pi$ and random function $F$.
\end{itemize}
Recall that at the basis of all oracles there are the $R$ cosets $\left( \matA(i), \vecB(i) \right)_{i \in [R]}$. We use $\matA(i), \vecB(i)$ as is in the oracles $\Ps$, $\Ps^{-1}$, but use $\matA(i) \cdot \matM_{i}$ in $\Ds$. We will resolve this discrepancy in this hybrid.
This hybrid is the same as the previous, with one change: For every $i \in [R]$, after sampling the coset $\left( \matA(i), \vecB(i) \right)$, instead of continuing to randomly sample $T^{\bot}_{i, 0}$, we simply define $T_{i, 0} := \colspan\left( \matA(i)^{(1)} \right)$ such that $\matA(i)^{(1)} \in \bbZ_{2}^{k \times (n - r - s)}$ is defined to be the last $n - r - s$ columns of the matrix $\matA(i) \in \bbZ_{2}^{k \times (n - r)}$. We will define intermediate hybrids $\Hyb_{3.1}, \Hyb_{3.2}$ and then explain why the previous hybrid is equivalent to $\Hyb_{3.1} \equiv \Hyb_{3.2} \equiv \Hyb_{4}$.

\begin{itemize}
    \item
    \textbf{Using the random permutation $\Pi$.} 
    For every $i \in [R]$ consider the superspace $T_{i, 0}^{\bot}$, which has $k - n + r + s$ dimensions. Recall the (invertible) matrix $\matM_{i} \in \bbZ_{2}^{(n - r) \times (n - r)}$ from previous hybrid such that $T_{i, 0}$ is the columns span of the $n - r - s$ rightmost columns of $\matT_{i} := \matA(i) \cdot \matM_{i}$. 

    In $\Hyb_{3.1}$, we define the permutation $\Gamma$ over $\{ 0, 1 \}^{n}$ defined as follows: For an input $x \in \{ 0, 1 \}^{n}$, it takes the left $r$ bits denoted $y \in \bbZ_{2}^{r}$, computes $i_{y} \in [R]$, then applies matrix multiplication by $\matM_{i_{y}}$ to the remaining right $n - r$ bits. Observe that since $\matM_{i}$ is invertible for all $i$, then $\Gamma$ is indeed a permutation. The change we make from $\Hyb_{3}$ to $\Hyb_{3.1}$ is that in the current hybrid we apply $\Gamma$ to the \emph{output} of $\Pi$ inside the execution of a query to $\Ps$, and apply $\Gamma^{-1}$ to the \emph{input} of $\Pi^{-1}$ inside the execution of a query to $\Ps^{-1}$. Note that for a truly random $n$-bit permutation $\Pi$, concatenating any fixed permutation $\Gamma$ like this is statistically equivalent to just computing $\Pi$ and $\Pi^{-1}$, thus the outputs (and in particular success probabilities) between $\Hyb_{3}$ and $\Hyb_{3.1}$ are identical.

    \item
    \textbf{Using the random function $F$.}
    In this intermediate $\Hyb_{3.2}$, instead of sampling $\matA_{i}$ and then using it as is, we will sample it multiplied by the inverse of $\matM_{i}$, that is, our sampler gives us $\matA_{i} \cdot \matM^{-1}_{i}$ instead of $\matA_{i}$. This is statistically equivalent due to the randomness of $F$, so perfectly indistinguishable, thus $\Hyb_{3.1} \equiv \Hyb_{3.2}$. Observe that the $\matM_{i}$ multiplication that $\Gamma$ applies now cancels with $\matM_{i}^{-1}$ and this is also true for the dual $\Ds$, because we sample $\matA_{i} \cdot \matM^{-1}_{i}$, and then when we multiply by $\matM_{i}$ in the the computation of $\matT$ this cancels again.

    \item
    Now we get to $\Hyb_{4}$, in which we stop applying the permutation $\Gamma$ to the output of $\Pi$ (and likewise stop applying $\Gamma^{-1}$ to the input of $\Pi^{-1}$), and also stop sampling $\matA_{i}$ multiplied by $\matM^{-1}_{i}$. The subspace $T_{i}^{\bot}$ is simply defined as $\colspan\left( \matA_{i}^{(1)} \right)^{\bot}$, i.e., $\widetilde{\matT}_{i} := \matA_{i}^{(1)}$. Due to our explanation of the cancellation of the matrix multiplication $\matM_{i}$, this is the same as what happens in $\Hyb_{3.2}$, so perfectly indistinguishable.
\end{itemize}
It follows that the success probability $\epsilon_{4}$ in $\Hyb_{4}$ equals the success probability from the previous hybrid.

\begin{itemize}
    \item $\Hyb_{5}$: Moving back to using an exponential number of cosets, by using small-range distribution again.
\end{itemize}
We rewind the process of sampling an $R$-small range distribution version of $F$, and use $F$ as a standard random function. By the same argument for the indistinguishability between $\Hyb_{0}$ and $\Hyb_{1}$, the output of the current process has statistical distance bounded by $\frac{300\cdot q^{3}}{R} = \frac{\epsilon}{2^{7} \cdot k^{2}}$, which means in particular that the outputs of this hybrid and the previous hybrid has statistical distance bounded by $\frac{\epsilon}{2^{7} \cdot k^{2}}$. It follows that the success probability of the current hybrid is
$$
:=
\epsilon_{5}
\geq
\epsilon_{4} - \frac{\epsilon}{2^{7} \cdot k^{2}}
\geq
\frac{ 3 \cdot \epsilon }{ 2^{7} \cdot k^{2} } - \frac{\epsilon}{2^{7} \cdot k^{2}}
=
\frac{ \epsilon }{ 2^{6} \cdot k^{2} }
\enspace .
$$

To conclude, note that the process $\Hyb_{5}$ is exactly the process where $\Adv$ executes on input oracle sampled from $\left( \Ps,\Ps^{-1},\Ds' \right) \gets \Oracle'_{n,r,k,s}$. This finishes our proof.
\end{proof}

\subsection{Simulating the Dual} \label{subsection:simulating_dual_oracle}
In this section we prove the following lemma. 

\begin{lemma} \label{lemma:simulating_dual_oracle}
Suppose there is an oracle aided $q$-query quantum algorithm $\Adv$ such that
\[
\Pr
\left[
\begin{array}{rl}
     & y_0 = y_1 := y , \\
     &\left( \vecU_0 - \vecU_1 \right) \notin \colspan \left( \matA(y)^{(1)} \right)
\end{array}
\;
:
\begin{array}
{rl}
\left( \Ps, \Ps^{-1}, \Ds' \right) & \gets \Oracle'_{n, r, k, s} \\
\left( x_0, x_1 \right) & \gets \Adv^{ \Ps, \Ps^{-1}, \Ds' } \\
(y_b, \vecU_b) & \gets \Ps(x_b)
\end{array}
\right]
\geq
\epsilon \enspace .
\]
Then, there is an oracle aided $q$-query quantum algorithm $\AdvB$ such that
\[
\Pr
\left[
\left( \overline{y}_{0} = \overline{y}_{1} \right) \land \left( \overline{x}_{0} \neq \overline{x}_{1} \right) \; :
\begin{array}{rl}
\left( \overline{\Ps}, \overline{\Ps}^{-1}, \overline{\Ds} \right) & \gets \Oracle_{ r + s, \: r, \: k - (n - r - s) } \\
\left( \overline{x}_{0}, \overline{x}_{1} \right) & \gets \AdvB^{ \overline{\Ps}, \overline{\Ps}^{-1} } \\
\left( \overline{y}_{b}, \overline{\vecU}_b \right) & \gets \overline{\Ps}\left( \overline{x}_{b} \right)
\end{array}
\right]
\geq
\epsilon
\enspace .
\] 
\end{lemma}

\begin{proof}
We first describe the actions of the algorithm $\AdvB$ (which will use the code of $\Adv$ as part of its machinery) and then argue why it breaks collision resistance with the appropriate probability. Given oracle access to $\overline{\Ps}, \overline{\Ps}^{-1}$ which comes from $\left( \overline{\Ps}, \overline{\Ps}^{-1}, \overline{\Ds} \right) \gets \Oracle_{ r + s, \: r, \: k - (n - r - s) }$, the algorithm $\AdvB$ does the following:
\begin{itemize}
    \item
    Sample a random function $F_{\matC}$ that outputs a uniformly random invertible $\matC(y) \in \bbZ_{2}^{ k \times k }$, and a uniformly random vector $\vecD(y) \in \bbZ_{2}^{n - r - s}$. Sample a random $n$-bit permutation $\Gamma$. Define the following oracles.

    \item 
    $\left( \; y \in \bbZ_{2}^{r}, \; \vecU \in \bbZ_{2}^{k} \; \right) \gets \Ps\left( x \in \bbZ_{2}^{n} \right)$:
    \begin{itemize}
        \item
        $\left(
        \; \overline{x} \in \bbZ_{2}^{r + s},
        \; \widetilde{x} \in \bbZ_{2}^{n - r - s} \;
        \right) \gets \Gamma(x)$.

        \item 
        $\left(
        \; y \in \bbZ_{2}^{r},
        \; \overline{\vecU} \in \bbZ_{2}^{k - (n - r - s)} \;
        \right) \gets \overline{\Ps}(\overline{x})$.

        \item 
        $\left(
        \; \matC(y) \in \bbZ_{2}^{ k \times k },
        \; \vecD(y) \in \bbZ_{2}^{n - r - s}
        \right)
        \gets
        F_{\matC}(y)$.

        \item 
        $\vecU
        \gets
        \matC(y) \cdot \left( \begin{array}{c} \overline{\vecU} \\ \widetilde{x} + \vecD(y) \end{array} \right)$.
    \end{itemize}

    \item 
    $\left( \; x \in \bbZ_{2}^{n} \; \right)
    \gets
    \Ps^{-1}\left( \; y \in \bbZ_{2}^{r}, \; \vecU \in \bbZ_{2}^{k} \; \right)$:
    \begin{itemize}
        \item 
        $\left(
        \; \matC(y) \in \bbZ_{2}^{ k \times k },
        \; \vecD(y) \in \bbZ_{2}^{n - r - s}
        \right) \gets F_{\matC}(y)$.

        \item 
        $\left( \begin{array}{c} \overline{\vecU} \\ \widetilde{x} \end{array} \right) \gets \matC(y)^{-1} \cdot \vecU - \left( \begin{array}{c} 0^{k-(n - r - s)} \\ \vecD(y) \end{array} \right)$.

        \item 
        $\left( \; \overline{x} \in \bbZ_{2}^{r + s} \; \right)
        \gets
        \overline{\Ps}^{-1}\left( y, \overline{\vecU} \right)$.

        \item
        $x
        \gets
        \Gamma^{-1}\left( \overline{x}, \widetilde{x} \right)$.
    \end{itemize}

    \item 
    $
    \vecC_{y, \vecV}
    \gets
    \Ds'\left( \; y \in \bbZ_{2}^{r}, \; \vecV \in \bbZ_{2}^{k} \; \right)$:
    \begin{itemize}
        \item 
        $\left(
        \; \matC(y) \in \bbZ_{2}^{ k \times k },
        \; \vecD(y) \in \bbZ_{2}^{n - r - s}
        \right)
        \gets
        F_{\matC}\left( y \right)$.

        \item 
        $\matA^{(1)}(y) := $ last $n - r - s$ columns of $\matC(y)$. 

        \item 
        Simulate the answer using $\matA^{(1)}(y)$, which is sufficient.
    \end{itemize}
\end{itemize}

The remainder of the reduction is simple: $\AdvB$ executes $\left( x_0, x_1 \right) \gets \Adv^{ 
\Ps, \Ps^{-1}, \Ds' }$ and then $\left( \overline{x}_{b}, \widetilde{x}_{b} \right) \gets \Gamma(x_{b})$ and outputs $\left( \overline{x}_{0}, \overline{x}_{1} \right)$. Assume that the output of $\Adv$ satisfies $y_{0} = y_{1} := y$ and also $\left( \vecU_0 - \vecU_1 \right) \notin \colspan \left( \matA(y)^{(1)} \right)$, and recall that $\matA(y)^{(1)} \in \bbZ_{2}^{k \times (n - r - s)}$ are the last $n - r - s$ columns of the matrix $\matA(y) \in \bbZ_{2}^{k \times (n - r)}$, which is generated by the reduction. We explain why it is necessarily the case that $\overline{x}_{0} \neq \overline{x}_{1}$.

First note that due to how we defined the reduction, $\matA(y) := \matC(y) \cdot \left( \begin{array}{cc} \overline{\matA}(y) & \\ & \Id_{n - r - s} \end{array} \right)$, where $\overline{\matA}(y) \in \bbZ_{2}^{(k - (n - r - s)) \times s}$ is the matrix arising from the oracles $\overline{\Ps}, \overline{\Ps}^{-1}$ and $\Id_{n - r - s} \in \bbZ_{2}^{(n - r - s) \times (n - r - s)}$ is the identity matrix of dimension $n - r - s$. Also note that because $\matC(y)$, $\overline{\matA}(y)$ are full rank then $\matA(y)$ is full rank. Now, since $\left( \vecU_0 - \vecU_1 \right) \notin \colspan \left( \matA(y)^{(1)} \right)$ and since $\matA(y)^{(1)}$ are the last $n - r - s$ columns of $\matA(y)$, it follows that if we consider the coordinates vector $\vecX \in \bbZ_{2}^{n - r}$ of $\left( \vecU_0 - \vecU_1 \right)$ with respect to $\matA(y)$, the first $s$ elements are not $0^{s}$. By linearity of matrix multiplication it follows that if we look at each of the two coordinates vectors $\vecX_{0}$, $\vecX_{1}$ (each has $n - r$ bits) for $\vecU_{0}$, $\vecU_{1}$, respectively, somewhere in the first $s$ bits, they differ.
Now, recall how we obtain the first $s$ bits of $\vecX_{b}$ -- this is exactly by applying $\overline{\Pi}$ (the permutation on $\{ 0, 1 \}^{r + s}$ arising from the oracles $\overline{\Ps}, \overline{\Ps}^{-1}$) to $\overline{x}_{b}$ and taking the last $s$ bits of the output. Since these bits differ in the output of the permutation, then the preimages have to differ, i.e., $\overline{x}_{0} \neq \overline{x}_{1}$.

Define $\epsilon_{\AdvB}$ as the probability that the output of $\Adv$ indeed satisfies $y_{0} = y_{1} := y$ and also $\left( \vecU_0 - \vecU_1 \right) \notin \colspan \left( \matA^{(1)}(y) \right)$, and it remains to give a lower bound for the probability $\epsilon_{\AdvB}$. We do this by a sequence of hybrids, eventually showing that the oracle which $\AdvB$ simulates to $\Adv$ is indistinguishable from an oracle sampled from $\Oracle'_{ n, r, k, s }$. More precisely, each hybrid describes a process, it has an output, and a success predicate on the output.

\begin{itemize}
    \item $\Hyb_{0}$: The above distribution $\left( \Ps, \Ps^{-1}, \Ds' \right) \gets \AdvB^{ 
\overline{\Ps}, \overline{\Ps}^{-1} }$, simulated to the algorithm $\Adv$.
\end{itemize}
The first hybrid is where $\AdvB$ executes $\Adv$ by the simulation described above. The output of the process is the output $(x_{0}, x_{1})$ of $\Adv$. The process execution is considered as successful if $y_{0} = y_{1} := y$ and $\left( \vecU_{0} - \vecU_{1} \right) \notin \colspan\left( 
\matA(y)^{(1)} \right)$.

\begin{itemize}
    \item $\Hyb_{1}$: Not applying the inner permutation $\overline{\Pi}$ (which comes from the oracles $\overline{\Ps}$, $\overline{\Ps}^{-1}$), by using the random permutation $\Gamma$.
\end{itemize}
Let $\overline{\Pi}$ the permutation on $\{ 0, 1 \}^{r + s}$ that's inside $\overline{\Ps}$. In the previous hybrid we apply the $n$-bit permutation $\Gamma$ to the input $x \in \bbZ_{2}^n$ and then proceed to apply the inner permutation $\overline{\Pi}$ to the first (i.e. leftmost) $r + s$ output bits of the first permutation $\Gamma$ (we also apply $\Gamma^{-1}$ to the output of the inverse of the inner permutation, in the inverse oracle $\Ps^{-1}$). The change we make to the current hybrid is that we simply apply only $\Gamma$ and discard the inner permutation and its inverse. Since a random permutation concatenated with any permutation distributes identically to a random permutation, the current hybrid is statistically equivalent to the previous and in particular the output of this process distributes identically to the output of the previous, and so does the success probability.

\begin{itemize}
    \item $\Hyb_{2}$: For every $y \in \bbZ_{2}^{r}$, taking $\matA(y)$ to be the direct output of $F$, by using the randomness of the random function.
\end{itemize}
In order to describe the change between the current and previous hybrid we first recall the structure of the oracles from the previous hybrid: Observe that in the previous hybrid, for every $y \in \bbZ_{2}^{r}$ we defined $\matA(y) := \matC(y) \cdot \left( \begin{array}{cc} \overline{\matA}(y) & \\ & \Id_{n - r - s} \end{array} \right)$, where $\matC(y) \in \bbZ_{2}^{k \times k}$ is the output of $F_{\matC}(y)$ and $\overline{\matA}(y) \in \bbZ_{2}^{(k - (n - r - s)) \times s}$ is the output of the inner random function $\overline{F}$ (which comes from the inside of the oracles $\left( \overline{\Ps}, \overline{\Ps}^{-1} \right)$). In the current hybrid we are going to ignore the inner random function $\overline{F}$, its generated matrix $\overline{\matA}(y)$ and also the pair $\matC(y)$, $\vecD(y)$, sample a fresh random function $F_{\matA}$ at the beginning of the process, and on query $y$ generate $\left( \matA(y), \vecB(y) \right) \gets F_{\matA}(y)$, for $\matA(y) \in \bbZ_{2}^{k \times (n - r)}$, $\vecB(y) \in \bbZ_{2}^{k}$.

To see why the two distributions are indistinguishable, note that the following two ways to sample $\matA(y)$, are statistically equivalent: (1) For every $y \in \bbZ_{2}^{r}$, the matrix $\matA(y)$ is generated by sampling a random full-rank matrix $\matC(y) \in \bbZ_{2}^{k \times k}$ and letting $\matA(y)$ be $\matC(y) \cdot \left( \begin{array}{cc} \overline{\matA}(y) & \\ & \Id_{n - r - s} \end{array} \right)$. (2) For every $y \in \bbZ_{2}^{r}$ just sample a full-rank matrix $\matA(y) \in \bbZ_{2}^{k \times (n - r)}$. Since we are using random functions and in the previous hybrid we are sampling $\matA(y)$ according to (1) and in the current hybrid we are sampling $\matA(y)$ according to (2), the outputs of the two hybrids distribute identically.

\paragraph{Finalizing the reduction.}
Observe that the distribution generated in the above $\Hyb_{2}$ is exactly an oracle sampled from $\Oracle'_{ n, r, k, s }$. From the lemma's assumptions, the success probability for $\Hyb_{2}$ is thus $\epsilon$. Since we also showed that the hybrids have identical success probabilities, it follows that $\epsilon_{\AdvB} = \epsilon$, which finishes our proof.
\end{proof}

We conclude this section by stating the following Theorem, which is obtained as a direct corollary from Lemmas \ref{lemma:bloating_dual_oracle} and \ref{lemma:simulating_dual_oracle}.

\begin{theorem} \label{thm:dualtodualfree}
Suppose there is an oracle aided $q$-query quantum algorithm $\Adv$ such that
\[
\Pr
\left[
x_0 \neq x_1
\land 
H(x_0) = H(x_1) 
\;
:
\begin{array}
{rl}
\left( \Ps, \Ps^{-1}, \Ds \right) & \gets \Oracle_{n,r,k} \\
\left( x_0, x_1 \right) & \gets \Adv^{\Ps, \Ps^{-1}, \Ds}
\end{array}\right]
\geq
\epsilon \enspace .
\]
Also, let $s \in \Nat$ such that $s \leq n - r - \secp$, define $s' := s - (n - r - s)$, and assume
\begin{enumerate}
    \item
    $\frac{ \lambda \cdot k^{8} \cdot q^8 \cdot s }{ \sqrt{ 2^{n - r - \secp - s} } } \leq o\left( \epsilon^4 \right)$,

    \item
    $\frac{ k^{2} \cdot q^3 \cdot (n - r - s) }{ 2^{s'} } \leq o\left( \epsilon^2 \right)$.
\end{enumerate}
Then, there is an oracle aided $q$-query quantum algorithm $\AdvB$ such that
\[
\Pr
\left[
x_0 \neq x_1 \land H(x_0) = H(x_1)
\;
:
\begin{array}{rl}
\left( \Ps, \Ps^{-1}, \Ds \right) \gets \Oracle_{r+s,\: r,\: k-(n - r - s)} \\
(x_0, x_1) \gets \AdvB^{\Ps,\Ps^{-1}}
\end{array}
\right]
\geq
\frac{ \epsilon }{ 2^{6} \cdot k^{2} }
\enspace .
\]
\end{theorem}

\subsection{Hardness of the Dual-free Case from Claw-free Permutations} \label{subsection:dual_free_to_two_to_one_oracle}
In this section we show that collision-finding in the dual-free case is provably hard with respect to an oracle. Our proof is different than the proof of \cite{C:ShmZha25} for collision resistance, which is based on random 2-to-1 functions, and we refer the reader to the technical overview for an intuitive explanation on the difference. 

We define coset partition functions (a notion defined in \cite{C:ShmZha25}) and claw-free permutations, two objects we will use in the proof.

\begin{definition} [Coset Partition Functions] \label{definition:coset_partition_function}
For $n, \ell \in \Nat$ such that $\ell \leq n$ we say a function $Q:\{0,1\}^n \rightarrow \{0,1\}^m$ is a $(n,m,\ell)$-\emph{coset partition function} if, for each $y$ in the image of $Q$, the pre-image set $Q^{-1}(y)$ has size $2^\ell$ and is a coset of a linear space of dimension $\ell$. We allow different pre-image sets to be cosets of different linear spaces.
\end{definition}

\begin{definition} [Claw-free Permutation] \label{definition:claw_free_permutation}
For security parameter $\lambda \in \Nat$, a sample from the distribution of claw-free permutations for parameter $\lambda$ is given as follows. Sample two uniformly random permutations $\Pi_0, \Pi_1 : \{ 0, 1 \}^{\lambda} \rightarrow \{ 0, 1 \}^{\lambda}$ and let $H^* : \{ 0, 1 \}^{\lambda + 1} \rightarrow \{ 0, 1 \}^{\lambda}$ the function that for input $\left( b \in \{ 0, 1 \}, x \in \{ 0, 1 \}^{\secp} \right)$ outputs $\Pi_{b}\left( x \right)$. The output sample of the claw-free distribution is $H^*$.
\end{definition}

\paragraph{From Dual-free to Claw-free Permutations.} 
We next show that finding collisions in the dual-free case is at least as hard as finding collisions in claw-free permutations, while crucially keeping the dimension that the cosets live in, denoted by $k$, linear in the security parameter. Specifically, in the below Theorem, $k$ need not be as large as $n$ and only requires $k \geq n - r + \lambda$, which is a key point of difference compared to the reduction from \cite{C:ShmZha25}.

\begin{theorem} [Reduction to Collision-finding in Claw-free Permutations] \label{thm:dualfreetocol}
Let $n, r, k \in \Nat$ with $n > r$ and furthermore such that $\frac{n}{n - r}$ is an integer (which implies that $\frac{r}{n - r}$ is also an integer). Denote $\lambda := \frac{r}{n - r}$ (which implies $\lambda + 1 = \frac{n}{n - r}$) and additionally assume $k \geq n - r + \lambda$. Suppose there is an oracle aided $q$-query quantum algorithm $\Adv$ such that
\[
\Pr
\left[
\left( y_{0} = y_{1} \right) \land \left( x_{0} \neq x_{1} \right) \; :
\begin{array}{rl}
\left( \Ps, \Ps^{-1}, \Ds \right) & \gets \Oracle_{ n, r, k } \\
\left( x_{0}, x_{1} \right) & \gets \Adv^{ \Ps, \Ps^{-1} } \\
\left( y_{b}, \vecU_{b} \right) & \gets \Ps(x_b)
\end{array}
\right]
\geq
\epsilon
\enspace .
\] 
Then there is an oracle aided $q$-query quantum algorithm $\AdvB$ that given oracle access to $H^* : \{ 0, 1 \}^{\lambda + 1} \rightarrow \{ 0, 1 \}^{\lambda}$, a uniformly random claw-free permutation, satisfies
\[
\Pr
\left[
\left( H^*(w_0) = H^*(w_1) \right) \land \left( w_0 \neq w_1 \right)
\;
:
\;
\left( w_0, w_1 \right) \gets \AdvB^{H^*}
\right]
\geq
\frac{\epsilon}{n - r}
\enspace .
\]
\end{theorem}

\begin{proof}
The reduction $\AdvB$ works in two main steps. First, it turns oracle access to $H^*$ into oracle access to some $(n, r, n - r)$-coset partition function $Q$ with a special property. Then, it will turn oracle access to $Q$ into oracle access to $\Ps,\Ps^{-1}$, while crucially keeping the parameter $k$ small.

$\AdvB$ samples $n - r - 1$ instances of claw-free functions $H_{1}, \cdots, H_{n - r - 1}$, and for each it also samples the "trapdoor" i.e. for every instance $i \in [n - r - 1]$ it samples not only the permutations $\Pi_{i, 0}, \Pi_{i, 1} : \{ 0, 1 \}^{\lambda} \rightarrow \{ 0, 1 \}^{\lambda}$, but also the inverses $\Pi^{-1}_{i, 0}, \Pi^{-1}_{i, 1} : \{ 0, 1 \}^{\lambda} \rightarrow \{ 0, 1 \}^{\lambda}$. Next, sample a uniformly random place $i^* \in [n - r]$ and consider the parallel repetition function $Q : \{ 0, 1 \}^{n} \rightarrow \{ 0, 1 \}^{r}$, defined as follows.
$$
Q\left( \vecW \in \{ 0, 1 \}^{n} \right) :=
H'_{1}\left( \vecW_1 \right), \cdots, H'_{n - r}\left( \vecW_{n - r} \right)
\enspace ,
$$
where $H^*$ is in place $i^*$, i.e., $H'_{i^*} = H^*$, and for every $i \in [n - r]$, $\vecW_{i}$ is derived by partitioning $\vecW \in \{ 0, 1 \}^{n}$ into consecutive, equally-sized $n - r$ parts, each one (accordingly) consisting of $\frac{n}{n - r} := \lambda + 1$ bits. We will later explain why $Q$ is a $(n, r, n - r)$-coset partition function and furthermore has an additional useful property of having "foldable" cosets.

First, we continue to describe the reduction. $\AdvB$ chooses a random permutation $\Gamma : \{0,1\}^n \rightarrow \{0,1\}^n$, and for each $y \in \{ 0, 1 \}^{r}$, it chooses a random full-column-rank matrix $\matC(y) \in \bbZ_{2}^{k \times (n - r + \lambda)}$ (which is possible since we assume $k \geq n - r + \lambda$) and random vector $\vecD(y) \in \bbZ_{2}^k$. It then runs $\Adv$, simulating the oracles $\Ps,\Ps^{-1}$ as follows:

\paragraph{The oracle $\Ps\left( x \in \{ 0, 1 \}^{n} \right)$:}
\begin{enumerate}
    \item 
    Compute $\vecW \gets \Gamma\left( x \right)$.

    \item
    $y \gets Q\left( \vecW \right)$.

    \item \label{CPF_reduction_fold}
    \textbf{Fold $\vecW \in \{ 0, 1 \}^{n}$ into $\widetilde{\vecW} \in \{ 0, 1 \}^{n - r + \lambda}$:}
    \begin{enumerate}
        \item 
        Generate a coordinates vector $\vecR \in \bbZ_{2}^{n - r}$ such that for $i \in [n - r]$, bit $i$ of $\vecR$ is the first bit of $\vecW_{i} \in \{ 0, 1 \}^{\lambda + 1}$, which in turn is the input to $H'_{i}$ inside the computation of $Q$. 

        \item 
        For each $\vecW_{i} \in \{ 0, 1 \}^{\lambda + 1}$ consider $\left( \vecW_{i} \right)_{-1} \in \{ 0, 1 \}^{\lambda}$, derived by discarding the first bit of $\vecW_{i}$. Then set
        $$
        \widetilde{\vecW} := \left( \vecR, \sum_{i \in [n - r]} \left( \vecW_{i} \right)_{-1} \right) \enspace .
        $$
    \end{enumerate}
    
    \item
    Output $\left( y, \matC(y) \cdot \widetilde{\vecW} + \vecD(y) \right)$.
\end{enumerate}

\paragraph{The oracle $\Ps^{-1}\left( y \in \{ 0, 1 \}^{r}, \vecU \in \bbZ_{2}^{k} \right)$:}
\begin{enumerate}
    \item
    $x \gets
    \begin{cases}
    \Gamma^{-1}\left( \mathsf{Unfold}\left( \widetilde{\vecW}, y \right) \right)
    &\text{ $\exists \; \widetilde{\vecW} \in \bbZ_{2}^{n - r + \lambda}$ such that $\matC(y) \cdot \widetilde{\vecW} + \vecD(y) = \vecU$} \\
    \bot
    &\text{ if no such $\widetilde{\vecW}$ exists}
    \end{cases}$

    \item \label{CPF_reduction_UNfold}
    $\vecW \gets \mathsf{Unfold}\left( \widetilde{\vecW}, y \right)$ executes as follows.
    \begin{enumerate}
        \item
        Recall the index $i^* \in [n - r]$ such that $H'_{i^*} := H^*$, where $H^*$ is the claw-free permutation which the reduction is trying to break. Also recall that for every $j \in [n - r] \setminus \{ i^* \}$ we have the inverse permutations $\Pi^{-1}_{j, 0}, \Pi^{-1}_{j, 1}$, as the reduction $\AdvB$ sampled them by itself.

        \item 
        Parse $\widetilde{\vecW} := \left( \vecR \in \bbZ_{2}^{n - r}, \overline{\vecW} \in \{ 0, 1 \}^{\lambda} \right)$. Write $y = \left( y_{1}, \cdots, y_{n - r} \right)$ where $y_{i} \in \{ 0, 1 \}^{\frac{r}{n - r}}$ for every $i \in [n - r]$. 

        \item 
        For every $j \in [n - r] \setminus \{ i^* \}$ compute $\vecW_{\left( j, \; y_{j}, \; \vecR_{j} \right)} \gets \Pi^{-1}_{j, \vecR_{j}}\left( y_{j} \right)$, where $\vecR_{j} \in \{ 0, 1 \}$ is the $j$-th bit of $\vecR$.

        \item 
        Subtract to get
        $$
        \vecW^* \gets \overline{\vecW} - \sum_{ j \in [n - r] \setminus \{ i^* \} } \vecW_{\left( j, \; y_{j}, \; \vecR_{j} \right)} \enspace .
        $$

        \item 
        Output
        \begin{align*}
        \vecW \gets \biggl(
        \left( \vecR_{1}, \vecW_{\left( 1, \; y_{1}, \; \vecR_{1} \right)} \right),
        \cdots, 
        &\left( \vecR_{i^* - 1}, \vecW_{\left( i^* - 1, \; y_{i^* - 1}, \; \vecR_{i^* - 1} \right)} \right), 
        \\
        &\left( \vecR_{i^*}, \; \vecW^* \right),
        \\
        &\left( \vecR_{i^* + 1}, \vecW_{\left( i^* + 1, \; y_{i^* + 1}, \; \vecR_{i^* + 1} \right)} \right),
        \cdots, 
        \left( \vecR_{n - r}, \vecW_{\left( n - r, \; y_{n - r}, \; \vecR_{n - r} \right)} \right)
        \biggr) \enspace .
        \end{align*}
    \end{enumerate}
    
    \item
    Output
    $\begin{cases}
    x &\text{ if $x \neq \bot$ and $Q\left( \Gamma(x) \right) = y$ } \\ \bot &\text{ if $x=\bot$ or $Q(\Gamma(x))\neq y$ }
    \end{cases}$
\end{enumerate}

Before we explain the rationale behind the above reduction, we observe a number of properties of the function $Q$.

\paragraph{$Q$ is a coset partition function.}
First, it can be verified by the reader that since the functions $H'_{i}$ are all 2-to-1, then $Q$ is a $(n, r, n - r)$-coset partition function. Let us be more formal about this: every output $y \in \{ 0, 1 \}^{r}$ of $Q$ is given by $n - r$ outputs $y_{i} \in \{ 0, 1 \}^{\lambda}$ of the functions $H'_{i}$, each of these consists of $\lambda := \frac{r}{n - r}$ bits. Since $H'_{i}$ is 2-to-1, for every $y_{i}$ there exist a pair $\vecW_{\left( i, \; y_{i}, \; 0 \right)}, \vecW_{\left( i, \; y_{i}, \; 1 \right)} \in \{ 0, 1 \}^{\lambda + 1}$ such that $H'_{i}\left( \vecW_{\left( i, \; y_{i}, \; 0 \right)} \right) = H'_{i}\left( \vecW_{\left( i, \; y_{i}, \; 1 \right)} \right) = y_{i}$.

We can extend each of $\vecW_{\left( i, \; y_{i}, \; 0 \right)}, \vecW_{\left( i, \; y_{i}, \; 1 \right)} \in \{ 0, 1 \}^{\lambda + 1}$ to $\vecW'_{\left( i, \; y_{i}, \; 0 \right)}, \vecW'_{\left( i, \; y_{i}, \; 1 \right)} \in \{ 0, 1 \}^{n}$ by padding with zeros, and placing the original $\vecW_{\left( i, \; y_{i}, \; 0 \right)}, \vecW_{\left( i, \; y_{i}, \; 1 \right)}$ at packet $i \in [n - r]$. Specifically, the resulting $\vecW'_{\left( i, \; y_{i}, \; b \right)} \in \{ 0, 1 \}^{n}$ will ultimately contain $n - r$ packets, each of size $\lambda + 1 = \frac{n}{n - r}$ and all but a single one (i.e., $n - r - 1$ packets) will contain only zeros, and the single non-zero packet will be at index $i \in [n - r]$, which will contain the original $\vecW_{\left( i, \; y_{i}, \; b \right)} \in \{ 0, 1 \}^{\lambda + 1}$.

Now, observe what are the cosets generated by $Q$: For every $y \in \{ 0, 1 \}^{r}$ consider the matrix $\overline{\matA}_{y} \in \bbZ_{2}^{n \times (n - r)}$ such that for every $i \in [n - r]$, its column $i$ is $\vecW'_{\left( i, \; y_{i}, \; 0 \right)} + \vecW'_{\left( i, \; y_{i}, \; 1 \right)}$. Also, consider the vector $\overline{\vecB}_{y} \in \{ 0, 1 \}^{n}$ defined as $\overline{\vecB}_{y} := \sum_{i \in [n - r]} \vecW'_{\left( i, \; y_{i}, \; 0 \right)}$. It remains to observe that for every $y \in \{ 0, 1 \}^{r}$ output of $Q$, the set of inputs that maps to $y$ is given by the coset $\colspan\left( \overline{\matA}_{y} \right) + \overline{\vecB}_{y}$.

\paragraph{For every preimage coset in $Q$ there exists a folded coset, with reversible mapping.}
We showed that for every output $y$ of $Q$, its preimage set $Q^{-1}\left( y \right)$ is a coset of the form $\colspan\left( \overline{\matA}_{y} \right) + \overline{\vecB}_{y}$ for some $\overline{\matA}_{y} \in \bbZ_{2}^{n \times (n - r)}$, $\overline{\vecB}_{y} \in \bbZ_{2}^{n}$. We next show that while the preimage sets of the outputs of $Q$ might need to be large (i.e. their elements take $n$ bits), for every $y$ there also exists \emph{a folded coset} $\colspan\left( \widetilde{\matA}_{y} \right) + \widetilde{\vecB}_{y}$ for some $\widetilde{\matA}_{y} \in \bbZ_{2}^{(n - r + \lambda) \times (n - r)}$, $\widetilde{\vecB}_{y} \in \bbZ_{2}^{n - r + \lambda}$, and not only that, the reduction $\AdvB$ above shows how to go \emph{reversibly} between the original and folded cosets.

The folded coset is defined as follows. For every $y_{i}$ consider the pair $\vecW_{\left( i, \; y_{i}, \; 0 \right)}, \vecW_{\left( i, \; y_{i}, \; 1 \right)} \in \{ 0, 1 \}^{\lambda + 1}$ such that $H'_{i}\left( \vecW_{\left( i, \; y_{i}, \; 0 \right)} \right) = H'_{i}\left( \vecW_{\left( i, \; y_{i}, \; 1 \right)} \right) = y_{i}$. Let $\widetilde{\matA}_{y} \in \bbZ_{2}^{(n - r + \lambda) \times (n - r)}$ the matrix such that its $i$-th column has two parts. The first part takes $n - r$ bits and is equal to the $i$-th column of the identity matrix $\matI_{(n - r) \times (n - r)}$. The remaining $\lambda$ bits are equal to $\left( \vecW_{\left( i, \; y_{i}, \; 0 \right)} + \vecW_{\left( i, \; y_{i}, \; 1 \right)} \right)_{-1}$. As for the coset shift $\widetilde{\vecB}_{y} \in \bbZ_{2}^{n - r + \lambda}$, it also has two parts. The first part is $0^{n - r}$, and the second part is $\sum_{i \in [n - r]} \left( \vecW_{\left( i, \; y_{i}, \; 0 \right)} \right)_{-1}$.

We conclude with two observations. First, the efficient and reversible mapping between the cosets: to fold $\vecW \rightarrow \widetilde{\vecW}$ one executes Step \ref{CPF_reduction_fold} in the reduction's simulation of $\Ps$, and to unfold, one needs the information of the image $y \in \{ 0, 1 \}^{r}$ and executes Step \ref{CPF_reduction_UNfold} in the above simulation of $\Ps^{-1}$. Finally, a key property which lets all of this connect, is that for any preimage $\vecW$ such that $Q\left( \vecW \right) = y$, the coordinates vector $\vecZ \in \bbZ_{2}^{n - r}$ of $\vecW$ in the original coset $\colspan\left( \overline{\matA}_{y} \right) + \overline{\vecB}_{y}$ and its folding $\widetilde{\vecW} \in \colspan\left( \widetilde{\matA}_{y} \right) + \widetilde{\vecB}_{y}$, are the same. Specifically, one can observe that this is exactly the vector $\vecR \in \bbZ_{2}^{n - r}$. Thus, when going back and forth between the cosets using the reduction $\AdvB$, the coordinates $\vecZ$ do not change.

\paragraph{Correct distribution of the simulated oracles $\Ps, \Ps^{-1}$.}
In a nutshell, what we'll do is use both cosets, the original and the folded one, to correctly simulate the two corresponding parts of the oracles $\Ps, \Ps^{-1}$, which are the random permutation $\Pi$ inside and the random cosets $\colspan\left( \matA(y) \right) + \vecB(y)$ for every output $y$. The original coset $\colspan\left( \overline{\matA}_{y} \right) + \overline{\vecB}_{y}$ will be used to simulate the permutation $\Pi$, and the folded coset $\colspan\left( \widetilde{\matA}_{y} \right) + \widetilde{\vecB}_{y}$ will be used to simulate the cosets $\colspan\left( \matA(y) \right) + \vecB(y)$, i.e., obliviously sample $\matA(y) \in \bbZ_{2}^{k \times (n - r)}$ and $\vecB(y) \in \bbZ_{2}^{k}$.

We start with showing that the simulated $\Ps$ distributes correctly. We define an augmented function $Q' : \{0,1\}^n \rightarrow \{0,1\}^n$. On input $\vecW$, the $n$-bit output of $Q'\left( \vecW \right)$ consists of two parts. The first $r$ bits are set to $y = Q\left( \vecW \right)$. The preimage set $Q^{-1}\left( y \right)$ is then a coset, which can be described as the set $\{ \overline{\matA}_y \cdot \vecR + \overline{\vecB}_y \}$ as $\vecR$ ranges over $\bbZ_2^{n-r}$ (where $\overline{\matA}_y$, $\overline{\vecB}_{y}$ are both unknown to the reduction algorithm $\AdvB$), as we described above. Define the function $\overline{J}\left( \vecW \right)$ that outputs the unique vector in $\bbZ_{2}^{n-r}$ such that $\vecW = \overline{\matA}_y \cdot \overline{J}\left( \vecW \right) + \overline{\vecB}_y$. Then define $Q'\left( \vecW \right) = \left( Q\left( \vecW \right) , \overline{J}\left( \vecW \right) \right)$. Note that $Q'$ is not efficiently computable without knowing $\overline{\matA}_y$ ,$\overline{\vecB}_y$, but here we will only need it to exist, and not need it to be efficiently computable. Notice that $Q'$ is a function from $\bbZ_2^{n}$ to $\bbZ_2^n$, and it is moreover a permutation with $\left( Q' \right)^{-1}\left( y, \vecR \right) = \overline{\matA}_y \cdot \vecR + \overline{\vecB}_y$. 

Since for every vector $\vecW$ with $Q\left( \vecW \right) = y$, its coordinates vector $\overline{J}\left( \vecW \right) \in \bbZ_{2}^{n - r}$ is the same in both, the original coset $\colspan\left( \overline{\matA}_{y} \right) + \overline{\vecB}_{y}$, and the folded coset $\colspan\left( \widetilde{\matA}_{y} \right) + \widetilde{\vecB}_{y}$, observe that $\AdvB$'s simulation of $\Ps$ is implicitly setting the following parameters
\begin{align*}
\Pi(x) & = Q'(\Gamma(x)) \; , & H(x) & = Q(\Gamma(x)) \; , & J(x)=\overline{J}(\Gamma(x)) \; , \\
\matA(y)  & =\matC(y)\cdot\widetilde{\matA}_y \; , & \vecB(y) & =\matC(y)\cdot\widetilde{\vecB}_y+\vecD(y) \enspace .
\end{align*}
Thus, we must check that these quantities have the correct distribution. Indeed, for every $Q$, which is in turn defines $\left( \overline{\matA}_y, \overline{\vecB}_y \right)_{ y \in \{ 0, 1 \}^{r} }$, the function $Q'(x)$ is a permutation: Given $y \in \{ 0, 1 \}^{r}$, $\vecR \in \bbZ_{2}^{n - r}$, one can recover $\vecZ \in \{ 0, 1 \}^{n}$ as $\vecZ = \overline{\matA}_y \cdot \vecR + \overline{\vecB}_y$. Hence $\Pi$ is a permutation since it is the composition of two permutations. Moreover, since one of the two permutations ($\Gamma$) is uniformly random, so is $\Pi$. 

Now we look at the distribution of $\matA(y),\vecB(y)$. Recall that $\widetilde{\matA}_y \in \Z_2^{(n - r + \lambda) \times (n-r)}$ is a full-column-rank matrix, and $\matC(y) \in \Z_2^{k\times (n - r + \lambda)}$ is a \emph{random} full-column-rank matrix. Thus, $\matA(y) = \matC(y) \cdot \widetilde{\matA}_y \in \Z_2^{k \times (n-r)}$ is also a random full-column-rank matrix. Also, we have that $\vecB(y) = \matC(y) \cdot \widetilde{\vecB}_y + \vecD(y)$ where $\vecD(y)$ is random, meaning $\vecB(y)$ is random. Thus, $\Ps$ has an identical distribution to that arising from $\Os_{n,r,k}$. 

For the inverse $\Ps^{-1}$, observe that $\Ps^{-1}\left( \Ps(x) \right) = x$, and for all pairs $\left( y \in \{ 0, 1 \}^{r}, \vecU \in \bbZ_{2}^{k} \right)$ that are not in the image of $\Ps$, we have $\Ps^{-1}\left( y, \vecU \right) = \bot$. Thus, $\Ps^{-1}$ is the uniquely-defined inverse of $\Ps$. Thus, since the distribution of $\Ps$ simulated by $\AdvB$ exactly matches the distribution arising from $\Os_{n,r,k}$, the same is true of the pairs $\left( \Ps, \Ps^{-1} \right)$.

\paragraph{Finishing touches.}
We saw that for $H^* : \{ 0, 1 \}^{\lambda + 1} \rightarrow \{ 0, 1 \}^{\lambda}$ a random claw-free permutation, the reduction $\AdvB$ first simulates a foldable $(n, r, n - r)$-coset partition function $Q$, and proceeds to perfectly simulate the view of $\Adv$, which consists of the pair of oracles $\Ps,\Ps^{-1}$ that distribute according to $\Os_{n,r,k}$. Hence, with probability $\epsilon$, the algorithm $\Adv$ will produce a collision $x_0\neq x_1$ such that $H(x_0)=H(x_1)$, where $H$ is the first $r$ output bits of $\Ps$. It remains to explain what $\AdvB$ does in order to obtain a collision in $H^*$, for every collision in the simulated $H$. Given $(x_{0}, x_{1})$, the reduction $\AdvB$ will then compute and output $\left( \vecW_0 = \Gamma(x_0) , \vecW_1 = \Gamma(x_1) \right)$. Observe that if $x_0\neq x_1$, then $\vecW_0 \neq \vecW_1$ since $\Gamma$ is a permutation. Moreover, if $H(x_0)=H(x_1)$, then 
$$
Q\left( \vecW_{0} \right) = Q\left( \Gamma(x_0) \right) = H(x_0) = H(x_1) = Q\left( \Gamma(x_1) \right) = Q\left( \vecW_{1} \right)
\enspace .
$$
Hence, with probability at least $\epsilon$, $\AdvB$ will output a collision for $Q$, which will constitute a collision $\vecW_{\left( i, \; y_{i}, \; 0 \right)}, \vecW_{\left( i, \; y_{i}, \; 1 \right)} \in \{ 0, 1 \}^{\lambda + 1}$ in at least one $H'_{i}$. Finally, we chose at random $i^* \gets [n - r]$, and for every choice of $i^*$, the distribution of the generated $Q$ is identically distributed, so perfectly indistinguishable. Consequently, the probability that $i = i^*$ is at least $\frac{1}{n - r}$, so overall the probability that $\AdvB$ gets out of $\Adv$ a collision is $\frac{\epsilon}{n - r}$. Notice that for each query that $\Adv$ makes to $\Ps(\cdot)$, the reduction $\AdvB$ needs to make exactly one query to $Q$ and the same goes for the inverse $\Ps^{-1}(\cdot)$. Each query to $Q$ constitutes exactly one query to $H^*$. Thus $\AdvB$ makes exactly $q$ queries to $H^*$. This completes the proof.
\end{proof}

\paragraph{Collision-resistance of claw-free permutations with respect to an oracle.}
It remains to explain why a uniformly random claw-free permutation is collision resistant with respect to a quantumly-queriable classical oracle.

\begin{lemma} \label{lemma:claw_free_permutation_collision_resistant}
For $\lambda \in \Nat$, a random claw-free permutation $H: \zo^{\lambda + 1} \rightarrow \zo^{\lambda}$ is collision resistant given quantum queries to $H$. In particular, any (unbounded) quantum algorithm making $q$ queries has a $O\left( \frac{q^3}{2^\lambda} \right)$ probability of producing a collision.
\end{lemma}

\begin{proof}
The sampling procedure of $H$ is given by sampling two uniformly random permutation $\Pi_{0}, \Pi_{1} : \zo^{\lambda} \rightarrow \zo^{\lambda}$, and defining
$$
H\left( b \in \zo, x \in \zo^{\secp} \right) :=
\Pi_{b}\left( x \right) \enspace .
$$
We next recall that for every $q$-query quantum algorithm, quantum oracle access to $\Pi : \zo^{\lambda} \rightarrow \zo^{\lambda}$ a random permutation (but not its inverse) is $O\left( \frac{q^3}{2^\lambda} \right)$-indistinguishable from $\Pi$ being a random function \cite{QIC:Zhandry15}. This means that the overall $H$ is now a random function from $\zo^{\lambda + 1}$ to $\zo^{\lambda}$. We conclude with recalling that random functions are quantumly collision-resistant: The probability of finding a collision in such random $H$ is at most $O\left( \frac{q^3}{2^{\lambda}} \right)$ by \cite{QIC:Zhandry15}. Thus, we can bound the overall probability for finding a collision by $O\left( \frac{q^3}{2^\lambda} \right)$.
\end{proof}

%% file: standard_construction.tex
\section{Short One-Shot Signatures in the Standard Model} \label{sec:standard}
Following our construction in an oracle model from Section \ref{sec:oss_oracle}, we present our standard model constructions. As in the oracle model, we have our base Construction \ref{constr:standard} of a non-collapsing CRH. Our standard model one-shot signature scheme and in particular its signing algorithm are identical to the oracle model, that is, it uses the base construction in a black box way, in the same way that the oracle model OSS uses the oracle model base construction.

\begin{construction}\label{constr:standard}
Let $\secp \in \Nat$ the statistical security parameter. Define $s := 18 \cdot \secp$ and let $n, r, k \in \Nat$ such that $r := s \cdot \secp$, $n := r + \frac{3}{2} \cdot s$, $k := \frac{3}{2}\cdot s + \lambda$. Let $d := \poly_{d}(\secp) \in \Nat$ the expansion parameter and $\kappa := \poly_{\kappa}(\secp) \in \Nat$ the cryptographic security parameter, for some sufficiently large polynomials in the statistical security parameter.

Let $\iO$ an iO scheme, $\left( \prf, \punc \right)$ a puncturable PRF, and $\left( \prp, \prp^{-1}, \permute \right)$ a permutable PRP for the class of all $\left( 2^{\poly\left( \kappa \right)}, \poly\left( \kappa \right) \right)$-decomposable permutations. Then we construct a hash function $\left( \gen, \hash \right)$ as follows:

\begin{itemize}
    \item
    $\gen\left( 1^\lambda \right)$:
    Sample $k_{\sf in}, k_{\sf out}, k_{\sf lin} \gets\{0,1\}^{\kappa}$. $\prp\left( k_{\sf in}, \cdot \right)$ is a permutation with domain $\{0,1\}^n$, $\prp\left( k_{\sf out}, \cdot \right)$ is a permutation with domain $\{0,1\}^d$, and $\prf\left( k_{\sf lin}, \cdot \right)$ is a PRF with inputs in $\{ 0,1 \}^d$. Let $H(\cdot)$ denote the first $r$ output bits of $\prp\left( k_{\sf in}, \cdot \right)$ and $J(\cdot)$ denote the remaining $n-r$ bits. For each $y \in \{ 0, 1 \}^d$, using the output pseudorandomness of $\prf\left( k_{\sf lin}, y \right)$ we sample $\matA(y) \in \Z_2^{k \times (n-r)}$ and $\vecB(y) \in \Z_2^k$. The matrix $\matA(y) \in \Z_2^{k \times (n-r)}$ is with full column rank $n - r$, and also such that its bottom $\lambda$ rows have full row rank $\lambda$. $\vecB(y) \in \Z_2^k$ is a pseudorandom vector.

    As the common reference string output $\crs = \left( \mathcal{P}, \mathcal{P}^{-1}, \mathcal{D} \right)$ where $\Ps \gets \iO\left( 1^{\kappa}, P \right)$, $\Ps^{-1} \gets \iO\left( 1^{\kappa}, P^{-1} \right)$, $\mathcal{D} \gets \iO\left( 1^{\kappa}, D \right)$ such that
        \begin{align*}
        P(x) & =
        \big( \; y\; , \; \matA(y)\cdot J(x)+\vecB(y) \; \big)\text{ where }y \gets \prp^{-1}\left( k_{\sf out} , \: H(x) || 0^{d-r} \right)
        \\
        \\
        P^{-1}\left( y, \vecU \right) & =
        \begin{cases}
        \Pi^{-1}\left( w || \vecZ \right) & \exists\; w, \vecZ : \left( \prp\left( k_{\sf out}, y \right) = w || 0^{d-r} \right) \land \left( \matA(y) \cdot \vecZ + \vecB(y) = \vecU \right)
        \\
        \bot & \text{ else }
        \end{cases}
        \\
        \\
        \Ds\left( y, \vecV \right) &=
        \begin{cases}
        \vecC_{y, \vecV} \;\;
        &\text{ if } \vecV^{T} \cdot \matA(y) \in \rowspan\left( \matA(y)_{[(k - \lambda) + 1]}, \cdots,  \matA(y)_{[(k - \lambda) + \lambda]} \right)
        \\
        \bot
        &\text{ otherwise }
        \end{cases}
    \end{align*} 
    where for $\matA \in \bbZ_{2}^{k \times (n - r)}$ and $j \in [k]$, $\matA_{[j]} \in \bbZ_{2}^{n - r}$ is the $j$-th row of $\matA$ (e.g., $k$-th row is bottom), and $\vecC_{y, \vecV} \in \bbZ_{2}^{\lambda}$ is the coordinates vector, of the vector $\vecV^{T} \cdot \matA(y) \in \bbZ_{2}^{n - r}$ with respect to the basis $\matA(y)_{[(k - \lambda) + 1]}, \cdots,  \matA(y)_{[(k - \lambda) + \lambda]}$ (i.e., element $j \in [\lambda]$ of $\vecC_{y, \vecV}$ is the coefficient of $\matA(y)_{[(k - \lambda) + j]}$).
    
    \item
    $\hash\left( \crs, x \right)$: Compute $( y,\vecU )\gets\mathcal{P}(x)$ and output $y$.
\end{itemize}
\end{construction}

\paragraph{Security in the Standard Model.}
In this section we prove the main Theorems \ref{theorem:standard_model_1}, \ref{theorem:standard_model_2}, \ref{theorem:standard_model_3}. As in the oracle model, we need to prove the collision resistance of $H$ in order to prove strong unforgeability of our OSS, which we will focus on for the rest of this section.

\subsection{Bloating the Dual} \label{section:bloating_dual_standard}
For $n', r', k' \in \Nat$, $s \in \Nat \cup \{ 0 \}$ such that $s \leq n' - r' - \lambda$, we define the modified generator $\widetilde{\gen}\left( 1^{\lambda}, n', r', k', s \right)$, as follows. It samples a distribution over $\Ps,\Ps^{-1},\Ds'$, where the values of "$n, r, k$" in Construction \ref{constr:standard} are replaced by $n', r', k'$ respectively. For $s$ we let $\matA(y)^{(0)} \in \bbZ_{2}^{k' \times s}$ denote the first $s$ columns of $\matA(y) \in \bbZ_{2}^{k' \times (n' - r')}$ and $\matA(y)^{(1)} \in \bbZ_{2}^{k' \times (n' - r' - s)}$ denote the remaining $n'-r'-s$ columns. Note that the standard generator is defined as $\widetilde{\gen}\left( 1^{\lambda}, n, r, k, 0 \right)$.
In the the setting of $\widetilde{\gen}\left( 1^{\lambda}, n, r, k, s \right)$, the functionality of $\Ps, \Ps^{-1}$ stays the same, and for the dual oracle, we do the same things as in the original $\Ds$, but with respect to $\matA(y)^{(1)} \in \bbZ_{2}^{k \times (n - r - s)}$ rather than with respect to $\matA(y) \in \bbZ_{2}^{k \times (n - r)}$. Formally:
\[
\Ds'\left( y, \vecV \right) =
    \begin{cases}
    \vecC_{y, \vecV} \;\;
    &\text{ if } \vecV^{T} \cdot \matA(y)^{(1)} \in \rowspan\left( \matA(y)^{(1)}_{[(k - \lambda) + 1]}, \cdots,  \matA(y)^{(1)}_{[(k - \lambda) + \lambda]} \right)
    \\
    \bot
    &\text{ otherwise }
    \end{cases}
\]
where here, similarly to the construction (but not quite the same), $\vecC_{y, \vecV}$ is the coordinates vector, of the vector $\vecV^{T} \cdot \matA(y)^{(1)} \in \bbZ_{2}^{n - r - s}$ with respect to the basis $\matA(y)^{(1)}_{[(k - \lambda) + 1]}, \cdots,  \matA(y)^{(1)}_{[(k - \lambda) + \lambda]}$.

\begin{lemma} \label{lemma:bloating_dual_standard_model}
Let $\secp, n, r, k \in \Nat$, and assume there is a quantum algorithm $\Adv$ with complexity $T_{\Adv}$ such that,
\[
\Pr
\left[
\left( y_0 = y_1 \right) \land \left( x_0 \neq x_1 \right) \; :
\begin{array}{rl}
\left( \Ps, \Ps^{-1}, \Ds \right) & \gets \widetilde{\gen}\left( 1^{\lambda}, n, r, k, 0 \right) \\
\left( x_0, x_1 \right) & \gets \Adv\left( \Ps,\Ps^{-1},\Ds \right) \\
\left( y_b, \vecU_b \right) & \gets \Ps\left( x_b \right)
\end{array}
\right]
\geq
\epsilon \enspace .
\] 
For our primitives in Construction \ref{constr:standard}, let the $\iO$ scheme be $\left( f_{\iO}\left( \cdot \right), \frac{1}{f_{\iO}\left( \cdot \right)} \right)$-secure, the puncturable PRF be $\left( f_{\PRF}\left( \cdot \right), \frac{1}{f_{\PRF}\left( \cdot \right)} \right)$-secure, the permutable PRP be $\left( f_{\PRP}\left( \cdot \right), \frac{1}{f_{\PRP}\left( \cdot \right)} \right)$-secure and assume that Lossy Functions (as in Definition \ref{definition:lf}) are $\left( f_{\LF}\left( \cdot \right), \frac{1}{f_{\LF}\left( \cdot \right)} \right)$-secure. Also assume that OWFs are $\left( f_{\OWF}\left( \cdot \right), \frac{1}{f_{\OWF}\left( \cdot \right)} \right)$-secure for $f_{\OWF}\left( \secp \right) := 2^{ \secp^{\delta} }$ for some constant real number $\delta > 0$ and let $w := \secp^{\frac{\delta}{2}}$, $s' := s - (n - r - s)$. 

Assume $f_{\iO}\left( \kappa \right), f_{\PRF}\left( \kappa \right), f_{\PRP}\left( \kappa \right), f_{\LF}\left( w \right) \geq T_{\Adv} + \poly'\left( \secp \right)$ for some fixed polynomial $\poly'\left( \cdot \right)$ and all of the following:
\begin{enumerate}
    \item \label{standard_model_bloating_dual_condition_1}
    For $f := \min\left( f_{\iO}, f_{\PRF}, f_{\PRP} \right)$, $\frac{ 2^r }{ f\left( \kappa \right) } \leq \frac{\epsilon}{k^2 \cdot 512}$,

    \item \label{standard_model_bloating_dual_condition_2}
    $\frac{ 1 }{ f_{\LF}\left( w \right) }\leq \frac{\epsilon}{k^2 \cdot 256}$ ,

    \item \label{standard_model_bloating_dual_condition_3}
    $\frac{ 2^{2\cdot w} \cdot k^4 \cdot \left( k^3 + T_{\Adv} \right) }{f_{\OWF}\left( n - r - \lambda - s \right) - \poly\left( n - r - \lambda - s \right)} \leq \epsilon^2$, and

    \item \label{standard_model_bloating_dual_condition_4}
    $\frac{ 4 \cdot (n - r - s) \cdot 2^w }{2^{s'}} \leq \epsilon$.
\end{enumerate}
Then, it follows that,
\[
\Pr
\left[
\begin{array}{rl}
     & y_0 = y_1 := y , \\
     &\left( \vecU_0 - \vecU_1 \right) \notin \colspan \left( \matA(y)^{(1)} \right)
\end{array}
\;
:
\begin{array}
{rl}
\left( \Ps, \Ps^{-1}, \Ds' \right) & \gets \widetilde{\gen}(1^{\lambda}, n, r, k, s) \\
\left( x_0, x_1 \right) & \gets \Adv\left( \Ps, \Ps^{-1}, \Ds' \right) \\
(y_b, \vecU_b) & \gets \Ps(x_b)
\end{array}
\right]
\geq
\frac{ \epsilon }{ 512 \cdot k^{2} }
\enspace .
\]
\end{lemma}

Note that $\vecU_{0} - \vecU_{1} \notin \colspan\left( \matA(y)^{(1)} \right)$ means in particular that $\vecU_{0}, \vecU_{1}$, and hence $x_0, x_1$, are distinct. Thus, the second expression means that $\Adv$ is finding collisions, but these collisions satisfy an even stronger requirement.

\begin{proof}
Let $\secp \in \Nat$ and assume there is a $T_{\Adv}$-complexity algorithm $\Adv$ and a probability $\epsilon$ such that $\Adv$ gets $\left( \Ps,\Ps^{-1},\Ds \right) \gets \gen\left( 1^{\lambda} \right)$ and outputs a pair $\left( x_{0}, x_{1} \right)$ of $n$-bit strings such that with probability $\epsilon$ we have $x_{0} \neq x_{1}$ and $y_{0} = y_{1}$.
We next define a sequence of hybrid experiments. Each hybrid defines a computational process, an output of the process and a predicate computed on the process output. The predicate defines whether the (hybrid) process execution was successful or not.

\begin{itemize}
    \item $\Hyb_{0}$: The original execution of $\Adv$.
\end{itemize}
The process $\Hyb_{0}$ is the above execution of $\Adv$ on input a sample from the distribution $\left( \Ps, \Ps^{-1}, \Ds \right) \gets \widetilde{\gen}\left( 1^{\lambda}, n, r, k, 0 \right)$. We define the output of the process as $\left( x_0, x_1 \right)$ and the process execution is considered as successful if $x_{0}$, $x_{1}$ are both distinct and collide in their $y$ values. By definition, the success probability of $\Hyb_{0}$ is $\epsilon$.

\begin{itemize}
    \item $\Hyb_{1}$: Preparing to switch to a bounded number of cosets $\left( \matA(y), \vecB(y) \right)$, by using an obfuscated puncturable PRF and injective mode of a lossy function.
\end{itemize}
Let $\left( \LFGen, \LFF \right)$ a $\left( f(\cdot), \frac{1}{f(\cdot)} \right)$-secure lossy function scheme (as in Definition \ref{definition:lf}). Sample $\pk_{\LF} \gets \LFGen\left( 1^{d}, 0, 1^{w} \right)$ where $w$ is as defined in our Lemma statement, and let $\LFF\left( \pk_{\LF}, \cdot \right) : \{ 0, 1 \}^{d} \rightarrow \{ 0, 1 \}^{m}$ the induced injective function.

We now consider two circuits in order to describe our current hybrid. $E_{0}\left( k_{\sf lin}, \cdot \right)$ is the circuit that given an input from $\{ 0, 1 \}^{d}$ applies $\prf\left( k_{\sf lin }, \cdot \right)$ to get $\matA(y)$, $\vecB(y)$. $E_{1}\left( \pk_{\LF}, k'_{ \sf lin }, \cdot \right)$ is the circuit that for a $\LF$ key $\pk_{\LF}$ and P-PRF key $k'_{ \sf lin }$ for a P-PRF with input size $m$ rather than $d$, given a $d$-bit input $y$, applies the lossy function with key $\pk_{\LF}$ and then the P-PRF to get $\matA'(y)$, $\vecB'(y)$.

Note that in the previous hybrid, the circuit $E_{0}$ is used in all three circuits $P$, $P^{-1}$, $D$ in order to generate the cosets per input $y$, and furthermore, each of these three circuits access $E_{0}$ only as a black box. The change that we make to the current hybrid is that we are going to use $E_{1}\left( \pk_{\LF}, k'_{ \sf lin }, \cdot \right)$ for a freshly sampled $\pk_{\LF}$, $k'_{ \sf lin }$, instead of $E_{0}$. Observe that in both cases, whether we are using $E_{0}$ or $E_{1}$, for each value in $\{ 0, 1 \}^{r}$ if we use real randomness to sample the coset, the distributions of the cosets are statistically equivalent between the two cases. Since we are sending obfuscations $\left( \Ps, \Ps^{-1}, \Ds \right)$ of the three circuits and due to the three circuits accessing the samplers $E_{0}$, $E_{1}$ only as black boxes, it follows by Lemma \ref{lem:distswap} that the output of the previous hybrid and the current hybrid are $\left( f(\kappa), \frac{|\Xs|}{f\left( \kappa \right)} \right)$-indistinguishable, where $f := \min \left( f_{\iO}, f_{\PRF} \right)$ and $\Xs$ is the set of all possible values of $y$, which has size $2^r$ (recall that while the $y's$ are of length $d >> r$, these are a sparse set inside $\{ 0, 1 \}^{d}$ because we are padding with zeros), and we also recall that $\kappa$, the cryptographic security parameter is some polynomial in the statistical security parameter $\secp$. It follows by Condition \ref{standard_model_bloating_dual_condition_1} from our Lemma statement, that the success probability of the current process is $:= \epsilon_{1} \geq \epsilon - \frac{2^r}{f\left( \kappa \right)} \geq \epsilon - \frac{\epsilon}{32} = \frac{31 \cdot \epsilon}{32}$.

\begin{itemize}
    \item $\Hyb_{2}$: Switching to a bounded number of cosets $\left( \matA(y), \vecB(y) \right)$, by using the lossy function.
\end{itemize}
The change from the previous hybrid to the current hybrid is that we are going to sample a lossy key $\pk_{\LF}^{1} \gets \LFGen\left( 1^{d}, 1, 1^{w} \right)$ and use it inside $E_{1}$ from the previous hybrid, instead of using an injective key $\pk_{\LF}^{0} \gets \LFGen\left( 1^{d}, 0, 1^{w} \right)$, which was used in the previous hybrid. Note that in this hybrid, there are at most $2^w$ cosets (that is, some different values of $y$ will have the same coset), by the correctness of the lossy function scheme. By the security of the lossy function scheme, the output of this hybrid is $\left( f_{\LF}\left( w \right), \frac{1}{ f_{\LF}\left( w \right) } \right)$-indistinguishable from the previous hybrid. It follows by Condition \ref{standard_model_bloating_dual_condition_2} from our Lemma statement that the success probability of the current process is
$$
:= \epsilon_{2}
\geq \epsilon_{1} - \frac{1}{ f_{\LF}\left( w \right) }
\geq \frac{31\cdot \epsilon}{32} - \frac{\epsilon}{32}
\geq \frac{30\cdot \epsilon}{32} \enspace .
$$

\begin{itemize}
    \item $\Hyb_{3}$: Relaxing dual verification oracle to accept a larger subspace, by using an obfuscated puncturable PRF and subspace hiding functions.
\end{itemize}
The change we make in this hybrid is that we make the membership check dual oracles more relaxed. Formally, we invoke Lemma \ref{lemma:subspace_hiding_function} with the following interface, for every value $m_{y} := \LFF\left( \pk_{\LF}, y \right)$. The smallest oracle "$S_{0}$" from the Lemma \ref{lemma:subspace_hiding_function} statement is simply $S_{y, 0}^{\bot} := \colspan\left( \matA(y) \right)^{\bot}$ here and "$S$" from the lemma is $S_{y}^{\bot}$, defined as the total set of vectors that $\Ds$ does not output $\bot$ on. The partition "$P$" from the lemma statement is the partition defined by the output of the oracle $\Ds$ (observe that this is indeed a partition into cosets of $S$ that are all parallel to $S_{0}$). That is, the oracle $\Ds$ computes the function $F_{S}$ for these subspaces.

Now, for every $y$, after sampling $\left( \matA(y), \vecB(y) \right)$, we sample superspace $T_{y, 0}^{\bot}$ and by the statement of Lemma \ref{lemma:subspace_hiding_function}, (1) $T_{y, 0}^{\bot}$ is a random superspace of $S_{y, 0}^{\bot}$ (which itself has dimension $k - (n - r)$) with $k - (n - r - s)$ dimensions, and (2) $T^{\bot}_{y}$ is the joint span of $T_{y, 0}^{\bot}$ and $S_{y}^{\bot}$. For concreteness and because it will later be useful to our proof, assume that our process of sampling $T_{y, 0}^{\bot}$ is by sampling its dual $T_{y, 0}$, which in turn is sampled by sampling a uniformly random invertible $\matM(y) \in \bbZ_{2}^{(n - r) \times (n - r)}$, multiplying by $\matA(y)$ and then taking the rightmost $n - r - s$ columns of the generated matrix. We denote by $C_{y}$ the circuit which gives access to the function $F_{S}$ and by $C'_{y}$ the circuit that gives access to the $F_{T}$. It follows there is a process that takes a value $y$, computes $m_{y} := \LFF\left( \pk_{\LF}, y \right)$ and than there is a process (which is accessed in a black-box fashion) that computes $C_{y} := C_{m_{y}}$ as a function of $m_{y}$ alone.

To argue indistinguishability formally, we now consider two circuits in order to describe our current hybrid. The first circuit $E_{S}\left( k_{S}, \cdot \right)$ is the circuit that given an input $m_y$ computes $C_{y}$ and returns an obfuscation of it. The second circuit $E_{T}\left( k_{T}, \cdot \right)$ is the circuit that given an input $m_{y}$ samples $C'_{y}$ and returns an obfuscation of it. In the previous hybrid, in order to compute the output of $\Ds$ we used $E_{S}$ and in the current hybrid we switch to using $E_{T}$.

Let $\Xs$ the set of all possible values $m_{y}$ that arise from the scheme, which has size $\leq 2^w$ by the lossy function. Per such value, by Lemma \ref{lemma:subspace_hiding_function} we have $\left( f_{\OWF}\left( n - r - \lambda - s \right), \frac{s}{f_{\OWF}\left( n - r - \lambda - s \right)} \right)$-indistinguishability between using $E_{S}$ or $E_{T}$ on input $y$, and this happens when the circuits (and the random algebraic objects inside them) are sampled with truly random bits. It follows by Lemma \ref{lem:distswap} that the output of the previous hybrid and the current hybrid are $\left( f_{\OWF}\left( n - r - \lambda - s \right), \frac{|\Xs|}{f_{\OWF}\left( n - r - \lambda - s \right)} \right)$-indistinguishable, where $\Xs$ is the set of all cosets derived from our scheme, which has size $\leq 2^w$ by the lossy function. It follows by Condition \ref{standard_model_bloating_dual_condition_3} that the success probability of the current process is
$$
:= \epsilon_{3}
\geq \epsilon_{2} - \frac{2^w}{f_{\OWF}\left( n - r - \lambda - s \right)}
\geq \frac{30 \cdot \epsilon}{32} - \frac{\epsilon}{32}
= \frac{29\cdot \epsilon}{32}
\enspace .
$$

\begin{itemize}
    \item $\Hyb_{4}$: Asking for sum of collisions to be outside of $T_{y, 0}$, using the hardness of concentration in dual of obfuscated subspace, the pseudorandomness of the puncturable PRF and the security of iO.
\end{itemize}
This hybrid is the same as the previous in terms of execution, but we change the definition of a successful execution, that is, we change the predicate computed on the output of the process. We still ask that $\left( y_0 = y_1 := y \right)$, but instead of only asking the second requirement to be $\left( x_0 \neq x_1 \right)$, we ask for a stronger condition: $\left( \vecU_{0} - \vecU_{1} \right) \in \left( S_{y, 0} \setminus T_{y, 0} \right)$. Note that we are not going to need to be able to efficiently check for the success of the condition, but we'll prove that it happens with a good probability nonetheless.

Let $\epsilon_{4}$ be the success probability of the current hybrid and note that $\Adv$ finds collisions with probability $\epsilon_{3}$ in the previous hybrid (and since this hybrid is no different and only change the definition of a successful execution, the same goes for the current hybrid). Let $\Xs \subseteq \{ 0, 1 \}^{m}$ the image of the lossy function $\LFF\left( \pk_{\LF}, \cdot \right)$ which we use to map our images $y$ to cosets $\left( \matA(y), \vecB(y) \right)$, that is, there are at most $|\Xs|$ cosets and by the lossyness we know that $|\Xs| \leq 2^w$. For every value $\vecX \in \Xs$ denote by $\epsilon_{3}^{(\vecX)}$ the probability to find a collision on value $\vecX$, or formally, that $y_{0} = y_{1} := y$, $x_{0} \neq x_{1}$ and $\vecX = \LFF\left( \pk_{\LF}, y \right)$. We deduce $\sum_{ \vecX \in \Xs } \epsilon_{3}^{(\vecX)} = \epsilon_{3}$. Let $L$ be a subset of $\Xs$ such that $\epsilon_{3}^{(\vecX)} \geq \frac{\epsilon_{3}}{2 \cdot |\Xs|}$ and note that $\sum_{\vecX \in L} \epsilon_{3}^{(\vecX)} \geq \frac{ \epsilon_{3} }{ 2 }$. Similarly, we define $\epsilon_{4}^{(\vecX)}$ as the probability to find a "strong" collision on value $\vecX$, or formally, that $y_{0} = y_{1} := y$, $\left( \vecU_{0} - \vecU_{1} \right) \in \left( S_{y, 0} \setminus T_{y, 0} \right)$ and $\vecX = \LFF\left( \pk_{\LF}, y \right)$. Note that $S_{y, 0}$, $T_{y, 0}$ are really functions of $\vecX$ rather than of $y$, so we can denote $\left( \vecU_{0} - \vecU_{1} \right) \in \left( S_{\vecX, 0} \setminus T_{\vecX, 0} \right)$ and also observe that $\sum_{\vecX \in \Xs} \epsilon_{4}^{(\vecX)} = \epsilon_{4}$.

We would now like to use Lemma \ref{lemma:dual_subspace_concentration}, so we make sure that we satisfy its requirements. Let any $\vecX \in L$, we know that by definition, $\epsilon_{3}^{(\vecX)} \geq \frac{\epsilon_{3}}{2 \cdot |\Xs|}$ and also recall that $\epsilon_{3} \geq \frac{28\cdot \epsilon}{32}$, $|\Xs| \leq 2^w$ and thus 
$$
\epsilon_{3}^{(\vecX)}
\geq
\frac{\epsilon_{3}}{2 \cdot |\Xs|}
\geq
\frac{29\cdot \epsilon}{64} \cdot \frac{1}{2^w}
>
\frac{ \epsilon }{ 2 \cdot 2^w }
\enspace .
$$
Let $s' := s - (n - r - s)$ and for any $\vecX \in L$ let $\ell_{\vecX} := \frac{k^2}{\epsilon_{3}^{(\vecX)}} \leq \frac{ k^2 \cdot 2^w \cdot 2 }{ \epsilon }$. Note that by our Lemma \ref{lemma:bloating_dual_standard_model} statement's Condition \ref{standard_model_bloating_dual_condition_4} we have (1) $\frac{(n - r - s)}{2^{s'}} \leq \frac{\epsilon_{3}^{(\vecX)}}{2}$, and by Condition \ref{standard_model_bloating_dual_condition_3} we have (2) $\ell_{\vecX} \cdot \left( k^3 + T_{\Adv} \right) \leq \frac{ f_{\OWF}\left( n - r - \lambda - s \right) - \poly\left( n - r - \lambda - s \right) }{ 2 \cdot \lambda \cdot \ell_{\vecX} }$. Since this satisfies Lemma \ref{lemma:dual_subspace_concentration}, it follows that for every $\vecX \in L$ we have $\epsilon_{4}^{(\vecX)} \geq \frac{ \epsilon_{3}^{(\vecX)} }{ 16 \cdot k^{2} } - \frac{1}{f(\kappa)}$ for $f := \min\left( f_{\iO}, f_{\PRF} \right)$. The expression $\frac{1}{f(\kappa)}$ is subtracted because for each $\vecX \in \Xs$, in order to use Lemma \ref{lemma:dual_subspace_concentration}, we need the randomness for the experiment to be genuinely random, which will necessitate us to invoke the security of the iO and puncturable PRF, which incurs the loss of $\frac{1}{f(\kappa)}$. It follows that
$$
\epsilon_{4}
=
\sum_{\vecX \in \Xs} \epsilon_{4}^{(\vecX)}
\geq 
\sum_{\vecX \in L} \epsilon_{4}^{(\vecX)}
\geq
\sum_{\vecX \in L} \frac{ \epsilon_{3}^{(\vecX)} }{ 16 \cdot k^{2} }
-
\frac{|\Xs|}{f(\kappa)}
\geq
\frac{ \left( \frac{ \epsilon_{3} }{ 2 } \right) }{ 16 \cdot k^{2} }
-
\frac{|\Xs|}{f(\kappa)}
\geq
\frac{ \left( \frac{ 29 \cdot \epsilon }{ 64 } \right) }{ 16 \cdot k^{2} }
-
\frac{2^w}{f(\kappa)}
\geq
\frac{ \epsilon }{ 64 \cdot k^{2} }
\enspace .
$$

\begin{itemize}
    \item 
    $\Hyb_{5}$: For every $y$, de-randomizing $T_{y, 0}$ and defining it as the column span of $\matA(y)^{(1)} \in \bbZ_{2}^{k \times \left( n - r - s \right)}$, the last $n - r - s$ columns of the matrix $\matA(y)$, by using permutable PRPs, obfuscated puncturable PRF and the security of iO.
\end{itemize}
Recall that at the basis of our scheme there are the $\leq 2^w$ cosets $\left( \matA(y), \vecB(y) \right)_{ y }$ and for each coset we also sample a pseudorandom subspace $T_{y, 0}$. Recall that we sample a pseudorandom matrix $\matM(y) \in \bbZ_{2}^{(n - r) \times (n - r)}$ which is explicitly used in the sampling of the subspace $T_{y, 0}$, that is, we define $T_{y, 0}$ as the column span of the rightmost $n - r - s$ columns of $\matA(y) \cdot \matM(y)$. The problem is that we use $\matA(y), \vecB(y)$ as is in the oracles $\Ps$, $\Ps^{-1}$, but use $\matA(y) \cdot \matM(y)$ in $\Ds$. We will resolve this discrepancy in this hybrid.
This hybrid is the same as the previous, with one change: For every $y$, after sampling the coset $\left( \matA(y), \vecB(y) \right)$, we will not continue to randomly sample $T^{\bot}_{y, 0}$, and simply define $T_{y, 0} := \colspan\left( \matA(y)^{(1)} \right)$ such that $\matA(y)^{(1)} \in \bbZ_{2}^{k \times (n - r - s)}$ is defined to be the last $n - r - s$ columns of the matrix $\matA(y) \in \bbZ_{2}^{k \times (n - r)}$. We will define intermediate hybrids and then explain why the previous hybrid is indistinguishable from the current.

\begin{itemize}
    \item
    \textbf{Using the security of the permutable PRP.}
    Assume we sample all components of our scheme, excluding the key $k_{\sf in}$ for the initial permutation $\Pi$ on $\{ 0, 1 \}^{n}$. Define the following permutation $\Gamma$ on $\{ 0, 1 \}^{n}$: Denote by $h$ the first $r$ bits of the input and by $j$ the last $n - r$ bits of the input to the permutation $\Gamma$. Recall that in the circuits $P, P^{-1}, D$, the value $y$, the coset $\matA(y), \vecB(y)$, the superspace $T_{y, 0}^{\bot}$ and the associated matrix $\matM(y) \in \bbZ_{2}^{ (n - r) \times (n - r) }$, are all computed as a function of $r$ bits, which in the construction take the role of $H(x)$. In fact, all of the above variables can be written as a function of $h \in \{ 0, 1 \}^{r}$ instead of as a function of $y$.
    
    The permutation $\Gamma$ takes $h \in \zo^{r}$ and computes $\matM(y)$, then interprets $j \in \zo^{n - r}$ as a vector in $\bbZ_{2}^{n - r}$, and applies $\matM(y)$ to $j$. For every value $h \in \{ 0, 1 \}^{r}$ observe that multiplication by $\matM(y)$ is a permutation (and moreover an affine permutation, which is decomposable efficiently). It follows that the overall $\Gamma$ is a controlled version of an affine permutation, and thus is controlled on decomposable permutations. Overall, we deduce that $\Gamma$ is a $\left( 2^{\poly(\secp)}, \poly(\secp) \right)$-decomposable permutation. We use the permutable PRP $\Pi$, and switch to a setting where we use the key $k_{\sf in}^{\Gamma}$ that applies $\Gamma$ to the output of $\Pi$ (and $\Gamma^{-1}$ to the input of $\Pi^{-1}$), instead of just applying $\Pi$ and its inverse. By the security of the permutable PRPs, this change is $\left( f_{\PRP}(\kappa), \frac{1}{f_{\PRP}(\kappa)} \right)$-indistinguishable. 

    \item
    \textbf{Using an obfuscated puncturable PRF and statistical equivalence of sampling the random matrix $\matA(y)$, for every $y$.}
    In this intermediate hybrid, instead of sampling $\matA(y)$ and then using it as is, we will sample it multiplied by the inverse of $\matM(y)$, that is, our sampler gives us $\matA(y) \cdot \matM^{-1}(y)$ instead of $\matA(y)$. Note that for every $y$, if we use a truly random $\matA(y)$, the two processes are statistically equivalent because $\matM$ is invertible. By a standard argument using an obfuscated puncturable PRF (i.e., we use lemma \ref{lem:distswap} and the fact that when using real randomness, the two ways to sample $\matA(y)$ are statistically equivalent), we get that this change is $\left( f(\kappa), 2^w \cdot \frac{1}{f(\kappa)} \right)$-indistinguishable, for $f := \min\left( f_{\iO}, f_{\PRF} \right)$.

    \item
    \textbf{Using the security of outside iO.}
    We now can more easily see why the current hybrid is indistinguishable from the previous. Looking into the functionality, we see that the matrix multiplications which include $\matM(y)$ all cancel out and can be ignored. Specifically, in the computation of $\Ps, \Ps^{-1}$, the concatenated permutation $\Gamma$ effectively induce a computation by $\matM(y)$, which then cancels out with the fact that $\matA(y)$ is sampled with $\matA(y) \cdot \matM^{-1}(y)$. Inside the computation of $\Ds$, we sample $\matA(y) \cdot \matM^{-1}(y)$ from the PRF as in $\Ps, \Ps^{-1}$, and then when we make the multiplication by $\matM(y)$ (which, as we recall, is used to derive the matrix $\matT$ which in turn imply the identity of the subspace $T_{y, 0}$), which cancels out again. Overall, the computation is logically equivalent to just sampling $\matA(y)$ from the puncturable PRF, and using that in all oracles, in particular the $n - r - s$ rightmost columns of $\matA(y)$ now yield $T_{y, 0}$. Due to functional equivalence we can invoke the security of the outside iO and deduce that this change is $\left( f_{\iO}(\kappa), \frac{1}{f_{\iO}(\kappa)} \right)$-indistinguishable.
\end{itemize}
Observe that after the last change we are exactly in the setting of $\Hyb_{5}$. It follows in particular that the success probability of the current process is
$$
:= \epsilon_{5}
\geq \epsilon_{4} - \frac{1}{f_{\PRP}(\kappa)} - 2^w \cdot \frac{1}{f(\kappa)} - \frac{1}{f_{\iO}(\kappa)}
\geq
\frac{ \epsilon }{ 64 \cdot k^{2} } - \frac{ \epsilon }{ 512 \cdot k^{2} }
=
\frac{ 7 \cdot \epsilon }{ 512 \cdot k^{2} }
\enspace .
$$

\begin{itemize}
    \item $\Hyb_{6}$: Going back to using $2^r$ cosets rather than $\leq 2^w$, by moving from lossy mode to injective mode in the lossy function.
\end{itemize}
We make the exact same change we made between $\Hyb_{1}$ to $\Hyb_{2}$, but in the opposite direction. That is we sample an injective key $\pk_{\LF}^{0} \gets \LFGen\left( 1^{d}, 0, 1^{w} \right)$ and use it instead of the previous lossy key $\pk_{\LF}^{1} \gets \LFGen\left( 1^{d}, 1, 1^{w} \right)$, which is used in the previous hybrid. By the exact same argument (which relies on the security of the lossy function), the output of this hybrid is $\left( f_{\LF}(w), \frac{1}{f_{\LF}(w)} \right)$-indistinguishable. It follows in particular the success probability of the current process is
$$
:= \epsilon_{6}
\geq \epsilon_{5} - \frac{1}{f_{\LF}\left( w \right)}
\geq \frac{ 7 \cdot \epsilon }{ 512 \cdot k^{2} } - \frac{ \epsilon }{ 256 \cdot k^{2} }
= \frac{ 5 \cdot \epsilon }{ 512 \cdot k^{2} }
\enspace .
$$

\begin{itemize}
    \item $\Hyb_{7}$: Stop using the lossy function, by an obfuscated puncturable PRF.
\end{itemize}
We make the exact same change we made between $\Hyb_{0}$ to $\Hyb_{1}$, but in the opposite direction. That is, we drop the lossy function $\LFF$ altogether and apply the PRF to $y$ directly and not to $\vecX$, the output of the lossy function on input $y$. By the exact same argument (which relies on Lemma \ref{lem:distswap}), the output of this hybrid is $\left( f(\kappa), \frac{|\Xs|}{f\left( \kappa \right)} \right)$-indistinguishable for $f := \min\left( f_{\iO}, f_{\PRF} \right)$, where $\Xs$ is the set of all possible values of $y$, which has size $2^r$ (recall that while the $y's$ are of length $d >> r$, these are a sparse set inside $\{ 0, 1 \}^{d}$ because we are padding with zeros). It follows that the success probability of the current process is
$$
:= \epsilon_{7}
\geq \epsilon_{6} - \frac{2^r}{f\left( \kappa \right)}
\geq \frac{ 5 \cdot \epsilon }{ 512 \cdot k^{2} } - \frac{ \epsilon }{ 512 \cdot k^{2} }
= \frac{ \epsilon }{ 128 \cdot k^{2} }
\enspace .
$$

To conclude, note that the generated obfuscations in the final hybrid form exactly the distribution $\left( \Ps, \Ps^{-1}, \Ds' \right)\gets \widetilde{\gen}\left( 1^{\secp}, n, r, k, s \right)$. This finishes our proof.
\end{proof}

\subsection{Simulating the Dual} \label{section:simulating_dual_standard}
Our next step is to show that an adversary which has access to the dual-free setting can simulate the CRS for an adversary in the restricted setting, where the dual verification check is bloated.

\begin{lemma} \label{lemma:simulating_dual_standard_model}
Let $\secp, n, r, k \in \Nat$ and assume there is a quantum algorithm $\Adv$ running in time $T_{\Adv}$ such that,
\[
\Pr
\left[
\begin{array}{rl}
     & y_0 = y_1 := y , \\
     &\left( \vecU_0 - \vecU_1 \right) \notin \colspan \left( \matA(y)^{(1)} \right)
\end{array}
\;
:
\begin{array}
{rl}
\left( \Ps, \Ps^{-1}, \Ds' \right) & \gets \widetilde{\gen}(1^{\lambda}, n, r, k, s) \\
\left( x_0, x_1 \right) & \gets \Adv\left( \Ps, \Ps^{-1}, \Ds' \right) \\
(y_b, \vecU_b) & \gets \Ps(x_b)
\end{array}
\right]
\geq
\epsilon \enspace .
\]
For our primitives in Construction \ref{constr:standard}, let the $\iO$ scheme be $\left( f_{\iO}\left( \cdot \right), \frac{1}{f_{\iO}\left( \cdot \right)} \right)$-secure, the puncturable PRF be $\left( f_{\PRF}\left( \cdot \right), \frac{1}{f_{\PRF}\left( \cdot \right)} \right)$-secure, the permutable PRP be $\left( f_{\PRP}\left( \cdot \right), \frac{1}{f_{\PRP}\left( \cdot \right)} \right)$-secure and let $f := \min\left( f_{\iO}, f_{\PRF}, f_{\PRP} \right)$. Assume $\frac{ 2^r }{f\left( \kappa \right)} \leq o\left( \epsilon \right)$. Then, there is a quantum algorithm $\AdvB$ running in time $T_{\Adv} + \poly\left( \secp \right)$ such that
\[
\Pr
\left[
\left( \overline{y}_{0} = \overline{y}_{1} \right) \land \left( \overline{x}_{0} \neq \overline{x}_{1} \right) \; :
\begin{array}{rl}
\left( \overline{\Ps}, \overline{\Ps}^{-1}, \overline{\Ds} \right) & \gets \widetilde{\gen}\left( 1^{\lambda}, r + s, r, k - (n - r - s), 0 \right) \\
\left( \overline{x}_{0}, \overline{x}_{1} \right) & \gets \AdvB\left( \overline{\Ps}, \overline{\Ps}^{-1} \right) \\
\left( \overline{y}_{b}, \overline{\vecU}_b \right) & \gets \overline{\Ps}(x_b)
\end{array}
\right]
\geq
\frac{\epsilon}{2}
\enspace .
\] 
\end{lemma}

\begin{proof}
We first describe the actions of the algorithm $\AdvB$ (which will use the code of $\Adv$ as part of its machinery) and then argue why it breaks collision resistance with the appropriate probability. Given $\overline{\Ps}, \overline{\Ps}^{-1}$ which comes from $\left( \overline{\Ps}, \overline{\Ps}^{-1}, \overline{\Ds} \right) \gets \widetilde{\gen}\left( 1^{\lambda}, r + s, r, k - (n - r - s), 0 \right)$, the algorithm $\AdvB$ does the following:
\begin{itemize}
    \item
    Sample a P-PRF key $k_{\matC}$ that outputs some sufficient (polynomial) amount of random bits on an $d$-bit input, and sample a permutable PRP key $k_{\Gamma}$ for a PRP on domain $\{ 0, 1 \}^{n}$. Define the following circuits.

    \item 
    $\left( \; y \in \bbZ_{2}^{d}, \; \vecU \in \bbZ_{2}^{k} \; \right) \gets P\left( x \in \bbZ_{2}^{n} \right)$:
    \begin{itemize}
        \item
        $\left(
        \; \overline{x} \in \bbZ_{2}^{r + s},
        \; \widetilde{x} \in \bbZ_{2}^{n - r - s} \;
        \right) \gets \Pi\left( k_{\Gamma}, x \right)$.

        \item 
        $\left(
        \; y \in \bbZ_{2}^{d},
        \; \overline{\vecU} \in \bbZ_{2}^{k - (n - r - s)} \;
        \right) \gets \overline{\Ps}(\overline{x})$.

        \item 
        $\left(
        \; \matC(y) \in \bbZ_{2}^{ k \times k },
        \; \vecD(y) \in \bbZ_{2}^{ n - r - s }
        \right)
        \gets
        \prf\left( k_{\matC}, y \right)$.

        \item 
        $\vecU
        \gets
        \matC(y) \cdot \left( \begin{array}{c} \overline{\vecU} \\ \widetilde{x} + \vecD(y) \end{array} \right)$.
    \end{itemize}

    \item 
    $\left( \; x \in \bbZ_{2}^{n} \; \right)
    \gets
    P^{-1}\left( \; y \in \bbZ_{2}^{d}, \; \vecU \in \bbZ_{2}^{k} \; \right)$:
    \begin{itemize}
        \item 
        $\left(
        \; \matC(y) \in \bbZ_{2}^{ k \times k },
        \; \vecD(y) \in \bbZ_{2}^{n - r - s}
        \right) \gets \prf\left( k_{\matC}, y \right)$.

        \item 
        $\left( \begin{array}{c} \overline{\vecU} \\ \widetilde{x} \end{array} \right) \gets \matC(y)^{-1} \cdot \vecU - \left( \begin{array}{c} 0^{k-(n - r - s)} \\ \vecD(y) \end{array} \right)$.

        \item 
        $\left( \; \overline{x} \in \bbZ_{2}^{r + s} \; \right)
        \gets
        \overline{\Ps}^{-1}\left( y, \overline{\vecU} \right)$.

        \item
        $x
        \gets
        \Pi^{-1}\left( k_{\Gamma}, \left( \overline{x}, \widetilde{x} \right) \right)$.
    \end{itemize}

    \item 
    $
    \vecC_{y, \vecV}
    \gets
    D'\left( \; y \in \bbZ_{2}^{d}, \; \vecV \in \bbZ_{2}^{k} \; \right)$:
    \begin{itemize}
        \item 
        $\left(
        \; \matC(y) \in \bbZ_{2}^{ k \times k },
        \; \vecD(y) \in \bbZ_{2}^{n - r - s}
        \right)
        \gets
        \prf\left( k_{\matC}, y \right)$.

        \item 
        $\matA(y)^{(1)} := $ last $n - r - s$ columns of $\matC(y)$. 

        \item 
        Simulate the answer of $D'$ using $\matA(y)^{(1)}$, which is sufficient.
    \end{itemize}

    \item 
    Use indistinguishability obfuscation in order to generate the input for $\Adv$: $\Ps \gets \iO\left( 1^{\kappa}, P \right)$, $\Ps^{-1} \gets \iO\left( 1^{\kappa}, P^{-1} \right)$, $\Ds' \gets \iO\left( 1^{\kappa}, D' \right)$.
\end{itemize}

The remainder of the reduction is simple: $\AdvB$ executes $\left( x_0, x_1 \right) \gets \Adv\left( \Ps, \Ps^{-1}, \Ds' \right)$ and then $\left( \overline{x}_{b}, \widetilde{x}_{b} \right) \gets \Pi\left( k_{\Gamma}, x_{b} \right)$ and outputs $\left( \overline{x}_{0}, \overline{x}_{1} \right)$. Assume that the output of $\Adv$ satisfies $y_{0} = y_{1} := y$ and also $\left( \vecU_0 - \vecU_1 \right) \notin \colspan \left( \matA(y)^{(1)} \right)$, and recall that $\matA(y)^{(1)} \in \bbZ_{2}^{k \times (n - r - s)}$ are the last $n - r - s$ columns of the matrix $\matA(y) \in \bbZ_{2}^{k \times (n - r)}$, which is generated by the reduction. We explain why it is necessarily the case that $\overline{x}_{0} \neq \overline{x}_{1}$.

\paragraph{Why if $\Adv$ finds a collision then $\AdvB$ finds a collision.}
First note that due to how we defined the reduction, $\matA(y) := \matC(y) \cdot \left( \begin{array}{cc} \overline{\matA}(y) & \\ & \Id_{n - r - s} \end{array} \right)$, where $\overline{\matA}(y) \in \bbZ_{2}^{(k - (n - r - s)) \times s}$ is the matrix arising from $\overline{\Ps}, \overline{\Ps}^{-1}$ and $\Id_{n - r - s} \in \bbZ_{2}^{(n - r - s) \times (n - r - s)}$ is the identity matrix of dimension $n - r - s$. Also note that because $\matC(y)$, $\overline{\matA}(y)$ are full rank then $\matA(y)$ is full rank. Now, since $\left( \vecU_0 - \vecU_1 \right) \notin \colspan \left( \matA(y)^{(1)} \right)$ and since $\matA(y)^{(1)}$ are the last $n - r - s$ columns of $\matA(y)$, it follows that if we consider the coordinates vector $\vecX \in \bbZ_{2}^{n - r}$ of $\left( \vecU_0 - \vecU_1 \right)$ with respect to $\matA(y)$, the first $s$ elements are not $0^{s}$. By linearity of matrix multiplication it follows that if we look at each of the two coordinates vectors $\vecX_{0}$, $\vecX_{1}$ (each has $n - r$ bits) for $\vecU_{0}$, $\vecU_{1}$, respectively, somewhere in the first $s$ bits, they differ.
Now, recall how we obtain the first $s$ bits of $\vecX_{b}$ -- this is exactly by applying $\overline{\Pi}$ (the permutation on $\{ 0, 1 \}^{r + s}$ arising from $\overline{\Ps}, \overline{\Ps}^{-1}$) to $\overline{x}_{b}$ and taking the last $s$ bits of the output. Since these bits differ in the output of the permutation, then the preimages have to differ, i.e., $\overline{x}_{0} \neq \overline{x}_{1}$.

Define $\epsilon_{\AdvB}$ as the probability that the output of $\Adv$ indeed satisfies $y_{0} = y_{1} := y$ and also $\left( \vecU_0 - \vecU_1 \right) \notin \colspan \left( \matA(y)^{(1)} \right)$, and it remains to give a lower bound for the probability $\epsilon_{\AdvB}$. We do this by a sequence of hybrids, eventually showing that the view which $\AdvB$ simulates to $\Adv$ is computationally indistinguishable from a sample from $\widetilde{\gen}\left( 1^{\lambda}, n, r, k, s \right)$. More precisely, each hybrid describes a process, it has an output, and a success predicate on the output.

\begin{itemize}
    \item $\Hyb_{0}$: The above distribution $\left( \Ps, \Ps^{-1}, \Ds' \right) \gets \AdvB\left( \overline{\Ps}, \overline{\Ps}^{-1} \right)$, simulated to the algorithm $\Adv$.
\end{itemize}
The first distribution is defined in the reduction above. The output of the process is the output $(x_{0}, x_{1})$ of $\Adv$. The process execution is considered as successful if $y_{0} = y_{1} := y$ and $\left( \vecU_{0} - \vecU_{1} \right) \notin \colspan\left( \matA(y)^{(1)} \right)$.

\begin{itemize}
    \item $\Hyb_{1}$: Not applying the inner permutation $\overline{\prp}_{\sf in}$ (which comes from the circuits $\overline{\Ps}$, $\overline{\Ps}^{-1}$), by using the security of an obfuscated permutable PRP.
\end{itemize}
Let $\overline{\prp}_{\sf in}$ the (first) permutable PRP that's inside $\overline{\Ps}$ (which is the obfuscation of the circuit $\overline{P}$). In the previous hybrid we apply the $n$-bit permutable PRP $\prp\left( k_{\Gamma}, \cdot \right)$ to the input $x \in \bbZ_{2}^n$ and then proceed to apply the inner permutation $\overline{\prp}_{\sf in}\left( \overline{k}_{\sf in}, \cdot \right)$ to the first (i.e. leftmost) $r + s$ output bits of the first permutation $\prp\left( k_{\Gamma}, \cdot \right)$. The change we make to the current hybrid is that we simply apply only $\prp\left( k_{\Gamma}, \cdot \right)$.

Recall two details: (1) By the results of \cite{C:ShmZha25} on permutable PRPs being decomposable, the inner permutable PRP $\overline{\prp}_{\sf in }\left( \overline{k}_{\sf in}, \cdot \right)$ is in and of itself $\left( 2^{\poly(\secp)}, \poly(\secp) \right)$-decomposable, and (2) the circuits $P$, $P^{-1}$ which apply the permutations are both obfuscated by iO to be generate the obfuscations $\Ps$, $\Ps^{-1}$. We can treat it as a fixed permutation that acts on the output of the permutation $\Pi\left( k_{\Gamma}, \cdot \right)$ and thus it follows by Lemma by the results of \cite{C:ShmZha25} that the current and previous hybrids are computationally indistinguishable, with indistinguishability $\frac{1}{f(\kappa)}$.

\begin{itemize}
    \item $\Hyb_{2}$: For every $y \in \bbZ_{2}^{d}$, taking $\matA(y)$ to be the direct output of the PRF $\prf$, by using an obfuscated punctured PRF.
\end{itemize}
In order to describe the change between the current and previous hybrid we first recall the structure of the circuits from the previous hybrid: In the previous hybrid, for every $y \in \bbZ_{2}^{r}$ we defined $\matA(y) := \matC(y) \cdot \left( \begin{array}{cc} \overline{\matA}(y) & \\ & \Id_{n - r - s} \end{array} \right)$, where $\matC(y) \in \bbZ_{2}^{k \times k}$ comes from the output $\prf\left( k_{\matC}, y \right)$ and $\overline{\matA}(y) \in \bbZ_{2}^{(k - (n - r - s)) \times s}$ is the output of the inner PRF $\overline{\prf}\left( \overline{k}_{\sf lin} \right)$ (which in turn comes from the inside of $\left( \overline{\Ps}, \overline{\Ps}^{-1} \right)$). In the current hybrid we are going to ignore the PRFs $\prf\left( k_{\matC}, y \right)$ and $\overline{\prf}\left( \overline{k}_{\sf lin} \right)$ and their generated values $\matC(y)$, $\vecD(y)$ and $\overline{\matA}(y)$ and instead, sample a fresh key $k_{\matA}$, and on query $y$ generate $\matA(y) \gets \prf\left( k_{\matA}, y \right)$, for $\matA(y) \in \bbZ_{2}^{k \times (n - r)}$.

First, note that the following two ways to sample $\matA(y)$, are statistically equivalent for every $y$: (1) The matrix $\matA(y)$ is generated by sampling a random full-rank matrix $\matC(y) \in \bbZ_{2}^{k \times k}$ and letting $\matA(y)$ be $\matC(y) \cdot \left( \begin{array}{cc} \overline{\matA}(y) & \\ & \Id_{n - r - s} \end{array} \right)$. (2) For every $y \in \bbZ_{2}^{r}$ just sample a full-rank matrix $\matA(y) \in \bbZ_{2}^{k \times (n - r)}$. This means that when truly random bits are used for generating $\matA(y)$ in the two cases, the distributions are statistically equivalent.

To see why the two distributions are computationally indistinguishable, a different description of the previous hybrid can be given as follows: We can consider a sampler $E_{0}$ that for every $y \in \{ 0, 1 \}^d$ samples $\matA(y)$  according to the first algorithm, and another sampler $E_{1}$ that samples $\matA(y)$ according to the second algorithm, and we know that for every $y$ (and recall there are $2^r$ actual values of $y$ which can appear as the output, and not $2^d$) the outputs of $E_{0}$ and $E_{1}$ are statistically indistinguishable. 

Since there are $2^{r}$ valid values for $y$, by Lemma \ref{lem:distswap}, the current hybrid is computationally indistinguishable from the previous, with indistinguishability $\frac{2^{r}}{f(\kappa)}$.

\begin{itemize}
    \item $\Hyb_{3}$: Discarding the inner obfuscations $\left( \overline{\Ps}, \overline{\Ps}^{-1} \right)$ completely, by using the security of the outer obfuscator.
\end{itemize}
The change between the current hybrid and the previous is that in the current hybrid we generate the circuits $P, P^{-1}, D'$ without using $\left( \overline{\Ps}, \overline{\Ps}^{-1} \right)$ at all. Note that this is possible, since in the previous hybrid, we moved to a circuit that did not use access to the circuits $\left( \overline{\Ps}, \overline{\Ps}^{-1} \right)$ any longer during the execution of any of the three circuits $P, P^{-1}, D'$, except from using the second permutation $\overline{\prp}_{\sf out}$, which acts on $\{ 0, 1 \}^d$ and does not need to act from inside the inner obfuscations $\left( \overline{\Ps}, \overline{\Ps}^{-1} \right)$ any more. This means that we can technically move the application of the inner permutation $\overline{\prp}_{\sf out}$ "outside of the inner circuits $\left( \overline{\Ps}, \overline{\Ps}^{-1} \right)$" and the functionality of the circuits $P, P^{-1}, D'$ did not change between the current and the previous hybrids, and thus, by the security of the indistinguishability obfuscator that obfuscates the three circuits, the current hybrid is computationally indistinguishable from the previous one, with indistinguishability of $\frac{1}{f(\kappa)}$.

\paragraph{Finalizing the reduction.}
Finally, observe that the distribution generated in the above $\Hyb_{3}$ is exactly a sample from $\widetilde{\gen}\left( 1^{\lambda}, n, r, k, s \right)$. Also observe that the outputs of $\Hyb_{0}$ and $\Hyb_{3}$ are $O\left( \frac{2^{r}}{f(\kappa)} \right)$-computationally indistinguishable. Recall that by the lemma's assumptions, with probability $\epsilon$, on a sample from $\widetilde{\gen}\left( 1^{\lambda}, n, r, k, s \right)$, the algorithm $\Adv$ outputs a pair $(x_{0}, x_{1})$ of $n$-bit strings such that $y_{0} = y_{1} := y$ and also $\left( \vecU_0 - \vecU_1 \right) \notin \colspan \left( \matA(y)^{(1)} \right)$. It follows that the probability for the same event when the input to $\Adv$ is generated by $\Hyb_{0}$, is at least $\epsilon - O\left( \frac{2^{r}}{f(\kappa)} \right) \geq \frac{\epsilon}{2}$, which finishes our proof.
\end{proof}

%% file: standard_model_2_to_1_reduction.tex
\subsection{Hardness of the Dual-free Case from Decomposable Claw-free Functions} \label{subsection:standard_model_collision_resistance}
Our main object in this section will be trapdoor claw-free functions $\hashL$, which may either be exact (i.e. where every output of $\hashL$ has exactly $2$ preimages) or approximate/noisy. We will also need the function to be "decomposable" in some sense -- we will gradually explain these properties below.

\begin{definition} [Trapdoor Claw-free Function] \label{definition:col_resis_2_to_1}
A trapdoor claw-free function scheme is given by classical algorithms $\left( \gen, L, L^{-1} \right)$ with the following syntax.

\begin{itemize}
    \item
    $\left( \pk, \sk \right) \gets \gen\left( 1^\secp \right)$: a probabilistic polynomial-time algorithm that gets as input the security parameter $\secp \in \Nat$, and samples a public and secret key pair.

    \item
    $y \gets L\left( \pk, x \right)$: a deterministic polynomial-time algorithm that gets as input the public key $\pk$ and an input $x \in \{ 0, 1 \}^{\inSize}$ and outputs a string $y \in \{ 0,1 \}^{\outSize}$ for some $\outSize \geq \inSize - 1$.

    \item 
    $x_{b} \gets L^{-1}\left( \sk, b \in \{ 0, 1 \}, y \right)$: a deterministic polynomial-time algorithm that gets as input the public key $\pk$, a string $y \in \{ 0, 1 \}^{\outSize}$ and a bit $b$. The algorithm outputs a string $x_b \in \{ 0,1 \}^{\inSize}$ or $\bot$.
\end{itemize}

The scheme satisfies the following guarantees.

\begin{itemize}
    \item 
    {\bf 2-to-1 mapping.}
    For every public key $\pk$ in the support of $\gen\left( 1^\secp \right)$, every output in the image of $L\left( \pk, \cdot \right) : \zo^{\inSize} \rightarrow \zo^{\outSize}$ has exactly either 2 or 1 preimages. In case it has 2, then these differ in their first input bit.

    \item 
    {\bf Efficient inversion given a trapdoor.}
    For every public-secret key pair $\pk, \sk$ in the support of $\gen\left( 1^\secp \right)$ and $y \in \zo^{\outSize}$ in the image of $L\left( \pk, \cdot \right)$, the inverse function $L^{-1}\left( \sk, \cdot \right) : \left( \zo \times \zo^{\outSize} \right) \rightarrow \zo^{\inSize}$, given input $\left( b, y \right)$, guarantees the following. In case $y$ has a preimage in $L\left( \pk, \cdot \right)$ that starts with $b$ then the inverse function outputs it, and otherwise it outputs $\bot$.
    
    \item
    {\bf Collision-resistance.}
    Let $f : \Nat \rightarrow \Nat$ a function. We say that $L$ is $\left( f(\cdot), \frac{1}{f(\cdot)} \right)$-collision resistant if for every quantum algorithm running in time $f(\lambda)$, the probability for the following experiment to succeed is bounded by $1/f(\lambda)$. The experiment samples $\left( \pk, \sk \right) \gets \gen\left( 1^\secp \right)$, gives $\pk$ to the quantum algorithm, and the algorithm needs to find a collision in $L\left(\pk, \cdot \right)$.

    \item 
    {\bf Effective input size.}
    There exists a number $\eSize \in \Nat$ which we refer to as the effective input size and it satisfies $\eSize \leq \inSize - 1$. Given an input $x \in \zo^{\inSize}$ we can map it to compressed input $x^{(\eSize)} \in \zo^{\eSize}$, by discarding the first bit of $x$, taking the following $\eSize$ bits of $x$ and discarding the rest, if there are any.
    There is also a public mapping $f_{\eSize \rightarrow \inSize}\left( \pk, \cdot \right)$ which uses the public key $\pk$, such that given the first bit of $x$, the output $y$ it maps to, and its effective input $x^{(\eSize)} \in \zo^{\eSize}$, recovers the original $x \in \zo^{\inSize}$.
\end{itemize}
\end{definition}

\paragraph{The parallel-repetition hash $\hashQ$.}
Based on the above object, we construct our hash function $\hashQ$ as follows.

\begin{construction} [Folding Coset Partition Function] \label{construction:hashQ}
Let $\hashL$ a function as in Definition \ref{definition:col_resis_2_to_1}, and let $n$, $r$ natural numbers.
We sample $n - r$ public keys $\pk_{1}, \cdots, \pk_{n - r}$ and their respective trapdoors $\sk_{1}, \cdots, \sk_{n - r}$. We define two functions
$$
\hashQ : \{ 0, 1 \}^{ (n - r) \cdot \inSize } \rightarrow \left( \{ 0, 1 \}^{(n - r)\cdot \outSize} \times \{ 0, 1 \}^{n - r + \eSize} \right) \enspace ,
$$
$$
\hashQ^{-1} : \left( \{ 0, 1 \}^{(n - r)\cdot \outSize} \times \{ 0, 1 \}^{n - r + \eSize} \right) \rightarrow \{ 0, 1 \}^{ (n - r) \cdot \inSize } 
$$
, as follows.

\paragraph{The computation $\left( y, \widetilde{\vecW} \right) \gets \hashQ\left( \vecW \right)$:}
\begin{enumerate}
    \item 
    We compute the first part of the output of $\hashQ$ by partitioning the input into $n - r$ equal-sized inputs to the instances of $\hashL$:
    $$
    y := \left( y_{1}, \cdots, y_{n - r} \right) \gets 
    \hashL_{1}\left( \pk_{1}, \vecW_1 \right),
    \cdots,
    \hashL_{n - r}\left( \pk_{n - r}, \vecW_{n - r} \right)
    \enspace .
    $$

    \item \label{Q_construction_fold}
    \textbf{Fold $\vecW \in \{ 0, 1 \}^{ (n - r) \cdot \inSize }$ into $\widetilde{\vecW} \in \{ 0, 1 \}^{n - r + \eSize}$:}
    \begin{enumerate}
        \item 
        Generate a coordinates vector $\vecR \in \bbZ_{2}^{n - r}$ such that for $i \in [n - r]$, bit $i$ of $\vecR$ is the first bit of $\vecW_{i} \in \{ 0, 1 \}^{\inSize}$, which in turn is the input to $\hashL_{i}$ inside the computation of $Q$. 

        \item 
        For each $\vecW_{i} \in \{ 0, 1 \}^{\inSize}$ consider $\vecW_{i}^{(\eSize)} \in \{ 0, 1 \}^{\eSize}$ the effective input of $\vecW_{i}$.
        Then set
        $$
        \widetilde{\vecW} := \left( \vecR, \sum_{i \in [n - r]} \vecW_{i}^{(\eSize)} \right) \enspace .
        $$
    \end{enumerate}
    
    \item
    Output $\left( y, \widetilde{\vecW} \right)$.
\end{enumerate}

\paragraph{The computation $\vecW \gets \hashQ^{-1}\left( y, \widetilde{\vecW} \right)$:}
\begin{enumerate}
    \item 
    Parse $\widetilde{\vecW} := \left( \vecR \in \bbZ_{2}^{n - r}, \overline{\vecW} \in \{ 0, 1 \}^{\eSize} \right)$. Write $y = \left( y_{1}, \cdots, y_{n - r} \right)$ where $y_{i} \in \{ 0, 1 \}^{\outSize}$ for every $i \in [n - r]$. 

    \item 
    For every $j \in [n - r]$ compute $\vecW_{\left( j, \; y_{j}, \; \vecR_{j} \right)} \gets \hashL^{-1}_{j}\left( \sk_{j}, \vecR_{j}, y_{j} \right)$, where $\vecR_{j} \in \{ 0, 1 \}$ is the $j$-th bit of $\vecR$.

    \item 
    Verify match between inverted values and input: for every $j$ compute the effective input $\vecW_{\left( j, \; y_{j}, \; \vecR_{j} \right)}^{(\eSize)} \in \{ 0, 1 \}^{\eSize}$. Check that $\sum_{j \in [n - r]} \vecW_{\left( j, \; y_{j}, \; \vecR_{j} \right)}^{(\eSize)} = \overline{\vecW}$ and otherwise terminate and reject.
    
    \item 
    Output $\vecW \gets \biggl(
    \vecW_{\left( 1, \; y_{1}, \; \vecR_{1} \right)},
    \cdots, 
    \vecW_{\left( n - r, \; y_{n - r}, \; \vecR_{n - r} \right)}
    \biggr)
    $.
\end{enumerate}
\end{construction}

\paragraph{Permutation-Indistinguishability of Trapdoor Functions.}
We would like for $\hashL$ to have an additional property of decomposability. The reason we need this property is that whenever we construct the function $\hashQ$, we will be able to embed it indistinguishably within the pair of oracles $\left( \Ps, \Ps^{-1} \right)$. The property itself will be a local property we need from $\hashL$, which will ask that when we put two permutable PRPs, one before the input and one after the output of $\hashL$, and then obfuscate the entire circuit with iO, this will be indistinguishable from simply obfuscating the two PRPs, concatenated to each other.

\begin{definition} [Permutation-Indistinguishable Trapdoor Function] \label{definition:permutation_indistinguishable_tdf}
Let $\hashL$ a trapdoor function as in Definition \ref{definition:col_resis_2_to_1}. Let $\prp\left( k_{\sf in}, \cdot \right)$ an OP-PRP with domain $\zo^{\inSize}$ that's output-permutable for all $\left( 2^{\poly(\lambda)}, \poly(\lambda) \right)$-decomposable permutations. Let $\prp\left( k_{\sf out}, \cdot \right)$ an IP-PRP with domain $\zo^{\inSize + \outSize}$ that's input-permutable for all $\left( 2^{\poly(\lambda)}, \poly(\lambda) \right)$-decomposable permutations. Let $\iO$ an iO scheme. We say that $\hashL$ is $\left( f_{D}, \frac{1}{f_{D}} \right)$-permutation-indistinguishable if the following two distributions are $\left( f_{D}, \frac{1}{f_{D}} \right)$-indistinguishable, when obfuscated using the $\iO$:
\begin{itemize}
    \item
    The first distribution is given by the pair $f_{0}, f_{0}^{-1}$. The function
    $$
    f_{0} : \zo^{\inSize} \rightarrow \left( \zo^{\inSize + \outSize} \times \zo \right)
    $$
    applies $\prp\left( k_{\sf in}, \cdot \right)$, then $\hashL$, then pads with sufficiently-many zeros, and then applies $\prp\left( k_{\sf out}, \cdot \right)$. It also outputs first output bit of $\prp\left( k_{\sf in} \right)$. The function $f_{0}^{-1}$ is the inverse of $f_{0}$, which can be computed using $\pk, \sk, k_{\sf in}, k_{\sf out}$.

    \item
    The second distribution is given by the pair $f_{1}, f_{1}^{-1}$, which is basically the same as the previous pair, only that we discard the use of $\hashL$. The function $f_{1} : \zo^{\inSize} \rightarrow \left( \zo^{\inSize + \outSize} \times \zo \right)$ applies $\prp\left( k_{\sf in}, \cdot \right)$ and then $\prp\left( k_{\sf out}, \cdot \right)$. It also outputs first output bit of $\prp\left( k_{\sf in} \right)$. The function $f_{1}^{-1}$ is the inverse of $f_{1}$, which can be computed using $k_{\sf in}, k_{\sf out}$.
\end{itemize}
\end{definition}

\paragraph{Constructions of Permutation-Indistinguishable Trapdoor Claw-free Functions.}
We mention two constructions of $\hashL$ based on two different assumptions. 

\paragraph{The LWE-based hash function $\hashL$.}
Here, we recall an \emph{approximate} 2-to-1 function $\hashL$ based on LWE which is a simplified version of the \emph{noisy claw-free trapdoor function} developed in~\cite{FOCS:BCMVV18}. Let $u,v,\sigma,B,\overline{B},q$ be parameters with the relationships described in Equation~\ref{eq:lweparams}.

\begin{align}\label{eq:lweparams}
\sigma&=u^{\Omega(1)} , &
\overline{B}&=\sigma\times u^{\Omega(1)} , \\
B&\geq \overline{B}\times u^{\omega(1)} , &
q&\geq B\times u^{\Omega(1)}\nonumber , \\
v&\geq \Omega(u\log q) , &
\exists&\delta \in (0, 1) : \frac{q}{\sigma} \leq 2^{u^\delta} \enspace .
\nonumber
\end{align}

The keys for the hash function have the form $\pk = \left( \matB, \vecC \right)$, where $\matB \gets \Z_q^{v \times u}$ and $\vecC \gets \matB \cdot \vecS + \vecE \bmod q$ where $\vecS \gets \Z_q^u$ and the entries of $\vecE \in \Z_q^v$ are i.i.d. sampled from discrete Gaussians of width $\sigma$, which in turn are guaranteed (w.h.p) to have entries in $( -\overline{B}, \overline{B} ]$.

We define the function $\hashL\left( \pk, \cdot \right) : \bbZ_q^u \times (-B,B]^v \times \{0,1\} \rightarrow \bbZ_{q}^{v}$ as follows.
$$
\hashL\left( \left( \matB \in \bbZ_{q}^{v \times u}, \vecC \in \bbZ_{q}^{v} \right), \left( \vecT \in \bbZ_{q}^{u}, \vecF \in (-B,B]^v \right), b \in \{ 0, 1 \} \right) =  \matB \cdot \vecT + \vecF + b\cdot \vecC \bmod q
\enspace .
$$
By choosing $B,q$ to be powers of 2, we can map the domain and range to bit-strings.

To match the above with the interface of Definition \ref{definition:col_resis_2_to_1}, note that $\inSize := 1 + u\cdot \log(q) + v \cdot \log(B)$, $\outSize := v \cdot \log(q)$. The effective input size can actually be reduced to $\eSize = u \cdot \log(q)$, by taking it to be $\vecT$. Observe that given the public key $\left( \matB \in \bbZ_{q}^{v \times u}, \vecC \in \bbZ_{q}^{v} \right)$, the effective input $\vecT$, the first bit of the original input and also the output $\matB \cdot \vecT + \vecF + b\cdot \vecC \bmod q$, we can subtract (or not, depending on the value of the first bit of the original input) from the output the vector $\vecC$ (which is part of the public key) then also subtract $\matB \cdot \vecT$ to obtain the noise $\vecF \in (-B,B]^v$. Now, $\vecF$ is the part of the original input which isn't included in the effective input, so we obtained the full original input. In \cite{C:ShmZha25} (Theorem 58) it is proved that the above function, when instantiated with modulus $q$ and LWE noise $B$, will be $\left( \frac{q}{B}, \frac{B}{q} \right)$-permutation-indistinguishable, for a sufficiently large $f := \min\left( f_{\iO}, f_{\OWF}, f_{\PRP} \right)$ for the corresponding hardness functions of $\iO$, OWFs and permutable PRPs. Also, for $u \geq \secp$, under the $2^{ \widetilde{\Omega}\left( \lambda \right) }$-hardness LWE, our LWE-based $\hashL$ is $2^{ \widetilde{\Omega}(\lambda) }$-collision-resistant.

\paragraph{Construction based on Permutable PRPs.}
Another plausible construction uses permutable PRPs. Specifically, the function takes as input $\lambda + 1$ bits and have inside two permutable PRPs $\prp\left( k_{0}, \cdot \right)$, $\prp\left( k_{1}, \cdot \right)$, each acting on domain $\zo^{\lambda}$. The function $\hashL$, for an input $\left( b \in \zo, x \in \zo^{\lambda} \right)$, outputs $\prp\left( k_{b}, x \right)$. We do not know how to prove that this construction is collision-resistant. We note that while the construction uses only an indistinguishability obfuscator and PRPs, and while there are impossibility results for getting collision-resistance from iO and OWFs alone, may still get collision resistance as long as \emph{there exists} a reduction and (possibly additional security assumptions) that works. This was the same situation, for example, for the distribution $\left( \Ps, \Ps^{-1} \right)$ when we use the LWE-based function $\hashL$, but only later in the hybrids of the security proof.

\paragraph{Collision-resistance of the dual-free distribution $\left( \Ps, \Ps^{-1} \right)$.}
Given the above properties of our base function $\hashL$ and the structure of $\hashQ$ given $\hashL$, we are now ready to prove the collision-resistance of our functions $\Ps$, $\Ps^{-1}$.

\begin{theorem} [Collision Resistance of $\left( \Ps,\Ps^{-1} \right)$] \label{thm:reduce2lwe}
For our primitives in Construction \ref{constr:standard}, let the $\iO$ scheme be $\left( f_{\iO}\left( \cdot \right), \frac{1}{f_{\iO}\left( \cdot \right)} \right)$-secure, the puncturable PRF be $\left( f_{\PRF}\left( \cdot \right), \frac{1}{f_{\PRF}\left( \cdot \right)} \right)$-secure, the permutable PRP be $\left( f_{\PRP}\left( \cdot \right), \frac{1}{f_{\PRP}\left( \cdot \right)} \right)$-secure and let $f := \min\left( f_{\iO}, f_{\PRF}, f_{\PRP} \right)$.

Let $n, r, k \in \Nat$ such that $r < n$ and also $\frac{n}{n - r}$ is an integer (which implies that $\frac{r}{n - r}$ is also an integer). Let $n/(n - r) := \inSize$ and suppose that $n - r + \eSize \leq k$ and that $\hashL$ is a trapdoor claw-free function that's $\left( f_{\hashL}\left( \eSize \right), \frac{1}{ f_{\hashL}\left( \eSize \right) } \right)$-collision-resistant and also $\left( f_{ D }\left( \eSize \right), \frac{1}{ f_{ D }\left( \eSize \right) } \right)$-permutation-indistinguishable (as in Definition \ref{definition:permutation_indistinguishable_tdf}).

Then, getting $\left( \Ps,\Ps^{-1} \right)$ sampled from $\left( \Ps,\Ps^{-1}, \Ds \right) \gets \widetilde{\gen}\left( 1^\secp, n, r, k, 0 \right)$, and finding a collision in $\hash$ (the function derived from $\Ps$), is $\left( \min\left( f_{ \hashL }\left( \eSize \right), f_{ D }\left( \eSize \right) \right), \frac{1}{ f_{\hashL}\left( \eSize \right) } + \frac{ 2^n }{ f(\kappa) } + \frac{n - r}{ f_{D}\left( \eSize \right) } \right)$-hard.
\end{theorem}

\begin{proof}
Suppose there exists an adversary $\As$ which, given $\left( \Ps,\Ps^{-1} \right)$ as sampled from $\left( \Ps,\Ps^{-1}, \Ds \right) \gets \widetilde{\gen}\left( 1^\secp, n, r, k, 0 \right)$ (recall that $\Adv$ gets the sampled circuits without the dual $\Ds$), finds a collision in the associated function $\hash(x) := \prp^{-1}\left( k_{\sf out}, \; H(x) || 0^{d-r} \right)$, with non-negligible probability $\epsilon$, where recall that $H(\cdot)$ applies $\prp\left( k_{\sf in}, \cdot \right)$ and outputs the first $r$ bits of the output. We will describe an adversary $\AdvB$ that violates the security of $\hashL$, using the adversary $\Adv$.

Consider the function $\hashQ$ from Construction \ref{construction:hashQ}. Now, given our definition of the hash functions $\hashL$ and $\hashQ$, we are ready to describe our reduction.

\paragraph{The reduction $\AdvB$ from collision finding in $\hashQ$ to collision finding in $\hash$.}
Given $\pk = \left( \pk_{1}, \cdots, \pk_{n - r} \right)$ a public key for $\hashQ$ (where $\pk_{i}$ is the i.i.d. sampled public key for $\hashL$), and a trapdoor $\td$ which consists of the secret keys of the corresponding instances of $\hashL$ (and is used to compute the inverse $\hashQ^{-1}$) the reduction $\AdvB$ samples permutable PRP keys $k_{\sf in}$, $k_{\sf out}$ and a puncturable PRF key $k'_{\sf lin}$. Denote by $\ell := \left( n - r \right) \cdot \outSize$ the output size of $\hashQ$, and we next define the functions $P$, $P^{-1}$ which we then obfuscate to get the circuits $\Ps$, $\Ps^{-1}$, which we will feed to $\As$.

\begin{itemize}
    \item $P\left( x \in \{ 0, 1 \}^n \right)$:
    \begin{enumerate}
        \item 
        $\vecW \gets \Pi\left( k_{\sf in}, x \right)$,
    
        \item
        $\left( \vecA, \widetilde{\vecW} \right) \gets \hashQ\left( \pk, \vecW \right)$,

        \item 
        $y \gets \Pi^{-1}\left( k_{\sf out}, \left( \vecA || 0^{d - \ell} \right) \right)$,

        \item 
        $\left( \matC_y \in \bbZ_{2}^{k \times \left( n - r + \eSize \right)}, \vecD_y \in \bbZ_{2}^{k} \right) \gets \prf\left( k'_{\sf lin}, y \right)$,

        \item 
        $\vecU \gets \matC_{y} \cdot \widetilde{\vecW} + \vecD_{y}$,

        \item 
        Output $\left( y \in \{ 0, 1 \}^{d}, \vecU \in \bbZ_{2}^{k} \right)$.
    \end{enumerate}

    \item $P^{-1}\left( y \in \{ 0, 1 \}^{d}, \vecU \in \bbZ_{2}^{k} \right)$:
    \begin{enumerate}
        \item
        $x \gets
        \begin{cases}
        \Pi^{-1}\left( k_{\sf in}, \hashQ^{-1}\left( \td, y, \widetilde{\vecW} \right) \right)
        &\text{ $\exists \widetilde{\vecW} \in \bbZ_{2}^{n - r + \eSize}$ such that $\matC_y \cdot \widetilde{\vecW} + \vecD_y = \vecU$} \\
        \bot
        &\text{ if no such $\vecW$ exists}
        \end{cases}$
        
        \item
        Output
        $\begin{cases}
        x &\text{ if $x \neq \bot$ and $y = \Pi^{-1}\left( k_{\sf out}, \left( \vecA || 0^{d - \ell} \right) \right)$, for  $\vecA \gets \hashQ\left( \pk, \vecW \right)$ } \\
        \bot &\text{ otherwise }
        \end{cases}$
    \end{enumerate}
\end{itemize}

$\AdvB$ obfuscates the two circuits to get $\Ps$, $\Ps^{-1}$ and executes $\left( x_{0}, x_{1} \right) \gets \Adv\left( \Ps, \Ps^{-1} \right)$. The output of $\AdvB$ is $\left( \vecW_{0} := \Pi\left( k_{\sf in}, x_{0} \right), \vecW_{1} := \Pi\left( k_{\sf in}, x_{1} \right) \right)$ as a collision in $\hashQ$. Observe that any collision in the simulated $\left( \Ps, \Ps^{-1} \right)$ implies a collision in $\hashQ$.
Our proof has two parts. First we show that the above distribution generated by the reduction is collision resistant. In the second part we show that it is indistinguishable from the original $\Ps, \Ps^{-1}$.

\begin{lemma} [Collision-resistance of $\hashQ$, $\hashQ^{-1}$ under obfuscation]
    Assume that $\hashL$ is $\left( f(\eSize), \frac{1}{f(\eSize)} \right)$-collision resistant.
    Then, the above distribution generated by $\AdvB$ is $\left( f(\eSize) - \poly(\lambda), \frac{2\cdot (n - r)}{f(\eSize)} \right)$-collision resistant. 
\end{lemma}

\begin{proof}
The key to our proof is the ability to choose any single index $i^* \in [n - r]$ in the parallel repetition of $\hashL$ inside $\hashQ$, and simulate both functionalities $\hashQ$, $\hashQ^{-1}$ with a logically equivalent circuit, without knowing the secret key of instance $i^*$.

Formally, assume there is a quantum algorithm $\Adv$ that given $\widehat{\hashQ}$, $\widehat{\hashQ}^{-1}$ obfuscations of the functions, finds a collision in $\hashQ$ with probability $\epsilon$. We describe an algorithm $\AdvB$ that will attack the collision resistance of $\hashL$. First, note that since $\Adv$ finds a collision, if we partition the $n$-bit input to $\hashQ$ into $n - r$ packets of $n/(n - r)$ bits each, then there exists a packet $i^* \in [n - r]$ such that we get a collision for it with probability at least $\epsilon/(n - r)$.

$\AdvB$ gets $\pk^*$ a public key for $\hashL$. It samples $n - r - 1$ public-secret key pairs $\left( \pk_{1}, \sk_{1}, \cdots, \pk_{n - r}, \sk_{n - r} \right)$ of claw-free functions $\hashL_{1}, \cdots, \hashL_{n - r - 1}$. We next denote $\hashL'_{1}\left( \pk'_{1}, \cdot \right)$, $\cdots, \hashL'_{n - r}\left( \pk'_{n - r}, \cdot \right)$, where instance number $i^*$ is given by $\hashL\left( \pk^*, \cdot \right)$, that is $\pk'_{i^*} := \pk^*$.

Next, consider the functions
$$
\hashQ_{sim} : \{ 0, 1 \}^{n} \rightarrow \left( \{ 0, 1 \}^{(n - r)\cdot \outSize} \times \{ 0, 1 \}^{n - r + \eSize} \right) \enspace ,
$$
$$
\hashQ_{sim}^{-1} : \left( \{ 0, 1 \}^{(n - r)\cdot \outSize} \times \{ 0, 1 \}^{n - r + \eSize} \right) \rightarrow \{ 0, 1 \}^{n} \enspace .
$$
, defined as follows.

\paragraph{The computation $\left( y, \widetilde{\vecW} \right) \gets \hashQ_{sim}\left( \vecW \right)$:}
\begin{enumerate}
    \item 
    Identical to the original $\hashQ$ computes its output, using the functions $\left( \hashL'_{j} \right)_{j \in [n - r]}$. Note that $\hashQ$ only uses the public keys so we have the information we need in $\hashQ_{sim}$.
\end{enumerate}

\paragraph{The computation $\vecW \gets \hashQ_{sim}^{-1}\left( y, \widetilde{\vecW} \right)$:}
\begin{enumerate}
    \item 
    Parse $\widetilde{\vecW} := \left( \vecR \in \bbZ_{2}^{n - r}, \overline{\vecW} \in \{ 0, 1 \}^{\eSize} \right)$. Write $y = \left( y_{1}, \cdots, y_{n - r} \right)$ where $y_{i} \in \{ 0, 1 \}^{\outSize}$ for every $i \in [n - r]$. 
    
    \item 
    For every $j \in [n - r] \setminus \{ i^* \}$ compute $\vecW_{\left( j, \; y_{j}, \; \vecR_{j} \right)} \gets \hashL^{-1}_{j}\left( \sk_{j}, \vecR_{j}, y_{j} \right)$, where $\vecR_{j} \in \{ 0, 1 \}$ is the $j$-th bit of $\vecR$.
    
    \item 
    For every $j \in [n - r] \setminus \{ i^* \}$ let $\left( \vecW_{\left( j, \; y_{j}, \; \vecR_{j} \right)} \right)_{-1} \in \{ 0, 1 \}^{\eSize}$ which is derived by discarding the first bit of $\vecW_{\left( j, \; y_{j}, \; \vecR_{j} \right)}$ and then taking the following $\eSize$ bits (and discarding the rest).

    \item 
    Subtract to get
    $$
    \vecW^* \gets \overline{\vecW} - \sum_{ j \in [n - r] \setminus \{ i^* \} } \vecW_{\left( j, \; y_{j}, \; \vecR_{j} \right)} \enspace .
    $$

    \item 
    Obtain full input from effective input:
    $$
    \left( \vecW^*, y_{i^*} \right) \mapsto_{\pk_{i^*}} \vecW_{\left( i^*, \; y_{i^*}, \; \vecR_{i^*} \right)} \enspace .
    $$

    \item 
    Verify that $y_{i^*} = \hashL\left( \pk_{^*}, \vecW_{\left( i^*, \; y_{i^*}, \; \vecR_{i^*} \right)} \right)$ and otherwise terminate and reject.
    
    \item 
    Output $\vecW \gets \biggl(
    \vecW_{\left( 1, \; y_{1}, \; \vecR_{1} \right)},
    \cdots, 
    \vecW_{\left( n - r, \; y_{n - r}, \; \vecR_{n - r} \right)}
    \biggr)
    $.
\end{enumerate}

\paragraph{Finalizing the reduction.}
One can observe that the functionality of $\hashQ$ and $\hashQ_{sim}$ is identical, and the functionality of $\hashQ^{-1}$ is also identical to that of $\hashQ_{sim}^{-1}$, for every choice of public keys $\pk_{1}, \cdots, \pk_{n - r - 1}$ and also $\pk^*$. In particular, the choice of $i^*$ did not change the functionality at all -- only the inner workings of the functions changed, but not the . This means that under obfuscation, we have indistinguishability by the security of the iO. Now, since we get collisions on packet $i^* \in [n - r]$ with probability $\epsilon/(n - r)$, this means our collisions will be on the packet that corresponds to the input for $\hashL\left( \pk^*, \cdot \right)$. This finishes our proof.
\end{proof}

We next prove that the distribution which is simulated by the reduction is indistinguishable from the riginal dual-free, which will prove that the dual-free distribution is collision-resistant.
Let $\epsilon_{\AdvB}$ the probability that $\AdvB$ outputs a collision in $\hashQ$. We next define a sequence of hybrid experiments. Each hybrid defines a computational process, an output of the process and a predicate computed on the process output. The predicate defines whether the (hybrid) process execution was successful or not.

\begin{itemize}
    \item
    $\Hyb_{0}$: The original execution of the reduction $\AdvB$. 
\end{itemize}
Here, we get the distribution generated by the reduction and we execute $\left( \vecW_{0}, \vecW_{1} \right) \gets \AdvB\left( \widetilde{\hashQ}, \widetilde{\hashQ}^{-1} \right)$. The output of this hybrid is $\left( \vecW_{0}, \vecW_{1} \right)$ and the process is defined as successful if the pair constitutes a collision in $\hashQ\left( \pk, \cdot \right)$. By definition, the success probability of this hybrid is $\epsilon_{0} := \epsilon_{\AdvB}$.

\begin{itemize}
    \item
    $\Hyb_{1}$: Generating $\vecU$ from a coordinates vector $\vecZ$ and the coset $\left( \overline{\matA}_{\vecA}, \overline{\vecB}_{\vecA} \right)$ instead of the preimage $\vecW$, by using the trapdoor of $\hashQ$ and the security of the iO. Also, changing back to asking for collisions in the original $\Ps$.
\end{itemize}
Here, the public key $\pk$ for $\hashQ$ is sampled together with a trapdoor $\td$. The change we make to the following hybrid is this: In the previous hybrid, we took $\vecA$ and computed $y \in \{ 0, 1 \}^{d}$, then computed $\left( \matC_y, \vecD_y \right) \gets \prf\left( k'_{\sf lin}, y \right)$ and set $\vecU \gets \matC_{y} \cdot \widetilde{\vecW} + \vecD_{y}$. Here, we compute $\left( \matC_y, \vecD_y \right)$ all the same, but instead of using $\widetilde{\vecW}$ to compute $\vecU$, we use (1) the coordinates vector $\vecZ$ of $\vecW$ and the fact that it is publicly computable, (2) the output $y$ and (3) the trapdoor $\td$.


Specifically, recall that given an image $\vecA \in \{ 0, 1 \}^{(n - r) \cdot\outSize}$ of $\hashQ\left( \pk, \cdot \right)$ and the trapdoor $\td$, we can efficiently compute the coset $\left( \overline{\matA}_{\vecA}, \overline{\vecB}_{\vecA} \right)$ and also, given an input $\vecW$ we can compute $\vecZ$ the coordinates vector of $\vecW$ with respect to the coset of the output $\vecA$ (publicly with no trapdoor). 
Given $y$ we use the key $k'_{\sf lin}$ to compute $\vecA$ and then use $\td$ to compute $\left( \overline{\matA}_{\vecA}, \overline{\vecB}_{\vecA} \right)$. Now, we take $\vecZ$ the last $n - r$ bits of $\vecW$ and set $\vecU \gets \matA_{y} \cdot \vecZ + \vecD_{y}$ for $\matA_{y} := \matC_{y} \cdot \overline{\matA}_{\vecA}$, $\vecD_{y} := \vecD_{y} + \matC_{y} \cdot \overline{\vecB}_{\vecA}$.

By the correctness of the trapdoor, the functionality of the circuits did not change. Thus, the circuits given to $\Adv$ in $\Hyb_{0}$ are computationally indistinguishable by the security of the iO that obfuscates the circuits $P$, $P^{-1}$. It follows in particular that the success probability of the current process is $:= \epsilon_{1}$ such that $\epsilon_{0} \geq \epsilon_{1} - \frac{1}{f\left( \kappa \right)}$.

Another change that we make to this hybrid the definition of successful execution: Instead of asking for collisions in $\hashQ$, we ask for collisions in the original $\Ps$. Since we move back and forth between collisions in $\hashQ$ and $\Ps$ using a permutation, any collision in $\hashQ$ can be translated into a collision in $\Ps$. Formally the reduction $\AdvB$ changes minimally: After executing $\Adv$ and obtaining $\left( x_{0}, x_{1} \right)$ we do not apply $\Pi\left( k_{\sf in}, \cdot \right)$ in the end to obtain $\vecW_{0}$, $\vecW_{1}$. The success probability still satisfies $\epsilon_{0} \geq \epsilon_{1} - \frac{1}{f\left( \kappa \right)}$.

\begin{itemize}
    \item
    $\Hyb_{2}$: Discarding the trapdoor $\td$ and computing the coset $\left( \matA_{y}, \vecB_{y} \right)$ as a function of $y$ alone, using an obfuscated puncturable PRF.
\end{itemize}
The change we will make to the following hybrid is to the circuit that samples the coset $\left( \matA_{y}, \vecB_{y} \right)$. Specifically, instead of sampling the coset through the process from previous hybrid (setting $\matA_{y} := \matC_{y} \cdot \overline{\matA}_{\vecA}$, $\vecD_{y} := \vecD_{y} + \matC_{y} \cdot \overline{\vecB}_{\vecA}$ for the pair $\left( \overline{\matA}_{\vecA}, \overline{\vecB}_{\vecB} \right)$ arising from the trapdoor $\td$ and the pseudorandomly generated $\left( \matC_y, \vecD_y \right) \gets \prf\left( k'_{\sf lin}, y \right)$), we just sample a fresh puncturable PRF key $k_{\sf lin}$ and sample the coset description from scratch $\left( \matA_y \in \bbZ_{2}^{k \times (n - r)}, \vecB_y \in \bbZ_{2}^{k} \right) \gets \prf\left( k_{\sf lin}, y \right)$.

Note that in the first distribution (arising from the sampling process of $\Hyb_{1}$), the matrix $\overline{\matA}_{\vecA}$ is full rank and thus for a truly random $\left( \matC_y, \vecD_y \right)$, the pair $\left( \matA_y, \vecB_{y} \right)$ is a truly random coset. To see why the two distributions are computationally indistinguishable, a different description of the previous hybrid can be given as follows: We can consider a sampler $E_{0}$ that for every $y \in \{ 0, 1 \}^d$ samples $\matA(y)$  according to the first algorithm, and another sampler $E_{1}$ that samples $\matA(y)$ according to the second algorithm, and we know that for every $y$ (and recall there are at most $2^{n}$ actual values of $y$ which can appear as the output, and not $2^d$) the outputs of $E_{0}$ and $E_{1}$ are statistically equivalent. 

Since there are $\leq 2^{n}$ valid values for $y$, by Lemma \ref{lem:distswap}, the current hybrid is $\frac{2^n}{f(\kappa)}$-indistinguishable from the previous. It follows in particular that the success probability of the current process is $:= \epsilon_{2}$ such that $\epsilon_{1} \geq \epsilon_{2} - \frac{2^n}{f(\kappa)}$.

\begin{itemize}
    \item
    $\Hyb_{3}$: The original distribution of the obfuscated dual-free construction $\left( P, P^{-1} \right)$, using the permutation-indistinguishability of $\hashL$ (which implies the permutation-indistinguishability of $\hashQ$).
\end{itemize}
Note that once we moved to the previous hybrid, the program that's obfuscated applies the original circuit $P$ from the Construction \ref{constr:standard}, only applying $\hashQ$ in the middle, between the two permutations $\Pi\left( k_{\sf in}, \cdot \right)$ and $\Pi\left( k_{\sf out}, \cdot \right)$.

In the current last hybrid we simply output $\left( \Ps, \Ps^{-1} \right)$ for $\left( \Ps,\Ps^{-1}, \Ds \right) \gets \widetilde{\gen}\left( 1^\secp, n, r, k, 0 \right)$. That is, we disregard the execution of $\hashQ$ in the middle of the permutations. Here, we use the permutation-indistinguishability of $\hashL$ (as per Definition \ref{definition:permutation_indistinguishable_tdf}). This in turn implies the permutation-indistinguishability of $\hashQ$, and finally, by the security of the two permutable PRPs (in the input and output of $P, P^{-1}$) and also the security of the iO that obfuscates the circuits $P, P^{-1}$, the output in this hybrid is $\left( \frac{n - r}{ f_{D}\left( \eSize \right) } + \frac{n - r}{ f(\kappa) } \right)$-indistinguishable from the previous. This means in particular that the success probability in the current hybrid is $:= \epsilon_{3}$ such that $\epsilon_{3} \geq \epsilon_{2} - \left( \frac{n - r}{ f_{D}\left( \eSize \right) } + \frac{n - r}{ f(\kappa) } \right)$. This finishes our proof.
\end{proof}